\documentclass[A4paper]{article}
\usepackage{amssymb}
\usepackage{amsmath}
\usepackage{amsthm}
\usepackage{graphicx}
\usepackage{rotating}
\usepackage{stmaryrd}
\usepackage{url}
\usepackage{subfigure}
\usepackage[mathscr]{eucal}

\usepackage[margin = 2.6cm]{geometry}

\usepackage{pgf,pgfarrows,pgfnodes,pgfautomata,pgfheaps,pgfshade}
\usepackage{tikz}
\usetikzlibrary{arrows,positioning,automata}

\def\arxiv{}

\theoremstyle{plain}
\newtheorem{theorem}{Theorem}[section]

\newtheorem{lemma}{Lemma}[section]
\newtheorem{proposition}{Proposition}[section]
\newtheorem{corollary}{Corollary}[section]

\newtheorem*{apptheorem}{Theorem}
\newtheorem*{applemma}{Lemma}
\newtheorem*{appproposition}{Proposition}

\theoremstyle{definition}
\newtheorem{definition}{Definition}[section]
\newtheorem{example}{Example}[section]
\newtheorem{assumption}{Assumption}

\theoremstyle{remark}
\newtheorem{remark}{Remark}[section]
\newtheorem{remarkstar}[remark]{Remark$^{\mathbf{*}}$}
\newtheorem*{exu}{Example}
\newtheorem{scaling}{Scaling}

\newcommand{\calH}{\mathcal{H}}

\newcommand{\calA}{\mathcal{A}}
\newcommand{\calR}{\mathcal{R}}
\newcommand{\calC}{\mathcal{C}}
\newcommand{\calT}{\mathcal{T}}

\newcommand{\calD}{\mathcal{D}}
\newcommand{\calP}{\mathcal{P}}
\newcommand{\calI}{\mathcal{I}}

\newcommand{\bbN}{\mathbb{N}}
\newcommand{\bbR}{\mathbb{R}}
\newcommand{\bbP}{\mathbb{P}}
\newcommand{\bbE}{\mathbb{E}}
\newcommand{\bbZ}{\mathbb{Z}}

\newcommand{\eps}{\varepsilon}

\newcommand{\vr}[1]{\mathbf{#1}}

\newcommand{\sspace}[1]{E_{#1}}
\newcommand{\ssborel}[1]{\mathcal{B}_{#1}}
\newcommand{\vfield}[1]{F_{#1}}
\newcommand{\boundary}[1]{\partial E_{#1}}

\newcommand{\flow}[1]{\phi_{#1}}

\newcommand{\defeq}{\stackrel{\mbox{\tiny def}}{=}}

\newcommand{\norm}[1]{\hat{#1}}
\newcommand{\n}[1]{\norm{#1}}
\newcommand{\asrep}[1]{\tilde{#1}}

\newcommand{\poisson}[2]{N_{#1}\left(#2\right)}
\newcommand{\cpoisson}[2]{\hat{N}_{#1}\left(#2\right)}

\newcommand{\X}[1]{\mathbf{X}_{#1}}
\newcommand{\Y}[1]{\mathbf{Y}_{#1}}
\newcommand{\Z}[1]{\mathbf{Z}_{#1}}
\newcommand{\W}[1]{\mathbf{W}_{#1}}
\newcommand{\x}[1]{\mathbf{x}_{#1}}
\newcommand{\y}[1]{\mathbf{y}_{#1}}
\newcommand{\z}[1]{\mathbf{z}_{#1}}
\newcommand{\w}[1]{\mathbf{w}_{#1}}
\newcommand{\Xb}[1]{\norm{\mathbf{X}}_{#1}}
\newcommand{\Yb}[1]{\norm{\mathbf{Y}}_{#1}}
\newcommand{\Zb}[1]{\norm{\mathbf{Z}}_{#1}}
\newcommand{\xb}[1]{\norm{\mathbf{x}}_{#1}}
\newcommand{\yb}[1]{\norm{\mathbf{y}}_{#1}}
\newcommand{\zb}[1]{\norm{\mathbf{z}}_{#1}}
\newcommand{\Xt}[1]{\asrep{\mathbf{X}}_{#1}}

\newcommand{\xt}[1]{\asrep{\mathbf{x}}_{#1}}

\newcommand{\T}[1]{\tau_{#1}}

\newcommand{\TC}{\mathfrak{TC}}
\newcommand{\TD}{\mathfrak{TD}}
\newcommand{\TS}{\mathfrak{TS}}
\newcommand{\init}[1]{init_{#1}}

\newcommand{\Priority}[1]{\mathbf{weight}[#1]}
\newcommand{\Guard}[1]{\mathbf{guard}[#1]}
\newcommand{\Reset}[1]{\mathbf{reset}[#1]}

\newcommand{\Srate}[1]{\mathbf{rate}[#1]}

\newcommand{\priority}[1]{\mathbf{p}_{#1}}
\newcommand{\guard}[1]{\mathbf{g}_{#1}}
\newcommand{\reset}[1]{\mathbf{r}_{#1}}
\newcommand{\ratef}[1]{\mathbf{f}_{#1}}
\newcommand{\cinc}[1]{\mathbf{k}_{#1}}
\newcommand{\rinc}[1]{\mathbf{\mu}_{#1}}
\newcommand{\stoich}[1]{\mathbf{\nu}_{#1}}
\newcommand{\srate}[1]{\mathbf{\lambda}_{#1}}

\newcommand{\priorityb}[1]{\n{\mathbf{p}}_{#1}}
\newcommand{\guardb}[1]{\norm{\mathbf{g}}_{#1}}
\newcommand{\resetb}[1]{\norm{\mathbf{r}}_{#1}}
\newcommand{\ratefb}[1]{\norm{\mathbf{f}}_{#1}}

\newcommand{\stoichb}[1]{\norm{\mathbf{\nu}}_{#1}}
\newcommand{\srateb}[1]{\norm{\mathbf{\lambda}}_{#1}}

\newcommand{\sCCP}{\textbf{sCCP}}

\newcommand{\Def}{\mathsf{Def}}

\newcommand{\acts}[1]{\mathsf{action}_s(#1)}
\newcommand{\acti}[1]{\mathsf{action}_i(#1)}
\newcommand{\disc}[1]{\mathsf{disc}(#1)}
\newcommand{\cont}[1]{\mathsf{cont}(#1)}

\newcommand{\ind}[1]{\mathbf{I}\{#1\}}

\newcommand{\cdim}{n}
\newcommand{\ddim}{k}
\newcommand{\fdim}{m}

\newcommand{\dt}[1]{\mathrm{d}#1}
\newcommand{\dxdt}[2]{\frac{\mathrm{d}#1}{\mathrm{d}#2}}

\newcommand{\integral}[4]{\int_{#1}^{#2} #3 \
\mathrm{d}#4}

\newcommand{\size}[1]{\gamma_{#1}}
\newcommand{\N}{^{(N)}}

\newcommand{\Np}{^{(N)+}}

\newcommand{\h}[1]{\calH_{#1}}

\begin{document}

\ifdefined\arxiv

\title{Hybrid Behaviour of Markov Population Models}

\author{Luca Bortolussi\\
Department of Mathematics and Geosciences\\
University of Trieste, Italy.\\
CNR/ISTI, Pisa, Italy.\\
\texttt{luca@dmi.units.it}
}

\date{}

\maketitle


\begin{abstract}
We investigate the behaviour of population models written in  Stochastic Concurrent
Constraint Programming (\sCCP), a stochastic extension of Concurrent Constraint Programming. In particular, we focus on models from which we can define a semantics of \sCCP\ both in terms of Continuous Time Markov Chains (CTMC) and in terms of Stochastic Hybrid Systems, in which some populations are approximated continuously, while others are kept discrete. We will prove the correctness of the hybrid semantics from the point of view of the limiting behaviour of a sequence of models for increasing population size.  More specifically, we prove that, under suitable regularity conditions, the sequence of CTMC constructed from \sCCP\
programs for increasing population size
converges to the hybrid system constructed by means of the hybrid semantics. We investigate in particular what happens for \sCCP\ models in which some transitions are guarded by boolean predicates or in the presence of instantaneous transitions.

\


\end{abstract}

\else

\begin{frontmatter}

\title{Hybrid Behaviour of Markov Population Models}

\author[TS,PI]{Luca Bortolussi}
\address[TS]{Department of Mathematics and Geosciences\\
University of Trieste, Italy.}
\address[PI]{CNR/ISTI, Pisa, Italy.}
\ead{luca@dmi.units.it}


\begin{abstract}
We investigate the behaviour of population models written in  Stochastic Concurrent
Constraint Programming (\sCCP), a stochastic extension of Concurrent Constraint Programming. In particular, we focus on models from which we can define a semantics of \sCCP\ both in terms of Continuous Time Markov Chains (CTMC) and in terms of Stochastic Hybrid Systems, in which some populations are approximated continuously, while others are kept discrete. We will prove the correctness of the hybrid semantics from the point of view of the limiting behaviour of a sequence of models for increasing population size.  More specifically, we prove that, under suitable regularity conditions, the sequence of CTMC constructed from \sCCP\
programs for increasing population size
converges to the hybrid system constructed by means of the hybrid semantics. We investigate in particular what happens for \sCCP\ models in which some transitions are guarded by boolean predicates or in the presence of instantaneous transitions.
\end{abstract}


\end{frontmatter}

\fi

%

\section{Introduction}\label{sec:intro}

Stochastic Process Algebras (SPA) are a powerful framework for quantitative modelling and
analysis of population processes~\cite{PA:Hillston:1996:CompositionalPerformanceModeling}.
They have been applied in a wide varieties of contexts, including computer systems \cite{PA:Hillston:1996:CompositionalPerformanceModeling}, biological systems \cite{SB:HillstonCiocchetta:2009:bioPEPA} \cite{My2008ConstraintsBioModelingSCCP} \cite{SB:Cardelli:2006:ChemicalPiCalculus}, ecological 
\cite{SB:Sumpter:2001:IndividualToPopulation} and crowd 
\cite{PA:HillstonMassink:FACS:2012:CrowdDynamics} modelling.

However, their standard semantics, given in terms of Continuous Time Markov Chain (CTMC,~\cite{STOC:Norris:1997:MarkovChains}), suffers
from the problem of state space explosion, which impedes the use of SPA to analyse models with a large state space. A recent technique introduced to tackle
this problem is fluid-flow approximation~\cite{PA:Hillston:2005:ODEandPEPA}, which describes
the number of system components by means of continuous variables and interprets rates as flows, thus providing a semantics in terms of Ordinary Differential Equations (ODE). 

The relationship between these two semantics is grounded on the law of large numbers for population processes \cite{STOC:Kurtz:1981:ApproxPopProc}, first proved by Kurtz back in the seventies \cite{STOC:Kurtz:1970:ODEandCTMC}.
Applying this theoretical framework to SPA models, one obtains that the fluid-flow ODE is the limit of the sequence of CTMC models
 \cite{PA:Tribastone:2012:FluidPEPA, My2009TCSBjournalSCCPandODE, PA:BradleyHaiden:2010:FluidFrameworkPEPA}, obtained by the standard SPA semantics for increasing system size, usually the total number of agents in the system. This also provides a link with a large body of mathematical literature on fluid and mean field approximation (see e.g. \cite{MyPEVAtutorial2012} for a recent review). 

These results provide the asymptotic correctness of the fluid semantics and justify the use of ODEs to analyse large collective SPA models. Fluid approximation is also entering into the analysis phase in a more refined way than just by numerical simulation. For instance, in \cite{PA:Hayden:2012:passageTimes}, the authors use fluid approximation for the computation of passage-times, while in \cite{My:CONCUR2012:FMC} the fluid approximation scheme is used to model check properties of single agents in a large population against CSL properties. 

Despite the remarkable success of fluid approximation of SPA models, its applicability is restricted to situations in which all components are present in large quantities, and all events depend continuously on the number of the different agent types. This excludes many interesting situations, essentially all those in which some sub-populations have a fixed and small size. This is the case in biological systems, when one considers gene networks, but also in computer systems when one models some form of centralized controller. Furthermore, the description of control policies is often simplified by using forced (or instantaneous) events, happening as soon as certain condition are met, and more generally guard predicates, modulating the set of enabled actions as a function of the global state of the system. 

These features of modelled systems are not easily captured in a fluid flow scheme, as they lead naturally to hybrid systems, in which discrete and continuous dynamics coexist. To deal with these situations, in \cite{My2010TCShybridDynamicsStochProg, My2011LogicsCSjournalHybridSCCPLattice} the authors proposed a hybrid semantics for a specific SPA, namely  stochastic Concurrent Constraint Programming
(\sCCP,~\cite{My2008ConstraintsBioModelingSCCP}), associating with a \sCCP\
model a hybrid system where continuous dynamics is interleaved with discrete Markovian jumps. In \cite{MyNETMAHIB2012}, also instantaneous transitions are incorporated in the framework.
In this way, one can circumvent the limits of fluid-flow
approximation, whilst keeping discrete only the portions of the system that cannot be safely described as continuous.
Roughly speaking, this hybrid semantics works by first identifying a subset of system variables 
to be approximated continuously, keeping discrete the remaining ones. The latter set of variables identifies the  discrete skeleton of the hybrid system, while the former defines the continuous state space. Then, each activity of agents, corresponding to a transition that modifies the state of the system, is classified as continuous, discrete/stochastic, or discrete/instantaneous. The first class of transitions is used to construct a vector field giving the continuous dynamics of the hybrid system (in each mode), while the other two transition classes define the discrete dynamics. 

The advantages of working with a hybrid semantics for  SPA are mainly rooted in the speed-up that can be achieved in the simulation, as discussed e.g. in \cite{My2011LogicsCSjournalHybridSCCPLattice} and \cite{SB:Pahle:2009:HybridSimSurvey}. Moreover, the hybrid semantics put at disposal of the modeller a broader set of analysis tools, like transient computation \cite{PA:Trivedi:1993:FSPN} or moment closure techniques \cite{SB:Hespanha:2006:momentClosure, STOC:Huisinga:HybridMasterEquation:2012}.

While the theory of deterministic approximation of CTMC is well developed, hybrid approximation has attracted much less attention. To the author's knowledge, the preliminary work \cite{My2010ASMTAkurtzForPDP} on which this paper is based was the first attempt to prove hybrid convergence results in a formal method setting. There has been some previous work on hybrid limits in \cite{SB:Kurtz:2006:asymptoticMultiscale}, restricted however to a specific biological example, and in the context of large deviation theory \cite{STOC:Shwartz:1995:LargeDeviationsPA}, where deterministic approximation of models with level variables has been considered (but in this case transitions between modes are fast, so that the discrete dynamics is always at equilibrium in the limit).
More recent work is \cite{STOC:Radulescu:2012:HybridConvergence}, which discusses hybrid limits for genetic networks (essentially the class of models considered in \cite{My2010ASMTAkurtzForPDP} with some extensions). 

The focus of this paper is to provide a general framework to infer consistence of hybrid semantics of SPA models in the light of asymptotic correctness. In doing this, we aimed for generality, proving hybrid limit theorems for a framework including instantaneous events, with guards possibly involving model time, random resets, and guards in continuous and stochastic transitions. 
The goal was to identify a broad set of conditions under which convergence holds, potentially usable in static analysis algorithmic procedures that check if a given model satisfies the conditions for convergence. We will comment on this issue in several points in the paper. To author's knowledge, this is the first attempt to discuss hybrid approximation in such generality.


We will start our presentation recalling \sCCP\ (Section \ref{sec:sCCP}) and the hybrid semantics (Section \ref{sec:sCCPtoTDHA}). We will formally define it in terms of Piecewise Deterministic Markov Processes (Section \ref{sec:PDMP}, \cite{STOC:Davis:1993:PDMP}), a  class of Stochastic Hybrid Processes in which the continuous dynamics is given in terms of Ordinary Differential Equations, while the discrete dynamics is given by forced transitions (firing as soon as their guard becomes true) and by Markovian jumps, firing with state dependent rate. The hybrid semantics is defined by introducing an intermediate layer in terms of an automata based description, by the so-called Transition-Driven Stochastic Hybrid Automata (TDSHA, Sections \ref{sec:TDHA} and \ref{sec:TDSHAtoPDMP}, \cite{My2010TCShybridDynamicsStochProg, My2011LogicsCSjournalHybridSCCPLattice}). 

After presenting the classic fluid approximation result, recast in our framework (Section \ref{sec:fluidLimit}), we turn our attention to \sCCP\ models that are converted to TDSHA containing only discrete/stochastic and continuous transitions, with no guards and no instantaneous transitions, but allowing random resets (general for discrete/stochastic transitions and restricted for continuous ones). In Section \ref{sec:hybridLimits}, we prove a limit theorem under mild consistency conditions on rates and resets, showing that the sequence of CTMC associated with a \sCCP\ program, for increasing system size, converges to the limit PDMP in the sense of weak convergence. Technically speaking, the appearance of weak convergence instead of convergence in probability, in which classic fluid limit theorems are usually stated, depends on the fact that the limit process is stochastic and can have discontinuous  trajectories.

We then turn our attention to the limit behaviour in the presence of sources of discontinuity, namely instantaneous transitions (Section \ref{sec:InstantaneousTransitions}) or guards in continuous (Sections \ref{sec:KurtzPWS} and \ref{sec:guardsContinuousTrans}) or discrete/stochastic transitions (Section \ref{sec:guardsDiscreteStochastic}).
 
In all these cases, the situation is more delicate and the conditions for convergence are more complex. Guards in continuous transitions introduce discontinuities in the limit vector fields, requiring us to define the continuous dynamics in terms of the so-called piecewise-smooth dynamical systems \cite{HA:Cortes:2008:DiscontinuousDynamicalSystems} or, more generally, in terms of  differential inclusions \cite{THMAT:AubinCellina:1984:DiffInclusion}. Here, however, we can exploit recent work in this direction \cite{MyQEST2011, STOC:Gast:2010:DifferentialInclusionMeanField}, and the hybrid convergence theorem extends easily, provided we can guarantee global existence and uniqueness of the solutions of the discontinuous differential equations. 
 
The situation with guards for discrete/stochastic transitions and with instantaneous events is even more delicate: subtle interactions between the continuous dynamics and the surfaces in which guards can  change truth status (called discontinuity or activation surfaces in the paper) can break convergence. We discuss this in detail first for instantaneous transitions (Section \ref{sec:InstantaneousTransitions}) and then for guards in discrete/stochastic transitions (Section \ref{sec:guardsDiscreteStochastic}). In these sections, we identify regularity conditions to control these subtle interactions, extending the convergence also to this setting. 
However, checking these conditions is more complicated, because they essentially impose restrictions on the global interactions between the vector fields and the discontinuity surfaces. 
%
%
%
A way out of this problem, hinted in the conclusions (Section \ref{sec:conc}) is to increase the randomness in the system by adding noise on resets and initial conditions or on the continuous trajectories, i.e. considering hybrid limits with continuous dynamics given by Stochastic Differential Equations or Gaussian Processes \cite{STOC:Kurtz:1986:MarkovProcesses}.
In the conclusions we will also comment on the applicability of our results to the stationary behaviour of the CTMC. Throughout the paper, starred remarks  contain material that can be skipped at a first reading.

\section{Preliminaries}\label{sec:preliminaries}
In this section, we introduce preliminary concepts needed in the following. We will start in Section \ref{sec:sCCP} by presenting \sCCP, the modelling language that will be used in the paper. We will then introduce Transition-Driven Stochastic Hybrid Automata (TDSHA, Section \ref{sec:TDHA}), an high level formalism to model the limit hybrid processes of interest, namely Piecewise Deterministic Markov Processes (PDMP, Section \ref{sec:PDMP}).  Finally, we will consider also how to define a hybrid semantics for \sCCP\ by syntactically transforming a \sCCP\ model into a TDSHA (Section \ref{sec:sCCPtoTDHA})
 and a TDSHA into a PDMP (Section \ref{sec:TDSHAtoPDMP}).

\subsection{Stochastic Concurrent Constraint Programming}\label{sec:sCCP}

We briefly present  stochastic Concurrent Constraint Programming (\sCCP, a stochastic extension of
CCP~\cite{PA:Saraswat:1993:CCP}). In the following we just sketch the basic notions and the concepts needed in the rest of the paper. More details on the language can be found in~\cite{PA:Bortolussi:2006:sCCP, 
My2008ConstraintsBioModelingSCCP}. 

\sCCP\ programs are defined by a set of agents interacting asynchronously and exchanging information through a shared memory called the\emph{ constraint store}. The constraint store consists of a set of variables plus a set of constraints, which are first order predicates restricting the admissible domain of variables. By adding constraints to the store, agents refine the available information. In this paper, we consider a restricted notion of constraint store, containing only \emph{stream variables}, i.e. variables ``\emph{a l\`a Von Neumann}'' which have a single value at any given time, and can be updated during the computation\footnote{Formally, one can view these variables as list, so that new values are appended at the end of the list, see~\cite{My2008ConstraintsBioModelingSCCP} for further details.}.
We further restrict the language by forbidding local variables. This restricted version of \sCCP has proved to be sufficiently expressive, compact, and especially easy to manipulate for our purposes, in particular for what concerns the definition of the fluid  \cite{My2009TCSBjournalSCCPandODE} and the hybrid semantics \cite{My2010TCShybridDynamicsStochProg, My2011LogicsCSjournalHybridSCCPLattice}. 
In this paper, however, we enlarge the primitives at our disposal with respect to \cite{My2009TCSBjournalSCCPandODE, My2011LogicsCSjournalHybridSCCPLattice}, as done in \cite{MyNETMAHIB2012}, by including also instantaneous transitions, random resets, and environment variables (which can take values in an uncountable set).

\begin{definition}\label{def:sCCP}
A \sCCP\ program is a tuple $\calA = (A,\Def,\vr{X},\calD,\vr{x_0})$,
where
\begin{enumerate}
\item The \emph{initial network of agents} $A$
and the \emph{set of definitions} $\Def$ are given by  the
following grammar:
$$\begin{array}{l}
\Def =  \emptyset~|~\Def \cup \Def~|~\{C\defeq M\}  \\
A =  \mathbf{0}~|~C~|~A\parallel A \ \hspace{2cm} M = \pi.A~|~M+M\\
\pi =  [g(\vr{X}) \rightarrow u(\vr{X},\vr{X'},\W{})]_{\lambda(\vr{X})}\  |\ [g(\vr{X}) \rightarrow u(\vr{X},\vr{X'},\W{})]_{\infty:p(\vr{X})}
\end{array}$$
\item $\vr{X}$ is the set of stream
variables of the store (with global scope). A variable $X\in \vr{X}$ takes values in $\calD_X$. Variables are divided into two classes: \emph{model} or \emph{system variables} whose domain $\calD_X$ has to be a countable subset of $\bbR$ (usually the integers), and \emph{environment variables}, whose domain can be the whole $\bbR$. 
The state space of the model is therefore $\calD = \prod_{X\in \vr{X}} \calD_X$;
\item $\vr{x_0} \in \calD$ is the \emph{initial value} of store variables.
\end{enumerate}
\end{definition}

System variables usually describe the number of individuals of a given population, like the number of molecules in a  biochemical mixture or the number of jobs waiting in a queue. Environment variables, on the other hand, are useful to describe properties of the ``environment'', like the temperature of a biochemical system, or the value of a controllable parameter that may change at run-time. Examples of the use of environment variables will be given in Section \ref{sec:hybridLimits}.

In the previous definition, a basic action $\pi$ (called throughout the paper also \emph{event} or \emph{transition}) is a \emph{guarded update}
of (some of the) store variables. In particular: 
\begin{itemize}
\item the \emph{guard} $g(\vr{X})$ is a quantifier-free first order formula whose atoms are inequality predicates on variables $\vr{X}$;
\item the \emph{update} $u(\vr{X},\vr{X'})$ is a predicate on $\vr{X},\vr{X'}$, a conjunction of \emph{atomic updates} of the form $\bigwedge X' = r(\vr{X},\W{})$ (where $X'$ denotes variable $X$ after the update), where each variable $X'$ appears only once. Here $r$ is a function with values in $\calD_X$, and can depend on the  store variables $\vr{X}$ and on a \emph{random} vector $\W{}$ in $\bbR^h$ (for some $h>0$), which can also depend on the current state of variables $\X{}$. Updates will be referred to also as \emph{resets}.
\item The \emph{rate function} $\lambda:\calD\rightarrow \bbR_{\geq 0}$ is the (state dependent) rate of the exponential distribution associated with $\pi$, which specifies the \emph{stochastic duration} of $\pi$;
\item if, instead of $\lambda$, an action $\pi$ is labelled by $\infty:p(\vr{X})$, it is an instantaneous action. In this case, $p:\calD\rightarrow\bbR_{\geq 0}$ is the \textit{weight} function associated with the action.
\end{itemize}
Updates can be seen as (random) functions from $\calD$ to itself, and they can be very general. However, in the following we will need to restrict them in order to define the fluid semantics. An atomic reset is a \emph{constant increment update} if it is of the form $X' = X + k$, with $k\in\bbR$ such that $X' \in \calD_X$ whenever $X\in\calD_X$ (usually $X,k\in\bbZ$) and it is a \emph{random increment update} if it is of the form $X' = X + \mu$, with $\mu$ a random number, such that $|\mu|$ has \emph{finite} expectation. An update is a constant/random increment if all its atomic updates are so.

\begin{example}
\label{ex:clientServerBasic}

We introduce now a simple example that will be used for illustrative purposes throughout the paper.
We will consider a simple client-server system, consisting of a population of clients which request a service and, after having obtained an answer, process it for some time before asking for another service, in a loop. The servers, instead, reply to client's request at a fixed rate. We ignore any internal behaviour of servers, for simplicity.  However, servers can break down and need some time to be repaired. 
We can model such system  in \sCCP\ by using 4 variables, two counting the number of clients requesting a service ($X_r$) and processing data ($X_t$), and two modelling the number of idle servers ready to reply to a request ($X_i$) and the number of broken servers ($X_b$). The initial network is then \texttt{client} $\parallel$ \texttt{server}, with initial conditions $X_r = X_b = 0$, $X_t = N_1$, and $X_i=N_2$.  The \texttt{client} and \texttt{server} agents are defined as follows ($*$ stands for true):
\begin{center}
{\tt
\begin{tabular}{lcl}
    client   & $\defeq$ & $[*\rightarrow X_r '= X_r - 1 \wedge X_t '= X_t + 1]_{\min\{k_r X_r, k_s X_i\}}$.client +\\
 & & $[* \rightarrow  X_r '= X_r + 1 \wedge X_t '= X_t - 1]_{k_t X_t}$.client\\
    server & $\defeq$ & $[*\rightarrow X_i '= X_i - 1 \wedge X_b '= X_b + 1]_{k_b X_i}$.server\\
    & + & $[*\rightarrow X_i '= X_i + 1 \wedge X_b '= X_b - 1]_{k_f X_b}$.server
\end{tabular}}
\end{center}

Note in the previous code how the rate at which information is processed by clients corresponds to the global rate of observing an agent finishing its processing activity. Observe also that we defined the service rate as the minimum between the total request rate of clients and the total service rate of servers. This use of minimum is consistent with the bounded capacity interpretation of queueing theory and of the stochastic process algebra PEPA \cite{PA:Hillston:1996:CompositionalPerformanceModeling}. 
This global interaction-based modelling style is typical of \sCCP, see~\cite{My2008ConstraintsBioModelingSCCP} for a discussion in the context of systems biology.
Furthermore, although we want to keep all variables $\geq 0$, we are not using any guard in the transitions. However, non-negativity is automatically ensured by rates, which, by being equal to zero, disallow transitions that would make one variable negative.
\end{example}

In order to simplify the definition of the fluid and hybrid semantics, we will work with a restricted subclass of \sCCP\ programs, that we will call \emph{flat}. A flat \sCCP\ program satisfies the following two additional restrictions: (a) each component $C$ is of the form $C = \pi_1.C + \ldots + \pi_h.C$, i.e. it always calls itself recursively, and (b) the initial network $A$ is the parallel composition of all components, i.e. $A = \parallel_{C\in\Def} C$. Note that the client-server model of Example \ref{ex:clientServerBasic} is flat.
 
The requirement of being flat is not a real restriction, as each \sCCP\ program respecting Definition \ref{def:sCCP} can be turned into an equivalent flat one, by adding fresh variables counting how many copies of each component $C$ are in parallel in the system. Guards, resets, rates and priorities have to be modified to update consistently these new variables. (see Appendix \ref{app:sccp} for an example)

In the following definitions, we will always suppose to be working with flat \sCCP\ models, possibly obtained by applying the flattening recipe. Given a (flat) \sCCP\ model $\calA = (A,\Def,\vr{X},\calD,\vr{x_0})$, we will denote by $\acts{C}$ the set of stochastic actions  of a component $C$ and by $\acti{C}$ the set of its instantaneous actions.
We will use the following notation:
\begin{itemize}
\item For an action $\pi\in\acts{C}\cup\acti{C}$, we denote by $\Guard{\pi}(\vr{X})$ or $\guard{\pi}(\vr{X})$ its guard.
\item  For an action $\pi\in\acts{C}\cup\acti{C}$, we denote by $\Reset{\pi}(\vr{X},\W{})$ or $\reset{\pi}(\vr{X},\W{})$ its update function (so that $\vr{X'} = \reset{\pi}(\vr{X},\W{})$).
\item For an action $\pi\in\acts{C}\cup\acti{C}$, if $\pi$ has a constant increment update, we will denote the increment vector by $\cinc{\pi}$ (so that $\vr{X'} = \vr{X} + \cinc{\pi}$), while if $\pi$ has a random increment update, we will denote it by $\rinc{\pi}$. 
We also let $\stoich{\pi}$ be either $\cinc{\pi}$ or $\rinc{\pi}$.  
\item  For an action $\pi\in\acts{C}$, we denote by $\Srate{\pi}$ or $\srate{\pi}$ its rate function.
\item For an action $\pi\in\acti{C}$, we denote by $\Priority{\pi}$ or $\priority{\pi}$ its weight.
\end{itemize}

A \sCCP\ program with all transitions stochastic can be given a standard semantics in terms of Continuous Time Markov Chains, in the classical Structural Operational Semantics style, along the lines of \cite{My2008ConstraintsBioModelingSCCP}. For a flat \sCCP\ model, the derivation of the labelled transition system (LTS) is particularly simple. 
First, the state space of CTMC corresponds to the domain $\calD$ of the \sCCP\ variables. Secondly, each stochastic action $\pi\in \acts{C}$ of a component $C$ defines a set of edges in the LTS. In particular, if in a point $\x{}$ it holds that $\guard{\pi}(\x{})$ is true and $\bbP\{\reset{\pi}(\x,\W{}) = \y{}\} = p_{\y{}} >0$,
then we have a transition from $\x{}$ to $\y{}$ with rate $p_{\y{}}\srate{\pi}(\x{})$. As customary, the rates of the edges of the LTS connecting  the same pair of nodes are summed up to get the corresponding rate in the CTMC. Instantaneous transitions, on the other hand, can be dealt with in the standard way as in \cite{PA:Balbo:1995:ModelingGSPN}, by partitioning states of $\calD$ into vanishing (in which there is an active instantaneous transition) and non-vanishing (in which there is no active instantaneous transition), and removing vanishing states in the LTS, solving probabilistically any non-deterministic choice between instantaneous transitions with probability proportional to their weight. 

We will indicate by $\X{}(t)$ the state at time $t$ of the CTMC associated with a \sCCP\ program $\calA$ with variables $\X{}$.

If all transitions of a \sCCP\ program are stochastic and have constant increment updates, they can be interpreted as flows, and a fluid semantics can be defined \cite{My2009TCSBjournalSCCPandODE}.
However, to properly deal with random resets and instantaneous transitions, it is more convenient to consider a more general semantics for \sCCP, in terms of stochastic hybrid automata \cite{My2010TCShybridDynamicsStochProg, My2011LogicsCSjournalHybridSCCPLattice, MyNETMAHIB2012}. This approach will also allow us to partition variables and transitions into discrete and continuous, so that only a portion of the state space will be approximated as fluid.

    \subsection{Transition-driven Stochastic Hybrid
Automata}\label{sec:TDHA}

\emph{Transition-Driven Stochastic Hybrid Automata}
(TDSHA, \cite{My2010TCShybridDynamicsStochProg, My2011LogicsCSjournalHybridSCCPLattice}) has
proved to be a convenient intermediate formalism to associate a Piecewise Deterministic Markov Process
with a \sCCP\ program. The emphasis of TDSHA is on \emph{transitions} which, as
always in hybrid automata, can be either discrete (corresponding to jumps) or continuous (representing flows acting on system variables). 
Discrete transitions can be of two kinds: either stochastic, happening at random jump times (exponentially distributed), or instantaneous, happening as soon as their guard becomes true. 

In this paper, we consider a slight variant of TDSHA, in which discrete modes of the automaton are described implicitly by a set of discrete variables (variables taking values in a discrete set), rather than explicitly. This syntactic variant is similar to the one used in \cite{My:QEST2010:hybridPEPA}, and is introduced in order to simplify the mapping from flat \sCCP\ models. 

\begin{definition}\label{def:TSHS}
A Transition-Driven Stochastic Hybrid Automaton (TDSHA) is a tuple\\
$\calT = (\vr{Z},Q,\vr{Y},\TC,\TD,\TS,\init{})$, where:
\begin{itemize}
\item $\vr{Z} = \{Z_1,\ldots,Z_k\}$ is the set of discrete variables, taking values in the countable set $Q\subset \bbR^k$. Each value $q\in Q$, $q=(z_1,\ldots,z_k)$ is a \emph{control mode} of the automaton. 
\item $\vr{Y} = \{Y_1,\ldots,Y_n\}$ is a set of real valued \emph{system variables}, taking values in $\bbR^n$. We let $\vr{X} = \vr{Z}\cup\vr{Y}$ be the vector of TDSHA variables, of size $m=k+n$.\footnote{Notation: the time
  derivative of $Y_j$ is denoted by $\dot{Y}_j$, while the value
of $Y_j$ after a change of mode is indicated by $Y_j'$.}

\item $\TC$ is the multi-set\footnote{Multi-sets are needed to take into account the proper multiplicity of transitions.} of \emph{continuous transitions or flows},
containing tuples $\eta=(\vr{k},f)$, where $\vr{k}$ is a real vector of size $m$ (identically zero on components corresponding to $\vr{Z}$), and
$f:Q\times\bbR^n \rightarrow \bbR$ is a piecewise continuous function for each fixed $q\in Q$ (usually, but not necessarily, Lipschitz continuous\footnote{A function $f:\bbR^m \rightarrow \bbR$ is Lipschitz continuous if and only if there is a constant $L>0$, such that $\| f(\vr{x_1}) - f(\vr{x_2})\| \leq L \|\vr{x_1} - \vr{x_2}\|$ }). We will denote them by  $\stoich{\eta}$, and
$\ratef{\eta}$, respectively.

\item $\TD$ is the multi-set of \emph{discrete or instantaneous transitions},
whose elements are tuples $\eta=(G,R,p)$, where: 
$p:Q\times\bbR^n\rightarrow \bbR_{\geq 0}$ is a \emph{weight} function used to resolve non-determinism between two or more active transitions,  $G$ is the \emph{guard}, a quantifier-free first-order formula with free variables among $\vr{X}$, and $R$ is the \emph{reset}, a conjunction of atoms of the form $X' = r(\X{},\W{})$, where $r:Q\times\bbR^n\times \bbR^h\rightarrow \bbR$, is the reset function of $X$, depending on variables $\vr{X}$ as well as on a random vector $\W{}$ in $\bbR^h$.  
Note that the guard can depend on discrete variables, and the reset can change the value of discrete variables $\vr{Z}$. 
In the following, we will interpret the reset as a vector of $k+n$ functions, $R:Q\times\bbR^n\times \bbR^h\rightarrow Q\times\bbR^n$, equal to  $X' = r(\X{},\W{})$ in the component corresponding to $X$ if $X' = r(\X{},\W{})$, and equal to the identity function for all those variables $X$ unchanged by the reset. 
The elements of a tuple $\eta$ are indicated by  $\guard{\eta}$, $\reset{\eta}$, and $\priority{\eta}$, respectively.

\item $\TS$ is the multi-set of \emph{stochastic transitions}, whose
elements are tuples $\eta = (G,R,\lambda)$, where $G$ and $R$ are as for transitions in $\TD$, while
$\lambda:Q\times\bbR^n\rightarrow \bbR^+$ is a function giving the state-dependent rate of the transition. Such a function is indicated by $\srate{\eta}$.

\item $\init{} = (\vr{z_0},\vr{y_0}) \in Q\times\bbR^n$ is the \emph{initial state} of the system.
\end{itemize}
\end{definition}

A TDSHA has three types of transitions. 
Continuous transitions represent flows and, for each $\eta \in \TC$, $\stoich{\eta}$ and
$\ratef{\eta}$   give the \emph{magnitude} and the \emph{ form} of the flow of $\eta$ on each variable $Y\in\vr{Y}$, respectively (see also Section~\ref{sec:TDSHAtoPDMP}). 
Instantaneous transitions $\eta\in\TD$, instead, are executed as soon as their guard $\guard{\eta}$ becomes true. When they fire, they can reset both discrete and continuous variables, according to the reset policy $\reset{\eta}$, which can be either deterministic or random. Weight $\priority{\eta}$ is used to resolve probabilistically the simultaneous activation of two or more instantaneous transitions, by choosing one of them with probability proportional to $\priority{\eta}$.
Finally, stochastic transitions $\eta\in\TS$ happen at a specific rate $\srate{\eta}$, given that their guard $\guard{\eta}$ is true and they can change system variables according to reset $\reset{\eta}$. Rates define a random race in continuous time, giving the delay for the next spontaneous jump.
\\
The dynamics of TDSHA will be defined in terms of PDMP, see Section~\ref{sec:TDSHAtoPDMP}
or~\cite{My2010TCShybridDynamicsStochProg, My2011LogicsCSjournalHybridSCCPLattice} for a more
detailed discussion.

\paragraph{Composition of TDSHA.}
We consider now an operation to combine two TDSHA with the same vectors of discrete and continuous variables, by taking the union of their transition multi-sets. 
Given two TDSHA $\calT_1 =
(\vr{Z},Q,\vr{Y},\TC_1,\TD_1,\TS_1,\init{})$ and $\calT_2 =
(\vr{Z},Q,\vr{Y},\linebreak\TC_2,\TD_2,\TS_2,\init{})$, agreeing on discrete and continuous variables and  on the initial state, their composition $\calT = \calT_1 \uplus \calT_2$ is simply $\calT =
(\vr{Z},Q,\vr{Y},\TC_1\cup\TC_2,\TD_1\cup\TD_2,\TS_1\cup\TS_2,\init{})$, where the union symbol $\cup$ refers to union of multi-sets.

    \subsection{From sCCP to TDSHA}\label{sec:sCCPtoTDHA}

In this section we recall the definition of the semantics for \sCCP\ in terms of TDSHA~\cite{My2011LogicsCSjournalHybridSCCPLattice}. 
We will assume to work with flat \sCCP\ models, so that we can ignore the structure of agents and focus our attention on system variables. In this respect, this approach differs from the one of \cite{My2011LogicsCSjournalHybridSCCPLattice}, but it provides a more homogeneous treatment.

The mapping proceeds in three steps. First we will partition variables into discrete and continuous. Then, we will convert each component into a TDSHA, and finally we will combine these TDSHA by the composition construction defined in the previous section.

The first step is to consider a flat \sCCP\ model $\calA = (A,\Def,\vr{X},\calD,\vr{x_0})$, and partition its set of variables $\X{}$. Recall that variables $\X{}$ are divided into model variables $\X{m}$ and environment variables $\X{e}$. Model variables $\X{m}$ are partitioned into two subsets: $\X{d}$, to be kept discrete, and $\X{c}$, to be approximated continuously. Hence $\X{} = \X{d} \cup \X{c} \cup \X{e}$. How to perform this choice depends on the specificity of a given model: some guidelines will be discussed in Remarks \ref{rem:partitionVariables} and \ref{rem:partitioningVariables}. We stress  here the double nature of environment variables: they will be treated like discrete variables in terms of the way they can be updated, but as continuous variables for what concerns their domain, i.e. they are part of the continuous state space of the TDSHA.

Once variables have been partitioned, we will 
process each component $C\in\Def$ separately,
subdividing its stochastic actions $\acts{C}$ into two subsets: $\disc{C}$, those to be maintained discrete, and $\cont{C}$, those to be treated continuously. This choice confers an additional degree of freedom to the mapping, but has to satisfy the following constraint:
\begin{assumption}
\label{assumption:continuousTransitions}
Continuous transitions must have a constant increment update or a random increment update, i.e. $\reset{\pi} = \vr{X} + \stoich{\pi}$. Furthermore, their reset cannot modify any discrete or environment variable, i.e. $\stoich{\pi}[X] = 0$, for each $X\in\X{d}\cup\X{e}$. 
\end{assumption}
We will now sketch the main ideas behind the definition TDSHA associated with a component $C$.

\begin{description}
  \item[Continuous transition.] With each $\pi\in\cont{C}$, we associate $\eta\in\TC$ with rate function $\ratef{\eta}(\vr{X}) = \ind{\guard{\pi}(\vr{X})} \cdot \srate{\pi}(\vr{X})$, where $\ind{\cdot}$ is the indicator function of the predicate $\guard{\pi}(\vr{X})$, equal to 1 if $\guard{\pi}(\vr{X})$ is true, and to zero if it is false. The update vector is $\cinc{\pi}$, if $\pi$ has a constant increment update. If $\pi$ has random increment $\rinc{\pi}$, we define the update vector as $\bbE[\rinc{\pi}]$, the expected value of the random vector $\rinc{\pi}$.\footnote{Alternatively, we could have considered the support $\{\rinc{\pi}^1,\ldots,\rinc{\pi}^k,\ldots\}$ of $\rinc{\pi}$, with probability density $P(\rinc{\pi}^1),\ldots,P(\rinc{\pi}^k),\ldots$, and generated a family of continuous transitions with rate $P(\rinc{\pi}^k) \ratef{\eta}(\vr{X})$ and update vector $\rinc{\pi}^k$. However, if we add up these transitions as required to construct the vector field (see Section \ref{sec:TDSHAtoPDMP}), we obtain $\bbE[\rinc{\pi}]\ratef{\eta}(\vr{X})$, i.e. the two approaches are equivalent.} 

  \item[Stochastic transitions.] Stochastic transitions are defined in a very simple way: 
  guards, resets, and rates are copied from the $\sCCP$ action $\pi\in\disc{C}$.

\item[Instantaneous transitions.] Instantaneous transitions are generated from \sCCP\ instantaneous actions $\pi\in\acti{C}$, by copying  guards, resets and priorities. 
\end{description}

We can define formally the TDSHA of a \sCCP\ component as follows:
\begin{definition}
\label{def:TDSHAofComponent}
Let $\calA = (A,\Def,\vr{X},\calD,\vr{x_0})$ be a flat \sCCP\ program and  $(\X{d},\X{c},\X{e})$ be a partition of the variables $\vr{X}$. Let $C$ be a component, with stochastic actions $\acts{C}$ partitioned into $\disc{C}\cup \cont{C}$, in agreement with Assumption~\ref{assumption:continuousTransitions}. The TDSHA associated with $C$ is $\calT(C,\disc{C},\cont{C}) = (\vr{Z},Q,\vr{Y},\TC,\TD,\TS,\init{})$, where  
\begin{itemize}
\item $\vr{Z}$ is equal to $\X{d}$, while $\Y{} = \X{c}\cup\X{e}$. $Q$ is the domain of $\X{d}$ in $\calA$, and $\init{} = \vr{x_0}$. 
\item With each $\pi\in\cont{C}$ with constant increment reset $\reset{\pi} = \vr{X} + \cinc{\pi}$, we associate $\eta = (\cinc{\pi},\ratef{\eta})\in\TC$, where $\ratef{\eta}(\vr{X}) = \vr{1}\{\guard{\pi}(\vr{X})\}\cdot \srate{\pi}(\vr{X})$.
\item With each $\pi\in\cont{C}$ with random increment reset $\reset{\pi} = \vr{X} + \rinc{\pi}$, we associate  $\eta = (\bbE[\rinc{\pi}],\ratef{\eta})\in\TC$, where  $\ratef{\eta}$  is defined as above. 
\item With each $\pi\in\disc{C}$ we associate $\eta = (\guard{\pi}(\vr{X}),\reset{\pi}(\vr{X}),\srate{\pi}(\vr{X}))\in\TS$.
\item With each $\pi\in\acti{C}$ we associate $\eta = (\guard{\pi}(\vr{X}),\reset{\pi}(\vr{X}),\priority{\pi}(\vr{X}))\in\TD$.
\end{itemize}
\end{definition}

Finally, the TDSHA of the whole \sCCP\ program is obtained by taking the composition of the TDSHA of each component, as follows: 

\begin{definition}
\label{def:TDSHAofComponent}
Let $\calA = (A,\Def,\vr{X},\calD,\vr{x_0})$ be a flat \sCCP\ program and  $(\X{d},\X{c},\X{e})$ be a partition of variables $\vr{X}$. The TDSHA $\calT(\calA)$ associated with $\calA$ is 
$$\calT(\calA) = \biguplus_{C\in\Def} \calT(C,\disc{C},\cont{C}).$$
\end{definition}

\begin{figure}[!t]
  \begin{center}
  \includegraphics[width=0.9\textwidth]{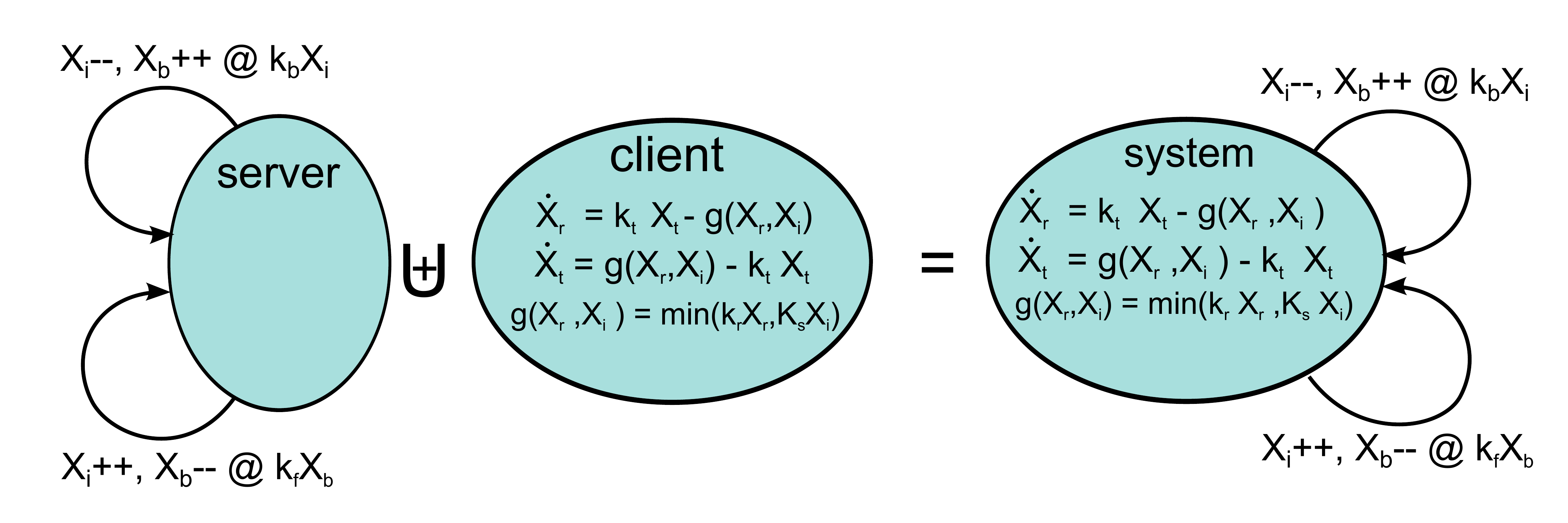}
  \end{center}
  
  \caption{TDSHA associated with \texttt{client} and
\texttt{server} components of Example~\ref{ex:clientServerBasic},
together with their composition. Continuous transitions are rendered into a set of ODEs, as explained in Section~\ref{sec:TDSHAtoPDMP}.}\label{fig:HAexample}
\end{figure}

\begin{exu}
Consider the \sCCP\ program of Example~\ref{ex:clientServerBasic}. The TDSHA associated with its two components (\texttt{client} and
\texttt{server}) and their composition are shown in
Figure~\ref{fig:HAexample}. In this case, we partitioned variables by making all client variables continuous, i.e. $X_r$ and $X_t$, and all server variables discrete, i.e. $X_i$ and $X_b$. This describes a situation in which there are few servers that have to satisfy the requests of many clients. Consequently, we considered all \texttt{client} transitions as continuous and all \texttt{server} transitions as discrete. 
\end{exu}

\begin{remark}
\label{rem:partitionVariables}
Choosing how to partition variables into discrete and continuous is a complicated matter, and  depends on specific features of the model under study. We postpone a more detailed discussion on this issue to Remark \ref{rem:partitioningVariables} in Section \ref{sec:hybridLimits}, as this choice can depend on the notion of system size, which has still to be introduced. Here we just note that a non-flat \sCCP\ model may naturally suggest a candidate subset of variables to be kept discrete, namely state variables of a sequential \sCCP\ component (i.e. an agent changing state but never forking or killing itself) present in a single copy in the network. This is the approach followed e.g. in \cite{My2010TCShybridDynamicsStochProg, MyNETMAHIB2012} to define the control modes of the TDSHA. However, the approach presented here is more general: different partitions of model variables and stochastic transitions lead to different TDSHA, which can be arranged in a lattice, as done in \cite{My2011LogicsCSjournalHybridSCCPLattice}.
\end{remark}

%
%

\subsection{Piecewise Deterministic Markov Processes}\label{sec:PDMP}


The dynamical evolution of Transition Driven Stochastic Hybrid Automata is defined by  mapping them to a class of stochastic processes known as Piecewise Deterministic Markov Processes
(PDMP,~\cite{STOC:Davis:1993:PDMP}). 
They have a continuous dynamics given by the solution of a set of ODE and a discrete and stochastic dynamics given by a Markov jump process. The following definition deviates slightly from the classical one for PDMP in the way the discrete state space is described.

\begin{definition}\label{def:PDMP}
A PDMP is a tuple $(\Z{},Q,\Y{},\sspace{},\flow{},\lambda,R)$, such
that:
\begin{itemize}
\item $\Z{}$ is a set of discrete variables, taking values in the countable set $Q\subset\bbR^\ddim$, the set of \emph{modes} or \emph{discrete states}. (Hence $q\in Q$ is of the form $(z_1,\ldots,z_\ddim)$.) $\Y{}$ is a vector of variables of dimension
$|\Y{}| = \cdim$. 
For each $q\in Q$, let $\sspace{q} \subset \bbR^\cdim$ be an open set, the continuous domain of mode $q$.
 $\sspace{}$, the \emph{hybrid state space}, is defined as the disjoint union of $\sspace{q}$ sets, namely $\sspace{} = \bigcup_{q\in Q} \{q\}\times \sspace{q}$. 
A point  $\x{}\in \sspace{}$ is  a pair $\x{}=(q,\y{})$, $\y{}\in\sspace{}$.\footnote{See Appendix \ref{app:background} for a brief discussion on metric and topological properties of hybrid state spaces.}
In the following, we will denote $\Z{}\cup\Y{}$ by $\X{}$, so that variables $\X{}$ range over $\sspace{}$.

  \item With each mode $q\in Q$ we associate a vector field $\vfield{q}:\sspace{q}\rightarrow \bbR^\cdim$.
 The ODE $\dot{\y{}} = \vfield{q}(\y{})$ is assumed to have a unique solution starting from each $\y{0} \in \sspace{q}$, globally existing in $\sspace{q}$ (i.e., defined until the time at which the trajectory leaves $\sspace{q}$). 
 The (semi)flow $\flow{q}:\sspace{q}\times\bbR^+ \rightarrow \bbR^\cdim$  of such vector field is assumed to be continuous in both arguments. $\flow{q}(t,\y{0})$ denotes the point reached at time $t$ starting from $\y{0}\in \sspace{q}$.\footnote{Usually, $\vfield{q}$ is  \emph{locally Lipschitz continuous}, hence the solution exists and is unique, provided trajectories do not explode in finite time. However, as in the paper we will consider also situations in which the vector field con be discontinuous due to the presence of guards, we have chosen this more general condition.}    
  
\item $\lambda: \sspace{}\rightarrow\bbR^+$ is the \emph{jump rate} and it gives  the hazard of executing a discrete transition. It is assumed to be (locally) integrable. 

\item $R:\sspace{}\cup\boundary{} \times \ssborel{}
\rightarrow [0,1]$ is the \emph{transition measure} or \emph{reset kernel}. It maps each $y\in \sspace{}\cup \boundary{}$ on a probability measure on $(\sspace{},\ssborel{})$, where $\ssborel{}$ is the Borel $\sigma$-algebra of $\sspace{}$. $R(\x{},A)$ is required to be measurable in the first argument and a probability measure for each $\x{}$. 
\end{itemize}

\end{definition}

The idea underlying the dynamics of PDMP is that, within each mode $q$, the process evolves along the flow $\flow{q}$. While in a mode, the process can jump spontaneously with hazard given by the rate function $\lambda_q$. Moreover, a jump
is immediately performed whenever the boundary of the state space of the current mode is hit. 

In order to formally capture the evolution, we need to define the sequence of jump times and target states of the PDMP, given by random variables $T_{1},\chi_1,T_{2},\chi_2,\ldots$. 
Let $t_*(\x{}) = \inf\{t>0~|~\flow{q}(t,\x{})\in\partial \sspace{q}\}$
(with $\inf\emptyset=\infty$) be the hitting time of the boundary $\boundary{q}$ starting from $\x{}=(q,\y{})\in \sspace{}$. We can define the
survivor function of the first jump time $T_1$, given that the
process started at $\x{}=(q,\y{})$, by $F(t,\x{}) = \bbP(T_{1}\geq t) =
I_{t<t_*(\x{})}\exp\left(-\int_0^t \lambda(q,\flow{q}(s,\vr{x}))
\dt{s}\right)$.
\\
This defines the probability distribution of the first jump time $T_{1}$, which can be simulated, as customary, by solving for $t$ the
equation $F(t,\x{}) = U_1^1$, with $U_1^1$ uniform random variable in $[0,1]$. Once the time of the first jump has been drawn, we can sample the target point $\chi_1$ of the reset map from the distribution $R(\x{T_{1}}^-,\cdot)$, with $\x{T_{1}}^- = \flow{q}(T_{1},\x{})$, using another independent uniform random variable $U_1^2$. From $\chi_1 = (q_1,\x{1})$, the process follows the flow $\flow{q_1}(t-T_{1},\x{1})$, until the next jump,
determined by the same mechanism presented above.

Using two independent sequences of uniform random variables $U_N^1$ and $U_N^2$, we are effectively construct a realization of the PDMP in the Hilbert cube $[0,1]^\omega$. 
A further requirement is that, letting $N_t = \sum_k \ind{t > T_{k}}$ be the random variable counting the number of jumps up to time $t$, it holds that $N_t$ is finite with probability 1, i.e. $\forall t, \bbE
N_t <\infty$, see~\cite{STOC:Davis:1993:PDMP} for further details. If this holds, then the PDMP is called \emph{non-Zeno}.

\begin{remark}
\label{rem:simplePDMP}
In \cite{My2010ASMTAkurtzForPDP}, we proved some limit results restricting the attention to the case in which no instantaneous jump can occur. This amounts to requiring that each $\sspace{q}$ has no boundaries, i.e.  $\sspace{q} = \bbR^\cdim$, or, more precisely, that $t^*(\x{}) = \infty$ for each $\x{}\in\sspace{}$. 
If, in addition to this description, we also require the vector field to be Lipschitz continuous and the stochastic jumps to be described by a finite set of transitions $\eta$ with rate $\srate{\eta}$ and reset given by a constant increment $\stoich{\eta}$, we obtain the so called \emph{simple PDMP} \cite{My2010ASMTAkurtzForPDP}. 


\end{remark}

    \subsection{From TDSHA to PDMP}\label{sec:TDSHAtoPDMP}

The mapping of TDSHA into PDMP is quite straightforward, with the exception of the definition of the reset kernel. 
Essentially, the problem lies in the fact that each discrete transition of a PDMP has to jump in the interior of the state space $\sspace{}$, which will be defined as the set of points in which no guard of any instantaneous transition is active. 
However, in a TDSHA we do not have any control over this fact, and we may define guards of transitions $\eta\in\TD$ in such a way that an infinite sequence of them can fire in the same time instant. 
For instance, the transitions $(X >= 1, X'=0, 1)$ and $(X <= 0, X'=1, 1)$ will loop forever if one of them is triggered. 
In order to avoid such bad behaviours, we will forbid by definition the possibility that two discrete transitions fire in the same time instant. We will call \emph{chain-free} a TDSHA with this property. 
This condition is unnecessarily restrictive, and can be relaxed allowing the firing of a finite number of finite sequences (\emph{loop-free} TDSHA), as done in \cite{My2011LogicsCSjournalHybridSCCPLattice}, but it allows a simpler definition of the reset kernel of the PDMP. The interested reader is referred to \cite{My2011LogicsCSjournalHybridSCCPLattice} for the construction of the reset kernel for loop-free TDSHA. The good news here is that all the results in this paper extend immediately to loop-free TDSHA. The bad news is that checking if a TDSHA is loop-free is in general undecidable, as one can easily encode an Unlimited Register Machine in a TDSHA \cite{My2011LogicsCSjournalHybridSCCPLattice}. 
However, some sufficient conditions in terms of acyclicity of a graph constructed from transitions in $\TD$ have been discussed in \cite{My2011HYPEjournal}. Practically, most models will satisfy the chain-free condition, as the discrete controller  described by instantaneous transitions is usually simple. More advanced controllers will perform some form of local computation, which can then result in a loop-free model. Violation of the loop-free property, instead, usually indicates an error in the model.

We now briefly introduce some notation, and then define chain-free TDSHA and the PDMP associated with a chain-free TDSHA.

Let $\calT = (\vr{Z},Q,\vr{Y},\TC,\TD,\TS,\init{})$ be a TDSHA. 
Given a transition $\eta\in\TD$, we let $G_\eta = \{\x{}\in Q\times\bbR^\cdim~|~\guard{\eta}(\x{})~true\}$, and $R_\eta = \{\vr{x}\in Q\times \bbR^n~|~\vr{x}\in \reset{\eta}(\bar{G}_\eta, \W{})\}$. 
$R_{\eta}$ is the set of points that can be reached after the firing of $\eta$, defined as the image under $\reset{\eta}$ of the closure $\bar{G}_\eta$ of the activation set $G_\eta$ of the guard. 
Similarly, for $\eta\in\TS$, we let $R_\eta = \{\vr{x}\in Q\times \bbR^n~|~\vr{x}\in \reset{\eta}(\{\x{1}~|~\srate{\eta}(\x{1})>0\}, \W{})\}$, the set of points that can be reached by a stochastic jump. 

\begin{definition}
\label{def:chainFreeTDSHA}
A  TDSHA is chain-free if and only if, for each $\eta_1 \in \TD\cup\TS$ and each $\eta_2\in\TS$, $R_{\eta_1}\cap \bar{G}_{\eta_2}=\emptyset$. 
\end{definition}

Consider now a chain-free TDSHA $\calT = (\vr{Z},Q,\vr{Y},\TC,\TD,\TS,\init{})$. Then, its associated PDMP $\calP = (\Z{},Q,\Y{},\sspace{},\flow{},\lambda,R)$ is defined by: 
\begin{itemize}
  \item Discrete and continuous variables, and discrete modes $Q$, are the same both
  in $\calT$ and in $\calP$.
  \item The state space of the PDMP, encoding the \emph{invariant region} of
continuous variables in each discrete mode,
   is defined as the set of points in which no instantaneous transition is active:
   $$\sspace{} = \bigcap_{\eta\in\TD} \overline{G}_{\eta}^c.$$
 Note that $\sspace{q}$ is open, because we are intersecting the complement of the closure of each set $G_{\eta}$. 
  \item The vector field is constructed from continuous  transitions, by adding their effects on system variables:
\begin{equation}\label{eqn:vectorFieldPDMP}
    \vfield{}(\x{}) = \sum_{\eta\in\TC}
  \stoich{\eta}\cdot\ratef{\eta}(\x{}).
\end{equation}
  \item The rate function $\lambda$ is defined by adding point-wise the rates of all active stochastic transitions:
  \begin{equation}\label{eqn:ratePDMP}
    \lambda(\x{}) = \sum_{\eta\in\TS} \ind{\guard{\eta}(\x{})}\srate{\eta}(\x{}).
  \end{equation}
  
\item The reset kernel $R$ for $\x{}\in \sspace{}$ is obtained by choosing
the reset of one active
stochastic transition in $\x{}$ with a probability
proportional to its rate. As all such resets jump to points in the interior of $\sspace{}$ by the chain-free property of the TDSHA, we have
\begin{equation}
\label{eqn:ResetKernelStoc}
    R(\x{},A) = \sum_{\eta\in\TS}\left(
  \frac{\ind{\guard{\eta}(\x{})}\srate{\eta}(\x{})}{\lambda(\x{})}\bbP\{\reset{\eta} (\x{},\W{})\in A\}\right),
\end{equation}
  where $A\in\ssborel{}$, the Borel $\sigma$-algebra of  $\sspace{}$. If the reset of $\eta$ is deterministic, then $\bbP\{\reset{\eta} (\x{},\W{})\in A\} = \delta_{\reset{\eta} (\x{},\W{})}$, where $\delta_{\x{1}}(A)$ is the Dirac measure on the point $\x{1}\in\sspace{}$, assigning probability 1 to  $\x{1}$ and 0 to the rest of the space. 

\item The reset kernel $R$ on the boundary $\boundary{}$ is defined
from  resets of instantaneous transitions. If more than one
transition is active in a point $\vr{x}\in \boundary{}$, we
choose one of them with probability proportional to their weight. Let $\priority{}(\x{}) = \sum_{\eta\in\TD~|~\guard{\eta}(\x{})~true} \priority{\eta}(\x{})$, then
\begin{equation}
\label{eqn:ResetKernelInst}
    R(\x{},A) = \sum_{\eta\in\TD~|~\guard{\eta}(\x{})~true}\left(
  \frac{\priority{\eta}(\x{})}{\priority{}(\x{})} 
  \bbP\{\reset{\eta} (\x{},\W{})\in A\}\right).
\end{equation}

\item The initial point is $\x{0}=\init{}$.
\end{itemize}

From now on, we implicitly assume that all the TDSHA obtained by the \sCCP\ models we consider are chain-free. In general this may not be true and has to be checked. However, the property will hold straightforwardly in all the examples of this paper, and it will also be true in many practical examples. Indeed, as it is enough to consider loop-free TDSHA \cite{My2011LogicsCSjournalHybridSCCPLattice}, this check may be automatically performed by the method of \cite{My2011HYPEjournal}.

\section{System Size and Normalisation}
\label{sec:scaling}

In this paper we are concerned with the correctness of the hybrid semantics of \sCCP\ in terms of approximation or limit results. Essentially, we want to show that ``taking the system to the limit'', the standard CTMC semantics of \sCCP\ converges (in a stochastic sense) to the PDMP defined by the hybrid semantics. 

Clearly, this idea of convergence requires us to have a sequence of models. This sequence will depend on the size $\size{}$ of the system. Hence, we will be concerned with the behaviour of a \sCCP\ program, when the system size goes to infinity.

The concrete notion of system size depends on the model under examination and the type of system being modelled. 
In general, it is related to the \emph{size of the population}, intended as the number of agents or entities in the system (which in flat \sCCP\ models are counted by the system variables). For instance, in the client/server example (Example \ref{ex:clientServerBasic}), this can be the total number of clients or the total number of clients and servers. In an epidemic model, this can be the size of the total population, or of the initial population, if we allow also birth and death events.  However, the notion of system size can also be connected to the size of the population in an area or a volume. In this case, when the size increases, both the number of agents and the area or volume increase, usually keeping constant the density (number to area or volume ratio). The classical examples here are biochemical systems, in which we consider molecules in a given volume. Furthermore, in a model of bacteria's growth (like that of Example \ref{ex:bacteria}), we may be interested in increasing the number of bacteria together with the area of the Petri dish in which the culture is grown.

In order to make the notion of size explicit, we will decorate a \sCCP\ model with the corresponding population size. 
\begin{definition}
A population-\sCCP\ program $(\calA,\size{})$ consists of a \sCCP\ program $\calA$ together with the population size $\size{}\in\bbR^+$.
\end{definition}
It is intended that rates of transitions, and even updates, of a population-\sCCP\ program can depend on the population size $\size{}$. We further stress that, in a population-\sCCP\ program, model variables usually take integer values.

\begin{example}
\label{ex:csTotalPop}
We go back to the client-server model of Example \ref{ex:clientServerBasic}, and consider the population-\sCCP\ model in which the size $\size{}$ corresponds to the total population of clients and servers, namely $\size{} = N_1 + N_2 = \X{r}(0)+\X{t}(0)+\X{i}(0)+\X{b}(0)$.  In this scenario, we are interested in what happens when the total population increases, maintaining constant the client-to-server ratio. 

A different notion of size can be envisaged,  corresponding to a different scaling law. More specifically, we can consider $\size{} = N_1 = \X{r}(0)+\X{t}(0)$, the total number of clients in the system. Increasing this notion of size, we are effectively increasing the number of clients requesting information to a fixed number of servers. Intuitively, these two different scalings for the client-server system should correspond to two different limit behaviours (taking $\size{}$ to infinity). 
\end{example}

In order to compare models for increasing values of the size $\size{}$, we need to normalize models to the same scale. This is done by the \emph{normalization} operation. Essentially, we will divide system variables by the system size (in fact, only those that will be approximated continuously), and express guards, rates, and resets in terms of such normalized variables.

We formalize now the operation of normalization. Consider a population-\sCCP\ program $(\calA,\size{})$, with $\calA = (A,\Def,\vr{X},\calD,\vr{x_0})$, let $\X{}(t)$ be the associated CTMC, and assume that variables $\X{}$ are partitioned into discrete $\X{d}$, continuous $\X{c}$, and environment variables $\X{e}$. Then the normalized CTMC $\Xb{}(t)$ is constructed as follows:
\begin{itemize}
\item Normalized variables are $\Xb{} = (\X{d},\Xb{c},\X{e})$, with $\Xb{c} = \size{}^{-1} \X{c}$;
\item Given a stochastic action $\pi = (\guard{\pi}(\X{}), \reset{\pi}(\X{}), \srate{\pi}(\X{})$, we define:
\begin{itemize}
\item $\guardb{\pi}(\Xb{}) = \guard{\pi}(\X{})$, the guard predicate with respect to normalized variables;
\item Let $X' = \reset{\pi}[X](\X{},\W{})$. If 
$X\in\X{d}\cup\X{e}$, then $\resetb{\pi}[X](\Xb{},\W{}) = \reset{\pi}[X](\X{},\W{})$. Otherwise, if $X\in\X{c}$, then 
$\resetb{\pi}[X](\Xb{},\W{}) = \size{}^{-1}\reset{\pi}[X](\X{},\W{})$ (hence, we replaced $\X{c}$ variables with their normalized counterpart in the reset function, but also rescaled the reset of $\Xb{c}$ variables by dividing the reset function for $\size{}$);
\item $\srateb{\pi}(\Xb{}) = \srate{\pi}(\X{})$.
\end{itemize} 
\item Instantaneous transitions are rescaled in the same way (expressing the weight function $\priorityb{}$ in terms of normalized variables like the rate of stochastic transitions);
\item Normalized initial conditions are $\xb{0} = (\x{d,0},\size{}^{-1} \x{c,0},\x{e,0})$. 
\end{itemize}
Applying this transformation to a \sCCP\ program, we can construct the normalized CTMC $\Xb{}(t)$ along the lines of the construction of Section \ref{sec:sCCP}. Furthermore, we can construct the TDSHA associated with a \sCCP\ program by considering normalized transitions and variables, instead of non-normalized ones. 
As we will always compare normalized processes, we will always assume that this construction has been carried out.

Given a population-\sCCP\ program $(\calA,\size{})$, our goal is to understand what will be the limit behaviour of a sequence of normalized CTMC $\Xb{}\N(t)$, constructed from $(\calA,\size{})$ and a sequence of system sizes $\size{N}\rightarrow\infty$ as $N\rightarrow\infty$. 
In order to properly do this, we need to get a better grasp on some related questions, namely:
\begin{enumerate}
\item how to split variables into discrete and continuous;
\item how rates and updates scale with the system size. 
\end{enumerate}
These two questions are somehow dependent; the last one, in particular, is crucial, as the correct form of the limit depends on the scaling of rates and updates. Investigating these issues, moreover, will give us some hints on how to choose discrete and continuous variables and transitions to define the hybrid semantics of \sCCP.

We will start by considering the fluid case, in which all variables are approximated as continuous. Rates and updates will be required to scale in a consistent way, and we will refer to these conditions as the \emph{continuous scaling}. Then, we will turn our attention to hybrid scaling and hybrid limits.

Before doing this, we stress that the normalization operation extends naturally to the TDSHA associated with a population-\sCCP\ 
program and, consequently, to the PDMP associated with the so-obtained TDSHA. In particular, if we have a sequence $(\calA,\size{N})$ of population-\sCCP\ models, we can construct its normalization for each $N$, and associate a TDSHA with each element of the sequence. We call $\n{\calT}(\calA,\size{N})$ such a TDSHA. However, in the rest of the paper we are interested in the limit behaviour, i.e. in models independent of $\size{N}$. The scaling conditions for each transition that we will introduce will naturally lead to the construction of a limit TDSHA, independent of any notion of size, referred to as $\n{\calT}(\calA)$ in the rest of the paper.

\section{Continuous Scaling and Fluid Limit}
\label{sec:fluidLimit}

We discuss now the standard fluid limit \cite{STOC:Kurtz:1970:ODEandCTMC} \cite{STOC:Kurtz:1981:ApproxPopProc} \cite{STOC:Kurtz:1986:MarkovProcesses} \cite{STOC:Darling:2002:PracticalFluid} \cite{STOC:DarlingNorris:2008:DifferentialEquationsCTMC} in our context. We will consider  a sequence of population-\sCCP\ programs $(\calA,\size{N})$ with divergent population size $\size{N}\rightarrow \infty$ as $N\rightarrow\infty$.

In the rest of this section, we will require the following assumptions: 
\begin{itemize}
\item All variables $\X{}$ \emph{are continuous} and thus normalized according to the recipe of the previous section (hence there are no discrete or environment variables).
\item There is  \emph{no instantaneous transition} in $\calA$.
\item All stochastic transitions are \emph{unguarded} and have \emph{constant or random increment updates}. 
\end{itemize}

In order to define the continuous scaling, we consider the domain $\sspace{}\subseteq \bbR^\fdim$ of normalized variables (note that here $E$ is not a hybrid state space), which depends on possible values that non-normalized variables can take in $\bbR^\fdim$ (usually in  $\bbZ^\fdim$, see also Remark \ref{rem:stateSpaceContinuousScaling} below). In particular, we assume that $E$ contains the domain of the normalized variables of a population-\sCCP\ program $(\calA,\size{N})$ for any $N\geq 0$, so that also the limit process will be defined in $\sspace{}$. 

Now we state the continuous scaling assumptions:
\begin{scaling}[Continuous Scaling]
\label{scaling:continuous}
A normalized \sCCP\ transition $\norm{\pi} = (true,\Xb{}' = \Xb{} + \stoichb{\pi}\N, \srateb{\pi}\N(\Xb{}))$ of a population-\sCCP\ program $(\calA,\size{N})$, with $\sspace{}\subseteq \bbR^m$ the domain of normalised variables $\Xb{}$, has \emph{continuous scaling} if and only if:
\begin{enumerate}
\item there is a function $g_{\pi}\N:E\rightarrow \bbR_{\geq 0}$ such that $\srateb{\pi}\N(\Xb{}) = \size{N} g_{\pi}\N(\Xb{})$.  Furthermore, $g_{\pi}\N$ converges uniformly to a locally Lipschitz continuous and locally bounded function $g_{\pi}:E\rightarrow \bbR_{\geq 0}$ (rates are $O(\size{N})$);
\item There is a constant or random vector $\stoich{\pi}\in \bbR^m$ such that the non-normalized increments $\stoich{\pi}\N$ converge weakly to $\stoich{\pi}$, $\stoich{\pi}\N\Rightarrow\stoich{\pi}$.\footnote{The concept of weak convergence is introduced in Appendix \ref{app:background}.} Furthermore,  $\stoich{\pi}\N$ and $\stoich{\pi}$ 
have \emph{bounded} and \emph{convergent} \emph{first order moments}, i.e. $\bbE[\|\stoich{\pi}\N\|]<\infty$, $\bbE[\|\stoich{\pi}\|]<\infty$, $\bbE[\stoich{\pi}\N] \rightarrow \bbE[\stoich{\pi}]$, and
$\bbE[\|\stoich{\pi}\N\|] \rightarrow \bbE[\|\stoich{\pi}\|]$.  In particular, it follows that normalized increments are $\Theta(\size{N}^{-1})$. 
\end{enumerate}
\end{scaling}

The intuition behind the previous conditions is that, as the system size increases, rates increase, leading to an increase of the density of events on the temporal axis. Furthermore, the increments become smaller and smaller, suggesting that the behaviour of the CTMC will become deterministic, with instantaneous variation equal to its mean increment.
This will produce a limit behaviour described by the solution of a differential equation.

\begin{remark}
\label{rem:densityDependence}
Scaling \ref{scaling:continuous} can be generalized in some way, see for instance \cite{STOC:DarlingNorris:2008:DifferentialEquationsCTMC, STOC:Darling:2002:PracticalFluid}. However, the version stated here is sufficiently general to deal with \sCCP\ programs. If we further restrict the previous scaling condition, requiring that $g_{\pi}\N(\Xb{}) = g_{\pi}(\Xb{})$, where $g_{\pi}(\Xb{})$ is a locally Lipschitz function independent of $\size{N}$, and $\stoich{\pi}\N = \stoich{\pi}$, then we obtain the so-called \emph{density dependent} scaling. 
For instance, all transitions in the client/server model of Example \ref{ex:clientServerBasic} are density dependent, as easily checked.
\end{remark}

\begin{remarkstar}
\label{rem:stateSpaceContinuousScaling}
The structure of the domain $\sspace{}$ of normalized variables depends mainly on conservation properties of the system modelled. For instance, a closed population model (i.e. without birth and death events) will preserve the total population (this is the case for the client/server model of Example \ref{ex:clientServerBasic}), hence the domain of the normalized variables will be contained in the unit simplex in $\bbR^\fdim$, which is a compact set.
For open systems, for instance a model of growth of a population of bacteria (see Example \ref{ex:bacteria}), in which the population can (in principle) become unbounded, the domain can be the whole $\bbR^\fdim$. However, it is unlikely that populations actually diverge (one may question the reliability of the model itself, if this happens), hence one can usually find a compact set that contains the interesting part of the state space (at least up to a finite time horizon). In particular, some of the hypotheses that we will state afterwards, like locally Lipschitzness or local boundedness, rely on this implicit assumption (i.e., that we can restrict our attention to a compact set in any finite time horizon). We will further discuss these issues while proving main theorems, once they emerge.
\end{remarkstar}

In order to state the fluid limit theorem, we need to construct the limit ODE. This is done according to the recipe of equation \ref{eqn:vectorFieldPDMP}. More specifically, for each $N$ we construct the drift or mean increment in $\xb{}$ as
\begin{equation}
\label{eqn:fluidVFN}
\vfield{}\N(\xb{}) = \sum_{\pi}  \bbE[\stoichb{\pi}\N] \srateb{\pi}\N(\xb{}) = \sum_{\pi} \bbE[\stoich{\pi}\N] g_{\pi}\N(\xb{}),
\end{equation}
where the sum ranges over all stochastic actions of the \sCCP\ program $\calA$. 
If all \sCCP\ transitions satisfy the continuous scaling assumption, $\vfield{}\N$ converges uniformly to 
\begin{equation}
\label{eqn:fluidVF}
\vfield{}(\xb{}) =  \sum_{\pi}  \bbE[\stoich{\pi}] g_{\pi}(\xb{}).
\end{equation}
The limit ODE is therefore $\dxdt{\xb{}(t)}{t} = \vfield{}(\xb{}(t))$, whose solution starting from $\x{0}$ is denoted by $\xb{}(t)$.
Note that this limit ODE can be obtained in terms of TDSHA with continuous transitions only, by the construction of Section \ref{sec:TDSHAtoPDMP}. In particular, the limit TDSHA corresponding to the fluid ODE has a continuous transition of the form $(\bbE[\stoichb{\pi}],\size{N}g_{\pi}) = (\size{N}^{-1}\bbE[\stoich{\pi}],\size{N}g_{\pi})$ for each normalized \sCCP\ transition $\n{\pi}$.

\begin{theorem}[Kurtz~\cite{STOC:Kurtz:1970:ODEandCTMC, STOC:Darling:2002:PracticalFluid, STOC:DarlingNorris:2008:DifferentialEquationsCTMC}]
\label{th:Kurtz}
Let $(\calA,\size{N})$ be a sequence of population-\sCCP\ models for increasing system size $\size{N}\rightarrow\infty$, satisfying the conditions of this section, and with all \sCCP-actions $\pi$ satisfying the continuous scaling condition. Let $\Xb{}\N(t)$ be the sequence of normalized CTMC associated with the \sCCP-program and $\xb{}(t)$ be the solution of the fluid ODE.
\\
If $\xb{0}\N \rightarrow \xb{0}$ almost surely, then 
for any $T<\infty$, $\sup_{t\leq T}\|\Xb{}\N(t) - \xb{}(t) \| \rightarrow 0$ as $N\rightarrow\infty$, almost surely. \qed
\end{theorem}

\begin{exu}
Consider again the client/server model of Example \ref{ex:clientServerBasic}, in which both the number of clients and of servers is increased. Therefore, consider a sequence of models with size $\size{N}$ equal to the total number of clients and servers. 
It is easy to see that its normalized models all live in the unit simplex $E$ in  $\bbR^4$, and that all its transitions are density dependent, hence satisfy the continuous scaling. Assume that $\xb{0} = (c,0,s,0)$, with $c+s=1$, so that $\X{0}\N = (cN,0,sN,0)$, meaning that we keep constant the client-to-server ratio.
The fluid ODE associated with this model is 
$$\left\{\begin{array}{rcl}
\dxdt{x_r}{t} &  = & k_t x_t - \min\{k_r x_r,k_s x_i\} \\
\dxdt{x_t}{t} &  = & \min\{k_r x_r,k_s x_i\} - k_t x_t \\
\dxdt{x_i}{t} &  = & k_f x_b - k_b x_i \\
\dxdt{x_b}{t} &  = & k_b x_i - k_f x_b  \\
\end{array} \right.$$ 
Hence, we can apply Theorem \ref{th:Kurtz} to infer convergence of the CTMC sequence $\Xb{}\N$ to its solution.
\end{exu}

%
%
%
%

\begin{remark}
\label{rem:miscellaneaKurtz}
The version of Kurtz theorem we presented here is similar to the one of \cite{STOC:Darling:2002:PracticalFluid}, but with scaling taken from \cite{STOC:DarlingNorris:2008:DifferentialEquationsCTMC}. 
The point of the scaling is to prove that noise goes to zero, which is usually shown either by some martingale inequality or by using the law of large number of Poisson random variables, using a Poisson representation of CTMC. 
In Appendix \ref{app:proofs}, we present a proof based on the Poisson representation.
\end{remark}

%
%

\begin{remarkstar}
\label{rem:randomIncrementesStateDependend}
In continuous transitions with random increments, we assumed for simplicity that the distribution of the increment is independent from the current state of the system. However, this restriction can be safely dropped, provided that we require uniform boundedness (in any compact $K\subset\sspace{}$) of the limit first order moments of the increments, i.e. $\sup_{\x{}\in K}\bbE[\stoich{}(\x{})] < \infty$ and $\sup_{\x{}\in K}\bbE[\|\stoich{}(\x{})\|] < \infty$, and uniform convergence of the expectation of $\stoich{}\N(\x{})$ to $\stoich{}(\x{})$, i.e. $\sup_{x\in K}\|\bbE[\stoich{}\N(\x{})] - \bbE[\stoich{}(\x{})]\|\rightarrow 0$ and $\sup_{x\in K}|\bbE[\|\stoich{}\N(\x{})\|] - \bbE[\|\stoich{}(\x{})\|]|\rightarrow 0$. Given these conditions, it is easy to check that the resulting sequence of CTMC still satisfy the conditions of \cite{STOC:Kurtz:1970:ODEandCTMC} (restricted to a suitable compact $K$), hence Theorem \ref{th:Kurtz} continues to hold.
\end{remarkstar}
\section{Hybrid Scaling and Hybrid Fluid Limits}
\label{sec:hybridLimits}

In this section we will introduce a scaling for transitions that cannot be approximated continuously, roughly speaking because their frequency remains constant as the population size grows. We will then prove that the sequence of normalized CTMC converges to the PDMP associated with the normalized \sCCP\ model. This proof will be first given under a suitable set of restrictions (essentially, restricting to unguarded stochastic actions with generic random resets for transitions kept discrete), in order to clarify the main ingredients that guarantee convergence. In the next sections, we will remove some of these restrictions, considering more complex hybrid limits. 

The first step in the construction of the hybrid limit, which coincides with the first step in constructing the \sCCP\ hybrid semantics, is the separation of model variables into discrete and continuous. This step is delicate and is model-dependent, as the same model can be interpreted in different ways.
For example, the client/server model of Example \ref{ex:clientServerBasic} can be interpreted continuously, assuming that the number of both clients and servers is increased with $\size{N}$, or in a hybrid way, assuming that only the number of clients increases, while the number of servers remains constant. In this case, the service rate has also to be increased in order to match the larger demand. This can be justified by thinking of an increased number of cores on the same machine, in such a way that the breakdown of a server will affect all its cores. We will discuss the partitioning of variables in Remark \ref{rem:partitioningVariables} below, after introducing the hybrid scaling conditions. 

To this end, we need to modify the conditions of Scaling \ref{scaling:continuous} for continuous transitions. In particular, we need to allow the possibility of activating a transition only in a subset of discrete modes. This is enforced by guards depending only on discrete (and environment) variables.

\begin{scaling}[Hybrid Continuous Scaling]
\label{scaling:continuousHybrid}
A normalized \sCCP\ transition $\norm{\pi} = (\guardb{\pi}(\Xb{}),\Xb{}' = \Xb{} + \stoichb{\pi}\N, \srateb{\pi}\N(\Xb{}))$ of a population-\sCCP\ program $(\calA,\size{N})$, with discrete variables $\X{d}$, continuous variables $\X{c}$, and environment variables $\X{e}$, and with $\Xb{}\in\sspace{}$, has \emph{hybrid continuous scaling} if and only if:
\begin{enumerate}
\item the rate $\srateb{\pi}\N(\Xb{})$ and the update $\Xb{}' = \Xb{} + \stoichb{\pi}\N$ satisfy the same conditions of Scaling \ref{scaling:continuous}
\item The guard predicate $\guardb{\pi}(\Xb{})$ depends only on discrete ($\X{d}$) and environment ($\X{e}$) variables.
\end{enumerate}
\end{scaling}

Additionally, we need to define the scaling for discrete stochastic transitions. Also in this case, we will assume that their guard depends only on discrete or environment variables.  

\begin{scaling}[Discrete Scaling for Stochastic Transitions]
\label{scaling:discreteStochastic}
A \emph{normalized} \sCCP\ transition with \emph{random reset} $\norm{\pi} = (\guardb{\pi}(\Xb{}),\Xb{}' = \resetb{\pi}\N(\Xb{}, \W{}\N(\Xb{}) ), \srateb{\pi}\N(\Xb{}))$ of a population-\sCCP\ program $(\calA,\size{N})$ with discrete variables $\X{d}$, continuous variables $\X{c}$, and environment variables $\X{e}$, with $\Xb{}\in\sspace{}$, has \emph{discrete scaling} if and only if:
\begin{enumerate}
\item the guard predicate $\guardb{\pi}(\Xb{})$ depends only on discrete ($\X{d}$) and environment ($\X{e}$) variables;
\item  $\srateb{\pi}\N(\Xb{}) = O(1)$, $\srateb{\pi}\N(\Xb{})$  converges uniformly  in each compact $K \subset \sspace{}$ to the continuous function $\srateb{\pi}(\Xb{})$; 
\item Resets converge weakly (uniformly on compact (sub)sets), i.e. for each $\xb{}\N\rightarrow\xb{}$ in $\sspace{}$, $\resetb{\pi}\N(\xb{}\N,\W{}\N(\xb{}\N)) \Rightarrow \resetb{\pi}(\xb{},\W{}(\xb{}))$, as random elements on $\sspace{}$.  
\end{enumerate}
\end{scaling}

\begin{remark}
\label{rem:partitioningVariables}

The choice on how to partition variables into discrete and continuous is a crucial step. This choice is usually model dependent, and relies heavily on the knowledge and intuition of the modeller. However, as a general  guideline, we can look at two aspects of the model:
\begin{description}
\item[Conservation Laws: ] Very often, the identification of discrete variables can be made by looking at conservation laws, i.e. at subsets of variables whose total mass is conserved during the evolution of the system, as pursued in \cite{My2009HybridPIJournal}. In fact, conserved variables usually are related to internal states of an agent which is present in one or very few copies. The identification of these sets can be carried out using algorithms like the Fourier-Motzkin elimination procedure \cite{PA:Colom:FourierMotzin:1991}, or using a constraint based approach \cite{SB:Soliman:InvariantsConstraints:2012}. In \sCCP, when describing non-flat models, these sets of variables, corresponding to state variables, are usually evident (cf. Remark \ref{rem:partitionVariables}). 
\item[Scaling of Rates: ] in describing a population-\sCCP\ model, a modeller is forced to make explicit the dependence of rates on the system size $\size{N}$. Given this knowledge, it is possible to identify some variables that cannot be continuous, otherwise both scaling \ref{scaling:continuous} and \ref{scaling:discreteStochastic} would be violated. For instance, if we have a rate like $kX_1X_2$, then at least one of $X_1$ and $X_2$ has to be discrete, otherwise the normalized rate would depend quadratically on $\size{N}$. On the contrary, $k\size{N}^{-1}X_1X_2$ is not compatible with both $X_1$ and $X_2$ discrete, otherwise the rate would vanish. Clearly, not all rate functions are informative; for instance, linear rates are compatible both with discrete and continuous scaling. 
\end{description}
The two previous arguments can be used to set up an algorithmic procedure to suggest a possible partition of variables into discrete and continuous, given a population-\sCCP\ model. However, we leave this for future work. 

We stress that, in general, if the modeller does not know how rates depend on the system size, she may choose a partition of variables and a scaling for each transition and impose a dependence of rates on system size that is correct with respect to the partition. This dependence has then to be validated a-posteriori, checking if it is meaningful in the context of the model. For instance, in a practical modelling scenario for the client/server example of Section \ref{sec:sCCP}, one usually has a fixed number of clients and servers and fixed parameters. To apply the convergence results of this paper, a specific scaling has to be assumed, and the parameters of the limit model have to be computed consequently. If, for instance, the number of servers is kept fixed, we obtain a meaningful limit if the \emph{service rate per client} is constant. If this cannot be assumed, namely if it is the global service rate of servers that remains constant, then the service rate per client depends on their number $N$, and goes to zero as $N$ increases. Hence, in the limit model the service rate is zero. However, for a fixed population size, we can still obtain a hybrid process that approximates closely the CTMC, using the size-dependent rates. This phenomenology (uninformative limit, but good size-dependent approximation) happens also in the fluid limit setting, see for instance \cite{PA:MassinkLBH:CrowdDynamics:FASE2011}. 
\end{remark}

Consider now a population-\sCCP\ model $(\calA,\size{N})$ with only stochastic actions, in which transitions satisfy either the continuous scaling \ref{scaling:continuous} or the discrete scaling \ref{scaling:discreteStochastic}. The \emph{limit TDSHA} $\n{\calT}(\calA)$ constructed from this model has continuous transitions of the form $(\bbE[\stoich{\pi}],g_\pi)$, for each \sCCP\ action $\pi$ satisfying continuous scaling and stochastic transitions of the form $(true,\resetb{\pi},\srateb{\pi})$, for each \sCCP\ action $\pi$ satisfying the discrete scaling. The limit PDMP is obtained from this TDSHA by the construction of Section \ref{sec:TDSHAtoPDMP}.

%

\begin{figure}[!t]

\subfigure[Client Server, Example \ref{ex:clientServerBasic}] {\label{fig:csBasic}
\includegraphics[width=.48\textwidth]{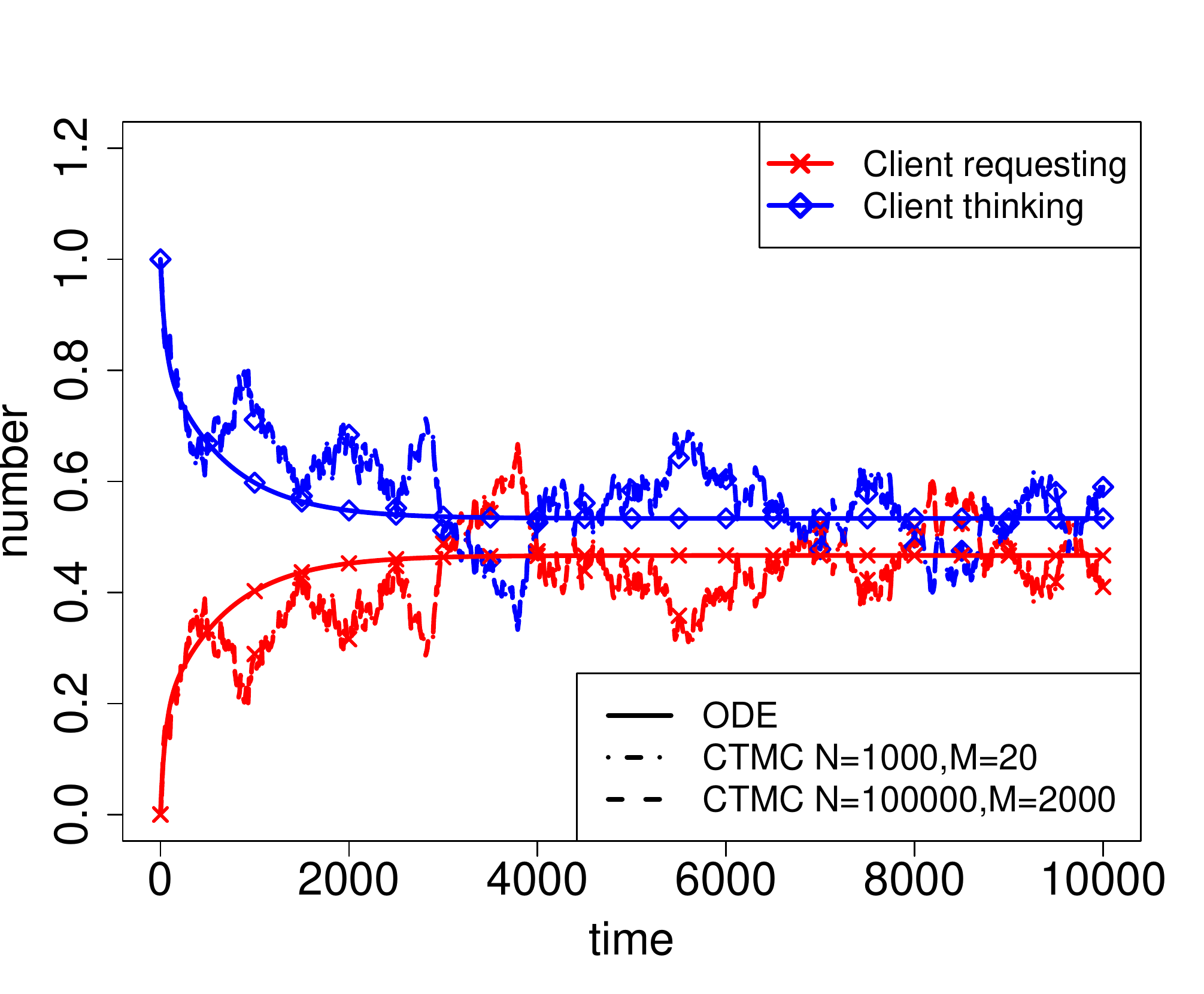} }
\subfigure[Gene Network, Example \ref{ex:geneNetwork}] {\label{fig:gene}
\includegraphics[width=.48\textwidth]{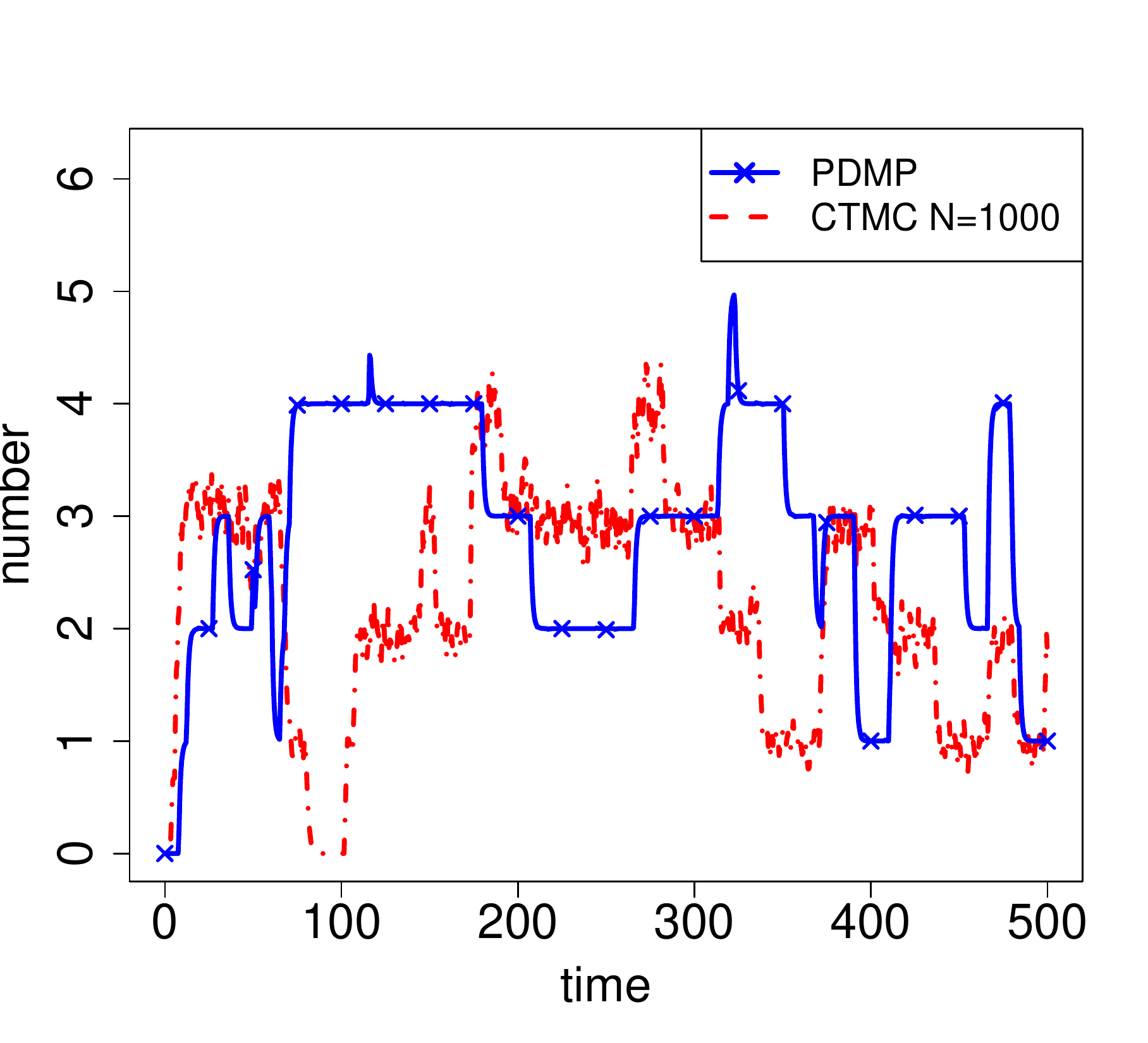} }

\caption{Left: comparison of stochastic trajectories and fluid ODE for the client-server model of Example \ref{ex:clientServerBasic}, with scaling discussed in Section \ref{sec:fluidLimit}. Parameters are $k_r = 2$, $k_s = 0.8$, $k_t = 1/50$, $k_b = 1/2000$, $k_f = 1/1000$, and initial conditions are $X_r\N(0) = N_1=100N$, $X_i(0) = N_2=2N$. In the plot, the CTMC trajectory for $N=100000$ and $M=2000$ fully overlap with the solution of the fluid ODE. Right: comparison of a trajectory of the limit PDMP and of the CTMC for the gene network model of Example \ref{ex:geneNetwork}. Parameters are $k_p = 0.1$, $k_t = 1$, $k_{d_p} = 1$, $k_{d_m} = 0.01$, $k_b = 0.1$, $k_u = 0.1$, and initial conditions are $P(0)=M(0)=G_{off}(0)=0$, $G_{on}(0)=1$. Note that both the stochastic and the hybrid system show a multi-modal behaviour.}
\label{fig:ODEandGENE}
\end{figure}

\begin{example}
\label{ex:geneNetwork}
We consider a new example with a biological flavour, namely a simple genetic network. Genes are the storage units of biological information: they encode in a string of DNA the information to produce a protein. Each cell has a biochemical machine that is capable of reading the information in a gene, first copying it into a mRNA molecule and then translating this molecule into a protein. Genes are in fact more than simple storage units: they are also part of the software that controls their own expression. In fact, expression is regulated by specific proteins, called transcription factors, which physically bind to the DNA close to a gene and activate or repress transcription. There are genes encoding for transcription factors that act as self-repressors. We model such a scenario here. 

To construct a population-\sCCP\ model, we need two integer-valued variables: $M$, counting the amount of mRNA, and $P$, counting the amount of protein. Here the size of the system $\size{}$ is the volume times the Avogadro number, so that normalized variables represent molar concentrations (see for instance \cite{SB:Voit:2000:SSystems, My2008ConstraintsBioModelingSCCP}). We will consider a model with one agent for the gene (which can be on or off), and agents for  translation of mRNA into protein and degradation of both protein and mRNA.

\begin{center}
{\tt
\begin{tabular}{lcl}
    gene\_on   & $\defeq$ & $[*\rightarrow M '= M + 1]_{k_p}$.gene\_on\\
 & + & $[* \rightarrow  P' = P - 1]_{ k_b \size{}^{-1} P}$.gene\_off\\
 gene\_off   & $\defeq$ & $[*\rightarrow P' = P + 1]_{k_u}$.gene\_on\\
  translate & $\defeq$ & $[*\rightarrow P' = P + 1]_{\size{} k_t M}$.translate\\
    degrade & $\defeq$ & $[*\rightarrow M' = M - 1]_{k_{dm} M}$.degrade\\
    & + & $[*\rightarrow P' = P - 1]_{k_{dp} P}$.degrade
\end{tabular}}
\end{center}

Inspecting the previous model, we can see that it is not 
flat. To convert it into a flat model, we need to add two additional variables, $G_{on}$ and $G_{off}$, with domain $\{0,1\}$, encoding the state of the gene agent. The structure of the gene agent itself reveals a conservation pattern in the system, namely that $ G_{on} + G_{off} = 1$, as they are indicator variables of the state of the gene.
Inspecting transitions, we can notice how translation has a rate depending on $\size{}$, suggesting that $M$ has also to be treated as a discrete variable. On the other hand, repression scales as $\size{}^{-1}$, i.e. it depends on the concentration of $P$, rather than on the number of molecules (repression depends only on the molecules close to the gene, the only ones that can bind to it). 
With this partitioning of variables, we obtain the following normalized TDSHA:
\begin{itemize}
\item Discrete variables are $G_{on}, G_{off}, M$, while $\norm{P}$ is the continuous variable. $Q = \{0,1\}\times\{0,1\}\times\bbN$ and $\norm{P}$ has domain $[0,\infty)$. 
\item Continuous transitions are $(*,\norm{P}'=\norm{P}+ \size{}^{-1},\size{} k_t M)$ and $(*,\norm{P}'=\norm{P}- \size{}^{-1},\size{} k_{dp} \norm{P})$;
\item Discrete transitions are $(*,G_{on} '= 0, G_{off}'=1, \norm{P}'=\norm{P} - \size{}^{-1},k_b \norm{P} G_{on})$, $(*,G_{on} '= 1, G_{off}'=0, \norm{P}'=\norm{P} + \size{}^{-1},k_u G_{off})$, 
$(*,M'=M + 1,k_p G_{on})$, $(*,M'=M - 1,k_{dm} M)$.
\end{itemize}
\end{example}

\begin{remarkstar}
\label{rem:fastDiscreteTransitions}
Scaling \ref{scaling:discreteStochastic} forbids discrete transitions to have a fast, $O(\size{})$ rate. If this would be the case, the dynamics of discrete transitions in the limit would be faster and faster, and one would expect that the discrete subsystem affected by these transitions reaches immediately the equilibrium (in a stochastic sense). This is what actually happens, under some regularity conditions on fast discrete dynamic, namely the possibility of isolating a discrete subsystem affected by fast discrete transitions, which is ergodic (considering only fast discrete transitions), and with fast rates depending continuously on continuous variables. In this case, one can compute the equilibrium distribution (as a function of other variables) of the fast discrete subsystem, remove the fast discrete variables and average the rate functions depending on fast discrete variables according to the equilibrium distribution. 
In case one has only  fast discrete variables, the fluid limit is given in terms of ODE \cite{PA:LeBoudec:2008:MeanFieldContinuousTime}. This scaling can be integrated quite easily in our framework, using the limit theorem of \cite{PA:LeBoudec:2008:MeanFieldContinuousTime} instead of Theorem \ref{th:Kurtz} and defining syntactically the averaging at the level of the TDSHA, given a method to compute the equilibrium distribution.  
\end{remarkstar}

We now turn to discuss the limit behaviour of a model showing hybrid scaling, i.e. with both discrete and continuous transitions. We will stick to further simplifying assumptions for the moment: the \sCCP\ program has no instantaneous transitions, all stochastic actions are unguarded and have continuous rates, variables and transitions have been partitioned into discrete and continuous,  discrete transitions have deterministic resets and satisfy discrete scaling \ref{scaling:discreteStochastic}, and  continuous transitions  satisfy continuous scaling \ref{scaling:continuous}.

We are now ready to state the main result of this section, namely that, under these restrictions, a normalized CTMC constructed from a \sCCP\ program converges weakly to the PDMP constructed from the normalized TDSHA associated with the \sCCP\ program.

\begin{theorem}[\cite{My2010ASMTAkurtzForPDP}]
\label{th:hybridBasic}
Let $(\calA,\size{N})$ be a sequence of population-\sCCP\ models for increasing system size $\size{N}\rightarrow\infty$, satisfying the conditions of this section, with variables partitioned into discrete $\X{d}$, continuous $\X{c}$, and environment ones $\X{e}$. 
Assume that discrete actions  satisfy scaling \ref{scaling:discreteStochastic} and continuous actions satisfy scaling \ref{scaling:continuousHybrid}.
Let $\Xb{}\N(t)$ be the sequence of normalized CTMC associated with the \sCCP\ program  and $\xb{}(t)$ be the PDMP associated with the limit normalized TDSHA $\norm{\calT}(\calA)$.

If $\xb{0}\N \Rightarrow \xb{0}$ (weakly) and the PDMP is non-Zeno, then $\Xb{}(t)$ converges weakly to $\xb{}(t)$,  $\Xb{} \Rightarrow \xb{}$, as random elements in the space of cadlag function with the Skorohod metric.\footnote{See Appendix \ref{app:background} for a breif introduction of these concepts.} 
\end{theorem}

\proof We just sketch the proof here. A detailed proof  can be found in Appendix \ref{app:proofs}. The main idea is to exploit the fact that we can restrict our attention to CTMC and PDMP that do at most $m$ discrete jumps. This is sufficient to obtain the weak convergence of the full processes, for two reasons. The first is related to the nature of the Skorohod metrics, which discounts the future (i.e. only $1/2^T$ of the distance comes from time instants greater than $T$), while the second is the non-Zeno nature of the limit PDMP, which implies that we can consider no more than $m$ jumps up to time $T$, with probability $1-\eps_m$, for $\eps_m\rightarrow 0$ as $m\rightarrow\infty$. 

In order to prove weak convergence of $\Xb{m}\N$, the CTMC with at most $m$ jumps of discrete transitions, to $\xb{m}$, the PDMP with at most $m$ jumps, we can exploit the piecewise deterministic nature of PDMP,  applying Theorem \ref{th:Kurtz} inductively. At the first step, we will prove that the time $\T{1}\N$ of the first stochastic jump for $\Xb{}\N$ converges weakly to $\T{1}$, the first jump time of $\xb{}$ (Lemma \ref{lemma:convergenceJumpTimes} in Appendix \ref{app:proofs}), and also the state $\Xb{}\N(\T{1}\N)$ after time $\T{1}\N$ converges weakly to $\xb{}(\T{1})$ (Corollary \ref{cor:weakKurtz}). This shows convergence of the processes up to the first stochastic jump.
Exploiting this and the strong Markov property,  we can restart $\xb{}(t)$ at time $\T{1}$ from $\xb{}(\T{1})$ and $\Xb{}(t)$ from $\Xb{}(\T{1}\N)$ at time $\T{1}\N$ and apply Theorem \ref{th:Kurtz} and its corollaries again (actually, a minor modification of Theorem \ref{th:Kurtz}, allowing to sample probabilistically the initial conditions of the ODE), to prove weak convergence of the CTMC to the PDMP up to the $m$-th jump, for any $m$.  Note that this argument is based on the continuity of vector fields, rates and resets, which holds in our setting as their guards depend only on discrete and environment variables, hence their values do not change  in each deterministic piece of the PDMP dynamics. \qed



\begin{figure}

\begin{center}

\subfigure[PDMP trajectory] {\label{fig:csHybrid}
\includegraphics[width=.45\textwidth]{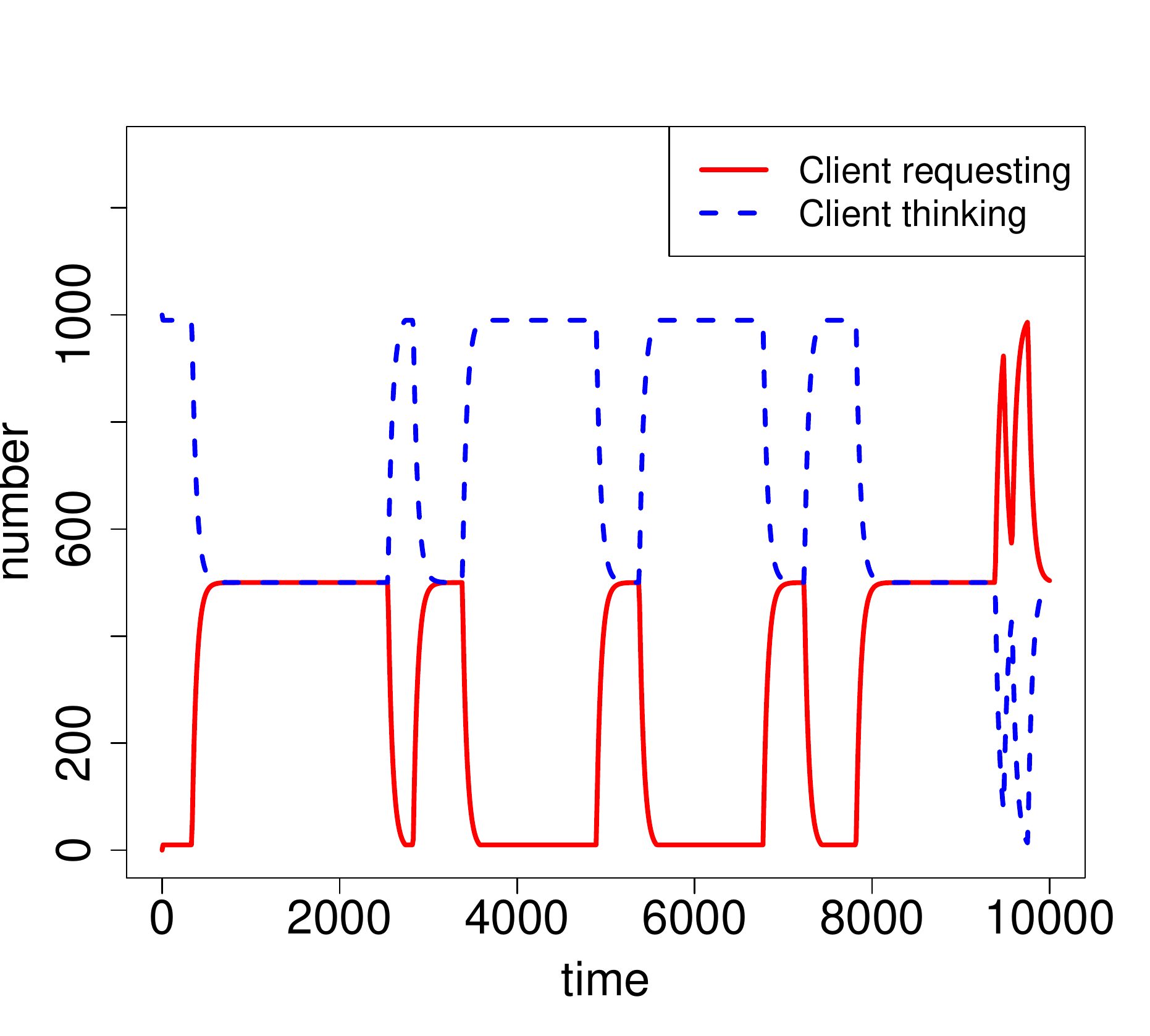} }
\subfigure[CTMC trajectory, $N=1000$] {\label{fig:csHstoc}
\includegraphics[width=.45\textwidth]{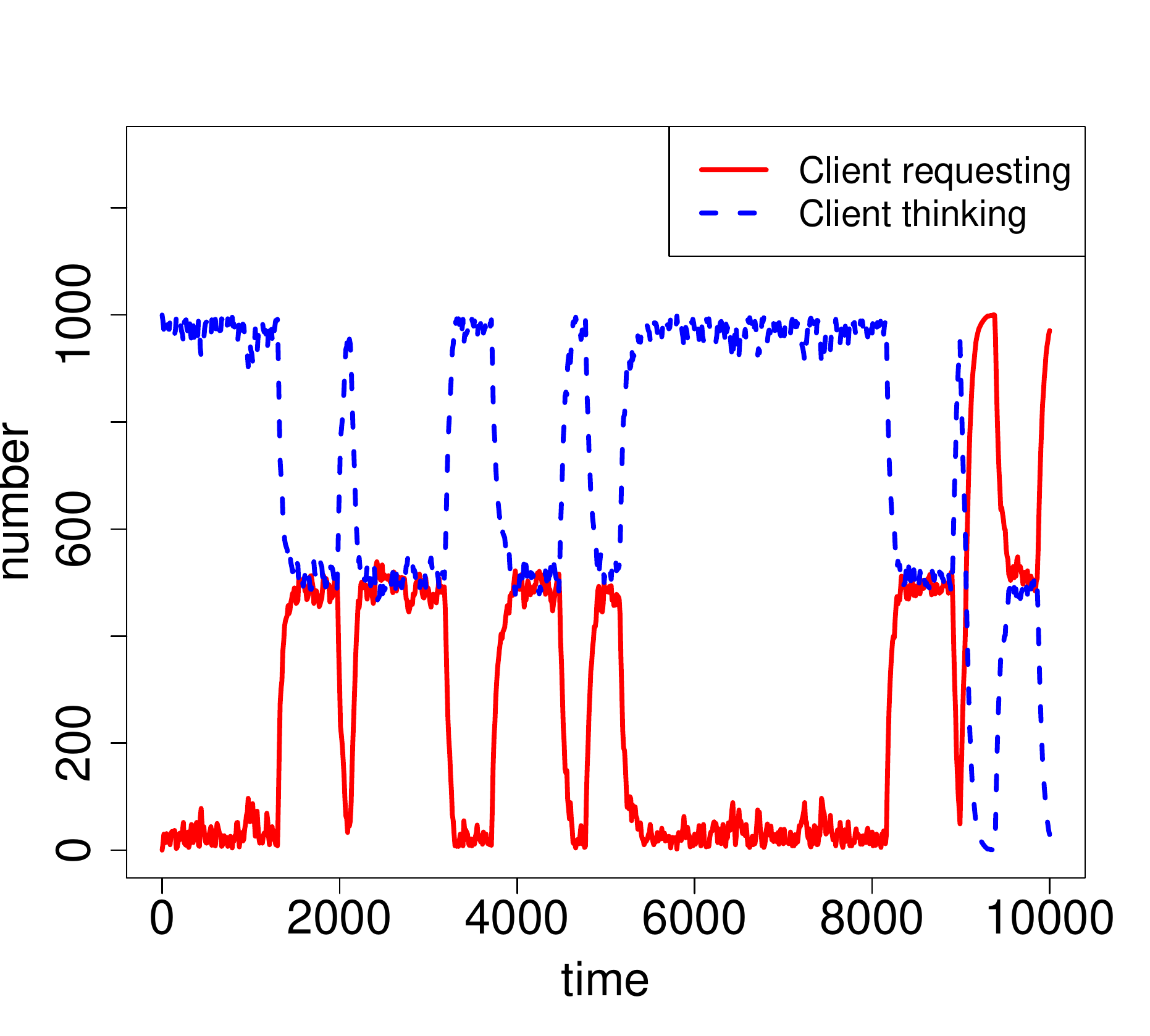} }


\subfigure[$t=10000$, PDMP and CTMC cumulative distributions] {\label{fig:csCDF}
\includegraphics[width=.7\textwidth]{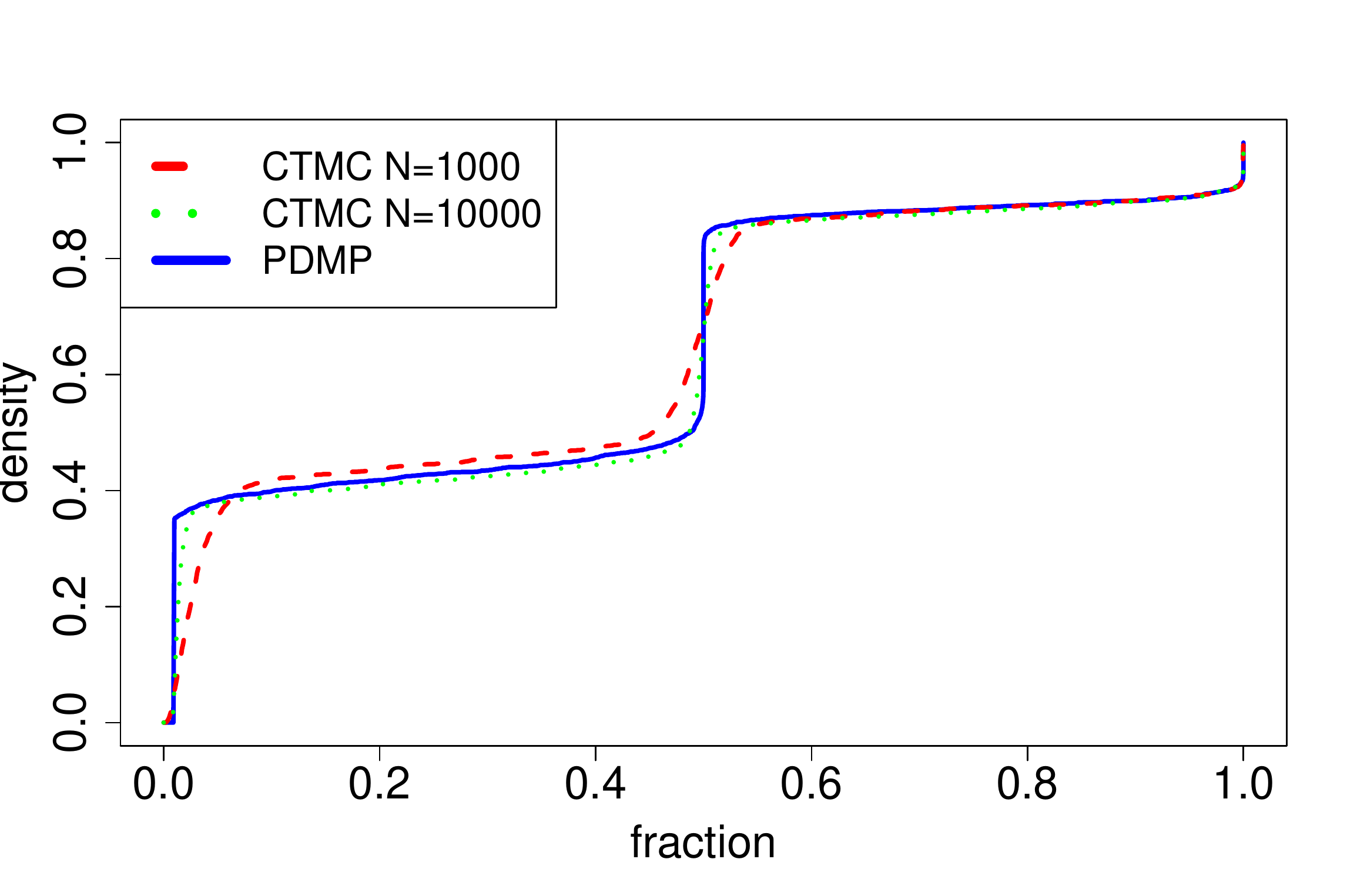} }

\end{center}

\caption{Client Server model of Example \ref{ex:clientServerBasic}, compared with the hybrid limit scaling keeping the number of server fixed to 2. Parameters are  $k_r = 2$, $k_s = 0.01$, $k_t = 1/50$, $k_b = 1/2000$, $k_f = 1/1000$, and initial conditions are $X_r\N(0) = N$, $X_i(0) = 2$. Figures \ref{fig:csHybrid} and \ref{fig:csHstoc} show one trajectory of the PDMP and the CTMC for $N=1000$, respectively. Figure \ref{fig:csCDF}, instead, compares the empirical cumulative distribution of the PDMP limit and the CTMC, for $N=1000$ and $N=10000$, at time $t=10000$, generated from 2500 sampled trajectories.}
\label{fig:csH}
\end{figure}

\begin{exu}
We consider again the simple client server network of Example \ref{ex:clientServerBasic}, but with a different scaling compared to Section \ref{sec:fluidLimit}. In particular, we consider as size $\size{N}$ the number of clients, assuming that the number of servers remains constant, but with service rate depending linearly on $\size{N}$. In this way, the rate of the request transition of \texttt{client} agents is  $\size{N} \min\{k_r \n{X}_r\N,k_sX_i\}$, and it satisfies the continuous scaling. Breakdown and repair transitions, on the other hand, will be kept discrete as they modify only the number of available servers. As their rate is independent of $\size{N}$ and their reset is constant and also independent of $N$, they both clearly satisfy the discrete scaling.
The limit TDSHA that we obtain in this way is shown in Figure \ref{fig:HAexample}. As the hypotheses of Theorem \ref{th:hybridBasic} are satisfied, the sequence of CTMC models obtained from \sCCP\ with the standard stochastic semantics converges (weakly) to the limit TDSHA. 

This can be seen in Figure \ref{fig:csH}, where we compare a trajectory of the CTMC with a trajectory of the PDMP, and the distribution of the number of clients requesting service at time $t=10000$.
\end{exu}

\begin{exu}
We reconsider now the genetic network model of Example \ref{ex:geneNetwork}. Also in this case, we can expect a bimodal behaviour for the CTMC semantics, due to the gene working as a discrete switch. If the binding strength of the repressor is large, meaning that the protein will remain bound to the gene for a long time, then the gene will be switched off for long periods, and we may expect to see a bursty behaviour. This is indeed the case, as can be seen in Figure \ref{fig:gene}. Moreover, the hybrid limit constructed in Example \ref{ex:geneNetwork} matches perfectly this behaviour, as can be seen in Figure \ref{fig:gene}. As the model satisfies the (scaling) assumptions of Theorem \ref{th:hybridBasic}, we can conclude that this is indeed the consequence of the (weak) convergence of the sequence of CTMC models to the hybrid limit. 
\end{exu}

\subsection{More on random resets}
\label{sec:randomResetsDiscrete}

The scaling condition \ref{scaling:discreteStochastic} requires us to check a convergence condition on resets that seems quite complicated at first glance, as it involves checking weak convergence of reset kernels for any possible convergent sequence of states.
We chose this condition because it is very general and it interfaces smoothly with the inductive proof technique that we use in the paper. However, in the following, we will briefly discuss several simpler conditions that can be checked more easily, and that should cover most practical cases.

We first start by observing that we can split the convergence condition in two parts, i.e. we can check that $\resetb{\pi}\N(\xb{},\w{}) \rightarrow \resetb{\pi}(\xb{},\w{})$ uniformly in $\xb{}$ and $\w{}$ and that 
$\W{\pi}\N(\xb{}\N)\Rightarrow\W{\pi}(\xb{})$ (weakly), for any $\xb{}\N\rightarrow \xb{}$.

We now focus attention on the weak convergence of random elements $\W{}\N(\xb{})$ to $\W{}(\xb{})$. 
First, note that the weak convergence condition is essentially equivalent to showing that \[\sup_{\xb{}\in K} \|\integral{\sspace{}}{}{g(\yb{})\bbP\{\W{\pi}\N(\xb{}) = \yb{}\}}{\yb{}} - \integral{\sspace{}}{}{g(\yb{})\bbP\{\W{\pi}(\xb{}) = \yb{}\}}{\yb{}}    \|\rightarrow 0,\] for any compact set $K\subseteq \sspace{}$ and any uniformly continuous function $g:\sspace{}\rightarrow\bbR$ and that $\integral{\sspace{}}{}{g(\yb{})\bbP\{\W{\pi}(\xb{}) = \yb{}\}}{\yb{}}$ is a continuous function \cite{STOC:Karr:1975:WeakConvergenceDTMP}, which may be sometimes easier to check. Moreover, in practice we can expect $\W{\pi}\N$ and $\W{\pi}$ to have a simple structure, which should facilitate the task of verifying the scaling condition. 

First of all, if $\W{\pi}\N$ and $\W{\pi}$ do not depend on $\xb{}$, then the condition reduces to $\W{\pi}\N\Rightarrow\W{\pi}$, which can be checked by showing one of the equivalent conditions of the Portmanteau theorem \cite{STOC:Billingsley:1999:convergenceProbabilities}. In particular, the condition is trivially true if $\W{\pi}\N$ does not depend on $N$, i.e. if $\W{\pi}\N = \W{\pi}$.

We consider now two  examples, to illustrate the use of random resets and the hybrid convergence in this case.

\begin{figure}
\subfigure[Geometric Breakdown: $t=10000$] {\label{fig:csBreakGeomHist}
\includegraphics[width=.48\textwidth]{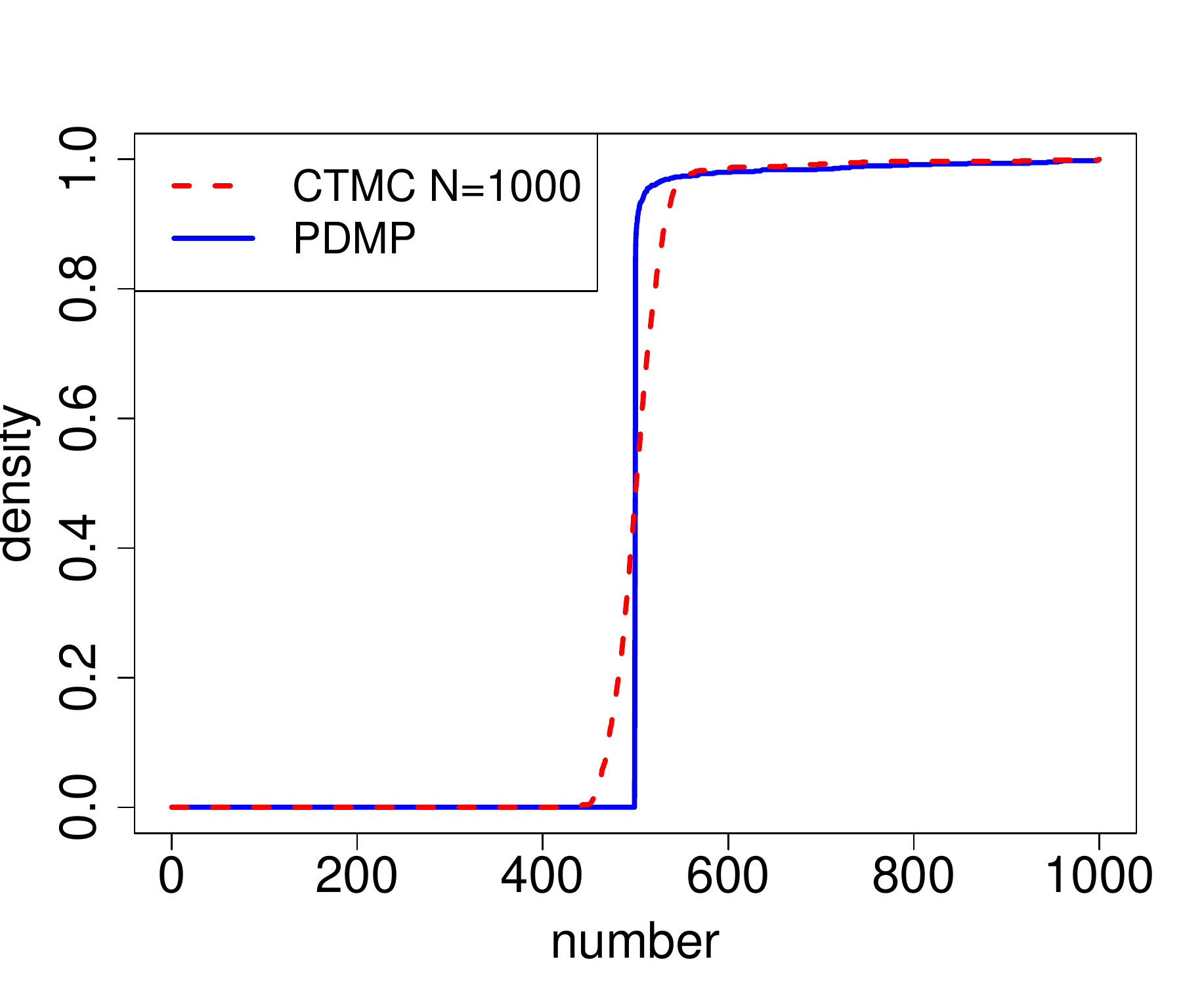}
}
\subfigure[Geometric Breakdown: average] {\label{fig:csBreakGeomAverage}
\includegraphics[width=.48\textwidth]{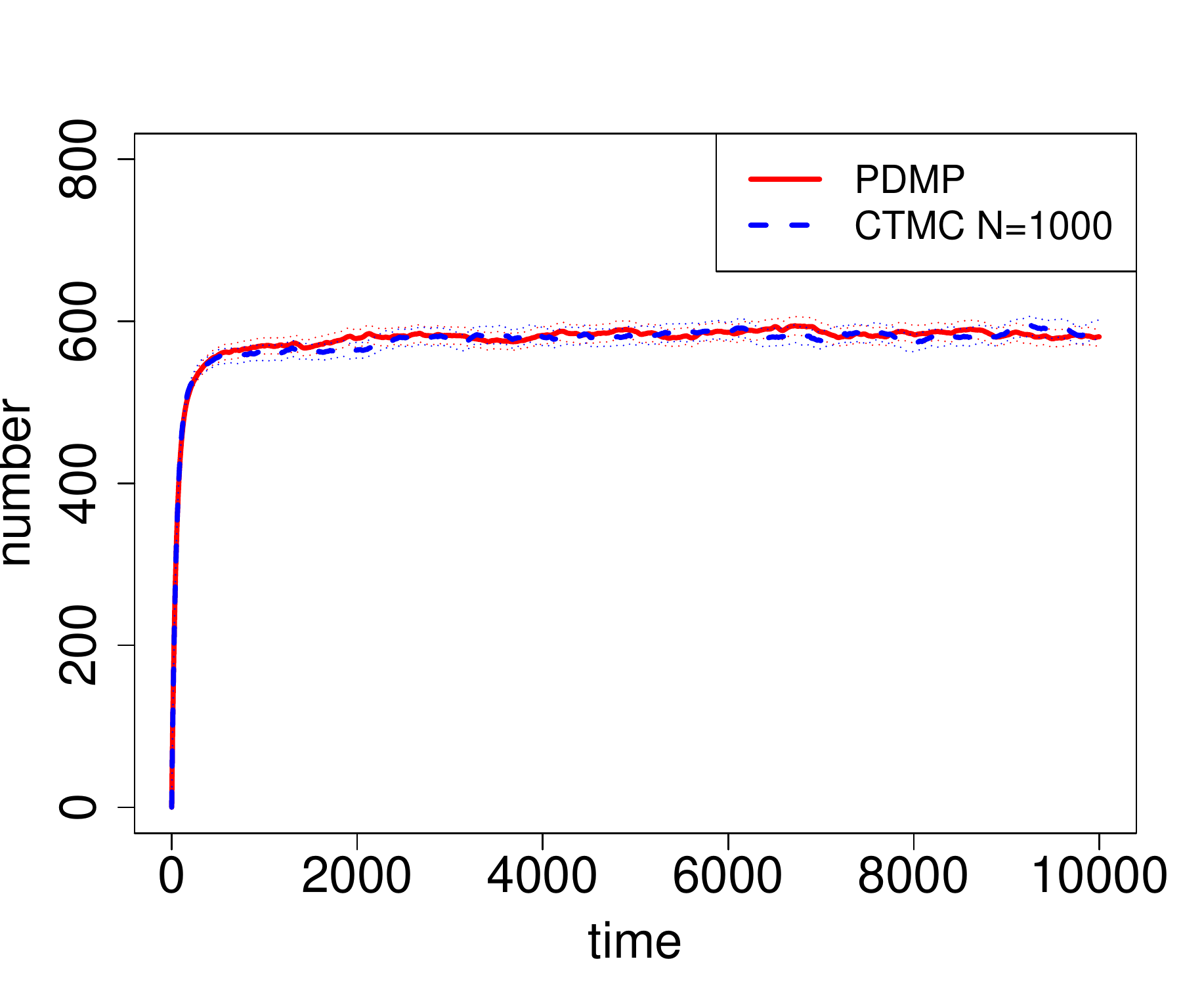} }

\subfigure[Lognormal Breakdown: $t=10000$] {\label{fig:csBreakLognormalHist}
\includegraphics[width=.48\textwidth]{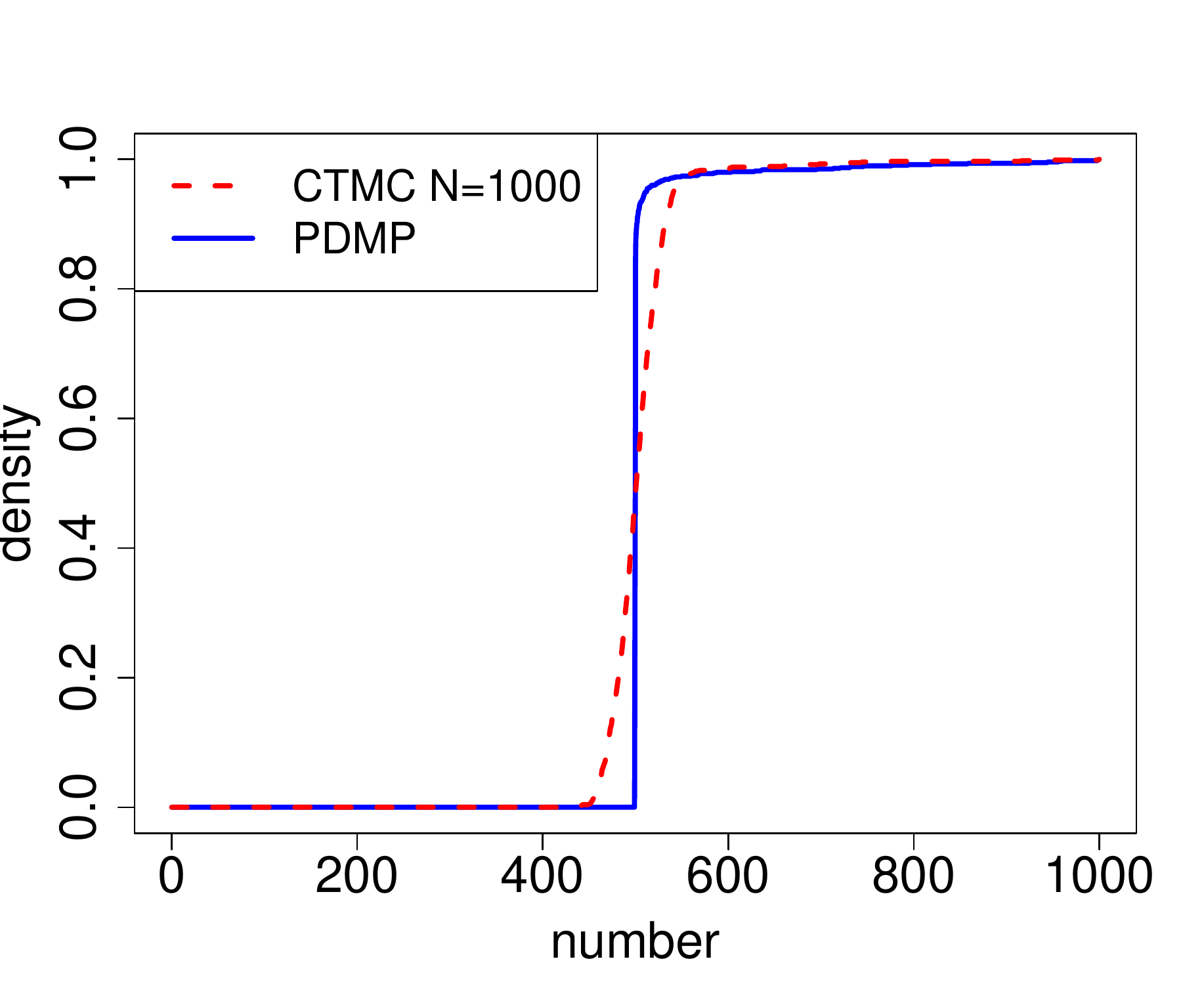} 
}
\subfigure[Lognormal Breakdown: average] {\label{fig:csBreakLognormalAverage}
\includegraphics[width=.48\textwidth]{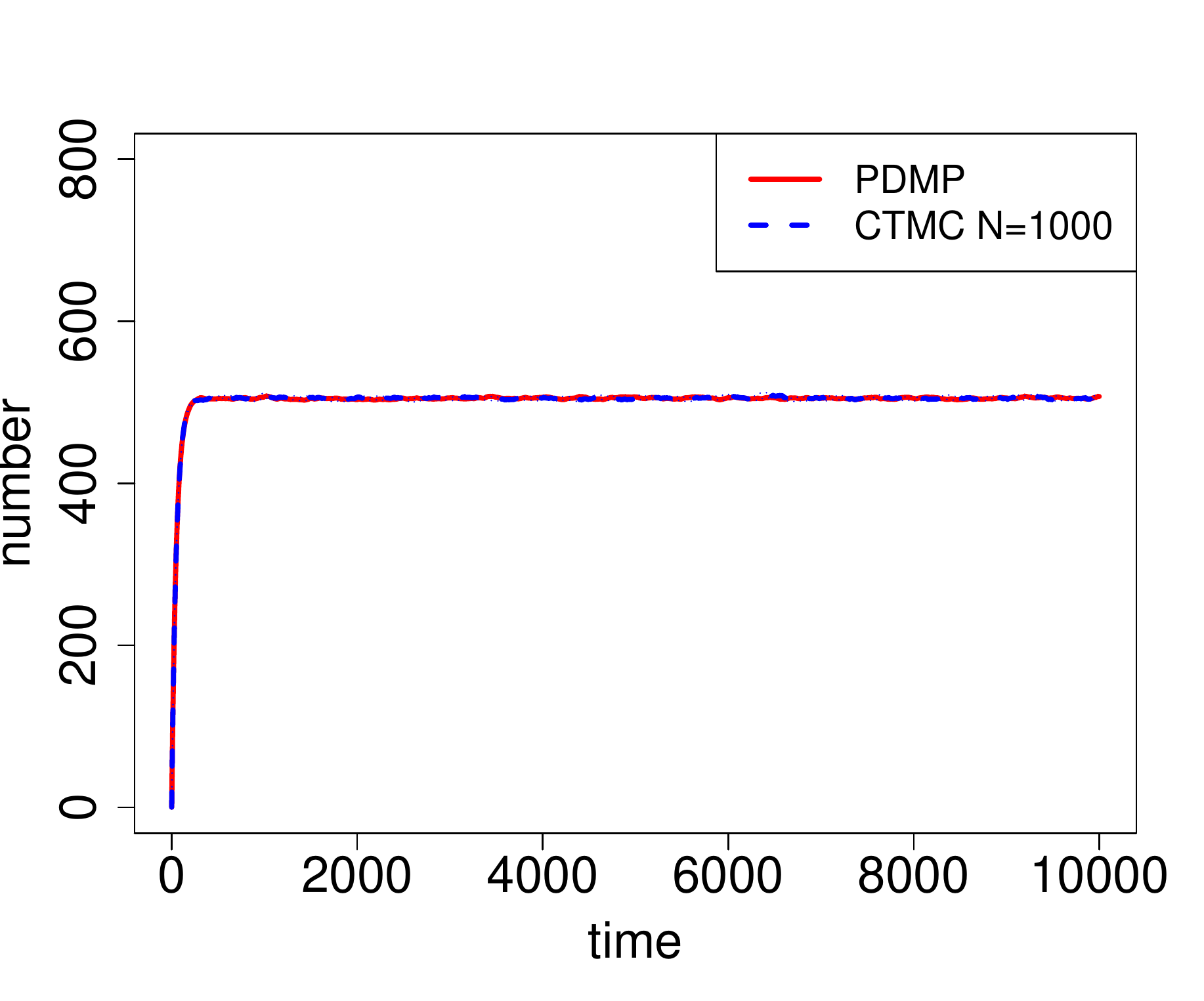} }

\caption{Empirical cumulative distribution of clients requesting service at time $t=10000$ and average number of clients request service for the client server models of Example \ref{ex:csBreakdownVariable}. The top row shows the model with severity level of breakdowns samples according to a geometric distribution with probability $p=0.5$. Parameters are as in caption of Figure \ref{fig:csH}, a part from $k_f = 1/200$. The bottom row shows the model with fix rate lognormally distributed with mean -2.5 and standard deviation 1.0. Note that both histograms present a similar pattern. The bimodality of the distribution is captured for the geometric breakdown. Moreover, the hybrid model has less variability  in the distribution (it has a sharper cumulative distribution function). The averages are almost indistinguishable.}
\label{fig:csBreakdown}
\end{figure}

\begin{example}
\label{ex:csBreakdownVariable}
We consider a small variation of the client server model of Example \ref{ex:clientServerBasic}. The difference is that we will assume different levels of severity of a breakdown, so that the repair time can be variable, depending on this level. For simplicity, we assume a single server, but a generalization to more than one server is straightforward. 
In order to model this situation in \sCCP, we can either 
increase the number of internal states of the server (one for each level of damage) or use an additional (discrete) variable. We chose this second approach, introducing $D$, the damage-level variable. We assume that $D$ takes values on the integers, and that each time a breakdown happens, its value is sampled from a geometric distribution with parameter $0.5$, so that we have a probability $1/2^k$ to see a damage of level $k$. We therefore let $W\sim Geom(0.5)$, so that $\bbP\{W=k\} = (1-0.5)^{k-1}\cdot 0.5$. We further assume that the repair time is proportional to the damage level, so that the rate of repair is $k_f/D$.  We therefore obtain the following \sCCP\ code, where variables $X_r,X_t,X_i,X_b$ are as in Example \ref{ex:clientServerBasic}: 
\begin{center}
{\tt
\begin{tabular}{lcl}
    client   & $\defeq$ & $[*\rightarrow X_r '= X_r - 1 \wedge X_t '= X_t + 1]_{\min\{k_r X_r, \size{N} k_s X_i\}}$.client +\\
 & & $[* \rightarrow  X_r '= X_r + 1 \wedge X_t '= X_t - 1]_{k_t X_t}$.client\\
    server & $\defeq$ & $[*\rightarrow X_i '= X_i - 1 \wedge X_b '= X_b + 1, D' = W]_{k_b X_i}$.server\\
    & + & $[*\rightarrow X_i '= X_i + 1 \wedge X_b '= X_b - 1]_{k_f/D \cdot X_b}$.server
\end{tabular}}
\end{center}
In this case, we clearly have that $X_i$, $X_b$, and $D$ are discrete variables, while $X_r$ and $X_t$ can be approximated continuously. 
$D$ can also be seen as an environment variable, as it is used to modify a parameter of the model. 
Therefore, the transitions of the client agent become continuous transitions in the associated TDSHA, while the transitions of the server agent remain discrete and stochastic. Note that we made explicit the dependence on size in the rate functions. Clearly, all transitions satisfy the scalings of Theorem \ref{th:hybridBasic}. This is true also for the breakdown transition, as $W$ does not depend on the current state of the system. It follows that Theorem \ref{th:hybridBasic} applies also to this example, see also Figures \ref{fig:csBreakGeomHist} and \ref{fig:csBreakGeomAverage}.

A variation of this model is to replace the finite damage levels with a continuous level of damage, essentially sampling the repair rate  from a continuous distribution. This can be done in \sCCP\ by using a real-valued environment variable, call it $K$. For simplicity, here we assume that the fixing rate is sampled from a  lognormal distribution with mean $\mu$ and standard deviation $\sigma$. We can obtain this variant of the model by replacing the \texttt{server} agent with the following one:
\begin{center}
{\tt
\begin{tabular}{lcl}
    server & $\defeq$ & $[*\rightarrow X_i '= X_i - 1 \wedge X_b '= X_b + 1, K' = W]_{k_b X_i}$.server\\
    & + & $[*\rightarrow X_i '= X_i + 1 \wedge X_b '= X_b - 1]_{K \cdot X_b}$.server
\end{tabular}}
\end{center}
Also in this case, the hypotheses of Theorem \ref{th:hybridBasic} are satisfied, and convergence to the hybrid limit works (see Figures \ref{fig:csBreakLognormalHist} and \ref{fig:csBreakLognormalAverage}). 
\end{example}

We turn now to discuss convergence of $\W{\pi}\N(\xb{})$ to $\W{\pi}(\xb{})$ when they depend on $\xb{}$. The situation is more delicate, as convergence has to be uniform. 
In the following, however, we list some sufficient conditions to guarantee convergence, that are of practical relevance. 
\begin{enumerate}
\item  $\W{\pi}\N(\xb{}) = \W{\pi}(\xb{})$ and $\W{\pi}(\xb{})$ depends continuously on $\xb{}$;
\item $\W{\pi}\N(\xb{})$ and $\W{\pi}(\xb{})$ are discrete distributions with mass concentrated on points $\{\yb{1},\ldots,\yb{k},\ldots\}$, and $\bbP\{\W{\pi}\N(\xb{}) = \yb{k}\}$ converges to $\bbP\{\W{\pi}(\xb{}) = \yb{k}\}$ uniformly in any compact $K\subseteq \sspace{}$;
\item $\W{}\N(\xb{})$ and $\W{}(\xb{})$ are unidimensional real random variables, with cumulative distribution functions $F\N(y,\xb{})$ and $F(y,\xb{})$, such that, for each $\xb{}\N\rightarrow \xb{}$,  $F\N(y,\xb{}\N) \rightarrow F(y,\xb{})$ 
pointwise for any continuity point $y$ of $F(y,\xb{})$. 
\item $\W{}\N(\xb{})$ and $\W{}(\xb{})$ have values in $\bbR^h$, and they have continuous density functions $g\N(\y{},\xb{})$, $g(\y{},\xb{})$, and $\sup_{\y{}\in\bbR^k,\xb{}\in K}\|g\N(\y{},\xb{})- g(\y{},\xb{})\| \rightarrow 0$, for each compact set $K\subseteq \sspace{}$.

\item $\W{}\N(\xb{})$ and $\W{}(\xb{})$ can be decomposed into the product of marginal and conditional distributions that converge in the sense of Scaling \ref{scaling:discreteStochastic}, i.e. $\W{}(\xb{}) = \W{i_1}(\xb{})\W{i_2}(\xb{},\w{i_1})\linebreak\cdots \W{i_k}(\xb{},\w{i_1},\ldots,\w{i_{n-1}})$, $\W{}\N(\xb{}) = \W{i_1}\N(\xb{})\W{i_2}\N(\xb{},\w{i_1})\cdots \W{i_k}\N(\xb{},\w{i_1},\ldots,\linebreak\w{i_{n-1}})$, and $\W{i_j}\N(\xb{}\N,\w{}\N) \Rightarrow \W{i_j}(\xb{},\w{})$, as $\xb{}\N\rightarrow \xb{}$ and $\w{}\N\rightarrow \w{}$. 

\item $\W{}\N(\xb{})$ and $\W{}(\xb{})$ are mixtures of distributions $\W{j}\N(\xb{})$ and $\W{j}(\xb{})$ of one of the previous types, i.e. $\W{}\N(\xb{}) = \sum_j p\N_j(\xb{})\W{j}\N(\xb{})$ and $\W{}(\xb{}) = \sum_j p_j(\xb{})\W{j}(\xb{})$.
\end{enumerate}
It is straightforward to show that each of these conditions implies that
$\W{}\N(\xb{}\N) \Rightarrow \W{}(\xb{})$ as $\xb{}\N\rightarrow \xb{}$, hence they can be used whenever it is more appropriate.

\begin{example}
\label{ex:wormSync}
We consider again the client-server model of Example \ref{ex:clientServerBasic}, but modify it by including 
the spread of a worm epidemic. We consider a situation in which a worm has spread on the network and activates on a specific date, sending all infected clients into a non-working state, called $X_d$, from which they need some time to recover. We abstract from the epidemic spreading and model the effect of the epidemics as an event that affects synchronously all clients and infects each of them with probability $p$. Let $W_i(X)$, $i=1,2$,  be binomial distributions with success probability $p$ and size given by $X$. For simplicity, we ignore the breakdown and repair of servers, so that we need four variables, $X_r$, $X_t$, $X_d$, and $X_i$, and initial network \texttt{client} $\parallel$ \texttt{worm}, where \texttt{client} is as in Example \ref{ex:clientServerBasic}, while \texttt{worm} is given by the following code:
\begin{center}
{\tt\small
\begin{tabular}{lcl}
    worm & $\defeq$ & $[*\rightarrow X_r '= X_r - W_1(X_r) \wedge X_t '= X_t - W_2(X_t) \wedge X_d' = X_d + W_1(X_r) + W_2(X_t)]_{k_w}$.worm\\
    & + & $[*\rightarrow X_d '= X_d - 1 \wedge X_r '= X_r + 1]_{k_d \cdot X_d}$.worm
\end{tabular}}
\end{center}

In the limit TDSHA model, the infection action remains discrete and stochastic, while all others are approximated continuously (including the recovery). Here, the system size is clearly the number of clients (server dynamics are ignored, so the number of servers can be seen as a parameter).  When looking at the normalized model for system size $\size{N} = N$, then the reset of the infection transition, say for what concerns clients thinking, is $\n{x}_t - \frac{1}{N} W(\lfloor N\n{x}_t\rfloor)$, which can be also written as $\n{x}_t -  \frac{\lfloor N\n{x}_t\rfloor}{N} \frac{1}{\lfloor N\n{x}_t\rfloor} W(\lfloor N\n{x}_t\rfloor)$, provided $\lfloor N\n{x}_t\rfloor>0$. By the law of large numbers, this expression converges to $\n{x}_t - p \n{x}_t$, so this should be the reset of the limit PDMP. 
However, to apply the limit results of this section to this model, we have to prove that $\frac{1}{N} W(\lfloor N\n{x}\N\rfloor) \rightarrow \n{x}p$ for any $\n{x}\N\rightarrow \n{x}$ and then apply point 3 of proposition above. To show this, observe that if $\n{x}>0$, then $\n{x}\N>\n{x}/2$ ultimately, hence $\lfloor N\n{x}\N\rfloor\rightarrow \infty$, so that $\frac{1}{\lfloor N\n{x}_t\rfloor} W(\lfloor N\n{x}_t\rfloor)\rightarrow p$. When $\n{x}=0$, instead, observe that $\frac{1}{N} W(\lfloor N\n{x}\N\rfloor) \leq    \frac{\lfloor N\n{x}\N\rfloor}{N}\rightarrow 0$, which shows the desired convergence.
 
\end{example}

\begin{figure}
\subfigure[Example \ref{ex:wormSync}, clients requesting at time $t=10000$] {\label{fig:wormHist}
\includegraphics[width=.48\textwidth]{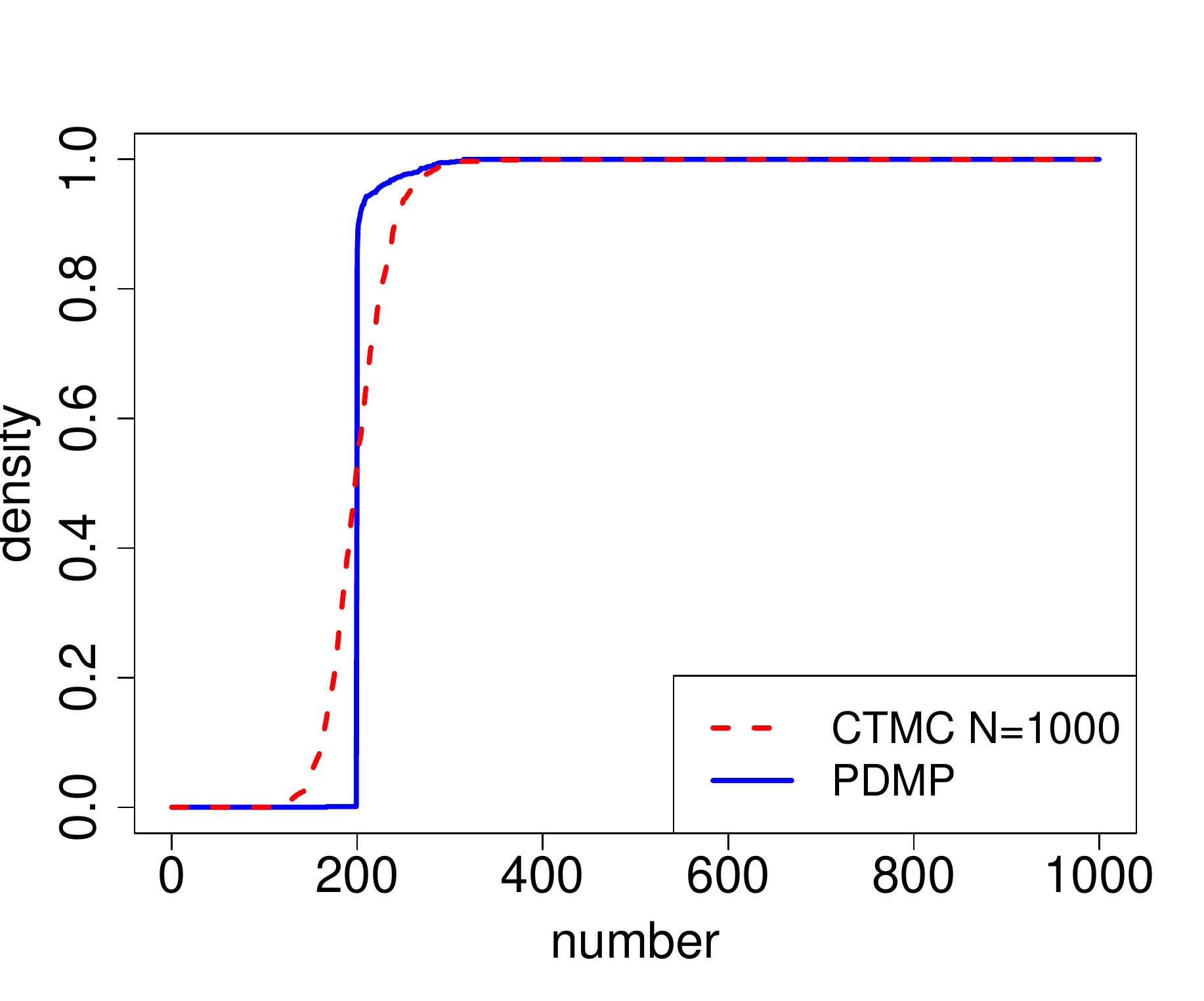} 
}
\subfigure[Example \ref{ex:wormSync}, average] {\label{fig:wormAverage}
\includegraphics[width=.48\textwidth]{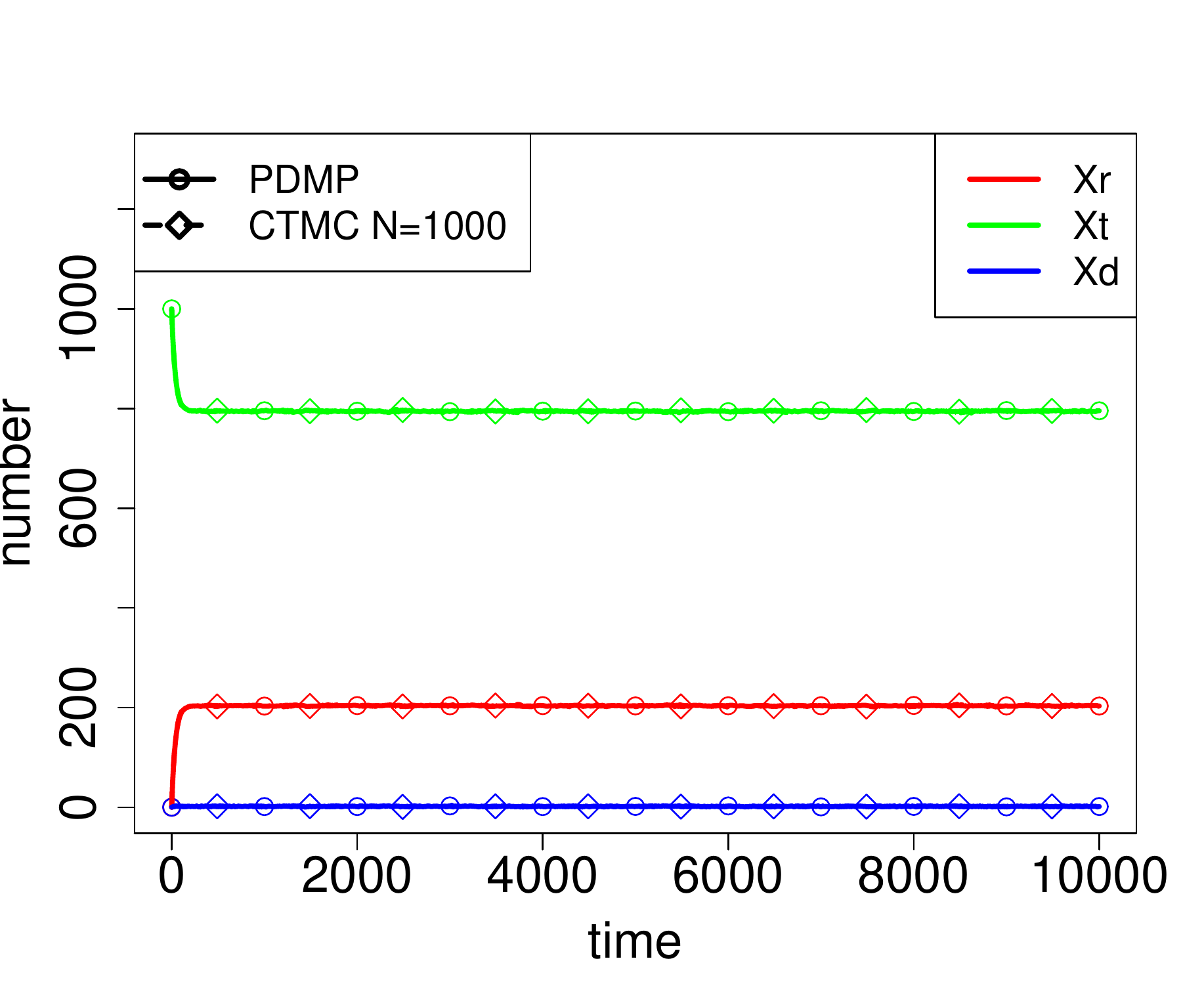} }

\caption{Comparison of the empirical cumulative distribution of clients requesting service at $t=10000$ (left) and average (right) of the limit PDMP and the CTMC at system size 1000, for the client-server model with worm infection of Example \ref{ex:wormSync}. There are 2 servers and $\size{N} = N$ clients. Parameters are $k_s = 0.01\size{N}$, $k_r = 2$, $k_t = 1/40$, $k_w = 1/2000$, $w_d = 0.1$, and the infection probability is $p = 0.33$.}
\label{fig:epiDisc}
\end{figure}


\begin{remarkstar}
\label{rem:checkingContinuousStochasticScaling}
The framework of population-\sCCP\ programs forces the modeller to explicitly consider the notion of system size and to incorporate it in the rate functions. 
This requirement greatly simplifies the manual verification of the scaling conditions, at least for what concerns rate functions. 

There are three kinds of conditions to check: convergence of rate functions, regularity of rate functions (local Lipschitzness), and convergence of reset kernels (or of increments).

Most of the time, these checks are easy to carry out: rates are often density dependent and differentiable and resets are constant increment updates. If rates depend on $\size{N}$, usually this dependence is simple and verifying convergence poses no challenges. For instance, in a biochemical system, the (normalized) mass action rate when two molecules of the same kind react together has the form $k\n{x}(\n{x}-\frac{1}{\size{N}})$, which is easily seen to converge uniformly in any compact set (i.e., whenever $\n{x}$ is bounded). As for the  regularity of rates, most of the time we will deal with functions constructed by algebraic operations, plus some other function like the exponential or the logarithm. All these functions are analytic \cite{THMAT:Kranz:2002:RealAnalyticFunctions}, hence locally Lipschitz. Also the use of minimum or maximum preserves this property. What can be more challenging is the case in which resets have a stochastic part depending on the current state of the model. However, the conditions discussed in this section should cover most of the practical cases. Indeed, we can expect in most models the use within resets of simple discrete or continuous distributions, like Gaussian or uniform ones.

What is undoubtedly more challenging is to make this check automatic. This is partly due to the generality of \sCCP\ as a modelling language, which allows a user to express very complex rates and updates. Hence, a malicious user can construct models that are very complicated to check. However, in most practical cases it may be possible to set up automatic routines that verify the scaling, by clever use of computer algebra systems. 

Another alternative is to identify a library of functions (for both rates and resets) which are guaranteed to satisfy the regularity and scaling conditions. This is  what happens in the process algebra PEPA \cite{PA:Tribastone:2012:FluidPEPA}, where the syntactic-derived restrictions on the possible set of rate functions and updates guarantee that the conditions of the fluid approximation theorem (Theorem \ref{th:Kurtz}) are always satisfied. Constructing a library of ``good'' functions restricts the expressive power of the language, but should be enough to cover most practical modelling activity. Furthermore, libraries can be extended when needed, and the user can also use  additional functions, if she also provides  a ``certificate of correctness''. 
We will pursue this line of investigation in the future, with the implementation of the framework in mind. 

%
%
%
%
%
%
%
%
%
%

\end{remarkstar}

\section{Dealing with instantaneous transitions} 
\label{sec:InstantaneousTransitions}

In this section we discuss convergence to the hybrid limit in presence of instantaneous events. These events remain discrete also in the limit process and can introduce  a discontinuity in the dynamics that is triggered as soon as their guard becomes true. The class of limit PDMP obtained in this way is more difficult to deal with than PDMP with just stochastic jumps. In fact, we cannot rely any more on the ``smoothing'' action in time of a continuous probability distribution like the exponential, but we need to track precisely the times at which instantaneous events happen. In particular, there can be time instants in which we can observe a jump in the limit process with probability greater than zero. This is particularly the case when the hybrid limit is a deterministic process, i.e. a process without discrete stochastic transitions and random resets. 

From the point of view of weak convergence, dealing with instantaneous transitions requires us to prove that
their firing times in the sequence of CTMC models converge to the firing time in the hybrid limit model. Furthermore, we need prove also convergence of the state after the reset. As we will see, both properties are not guaranteed to always hold. The problem resides in the intrinsic discontinuous nature of the exit times and resets on the activation region of guards. Thus, to prove convergence, we need to impose further regularity conditions on the PDMP, forcing its dynamics to avoid these discontinuous regions (with probability 1). We will start by discussing the issues with the exit time,  then turn to reset kernels, and finally move to the limit theorem. After having discussed examples, we will consider a small extension of \sCCP, allowing guards to depend on (simulation) time, and discuss limit theorems for this extended class of models.  

%
%

\subsection*{Convergence of Exit Times}
\begin{figure}
\begin{center}
\subfigure[$\n{x}_1$ trajectory] {\label{fig:tangential}
\includegraphics[width=.47\textwidth]{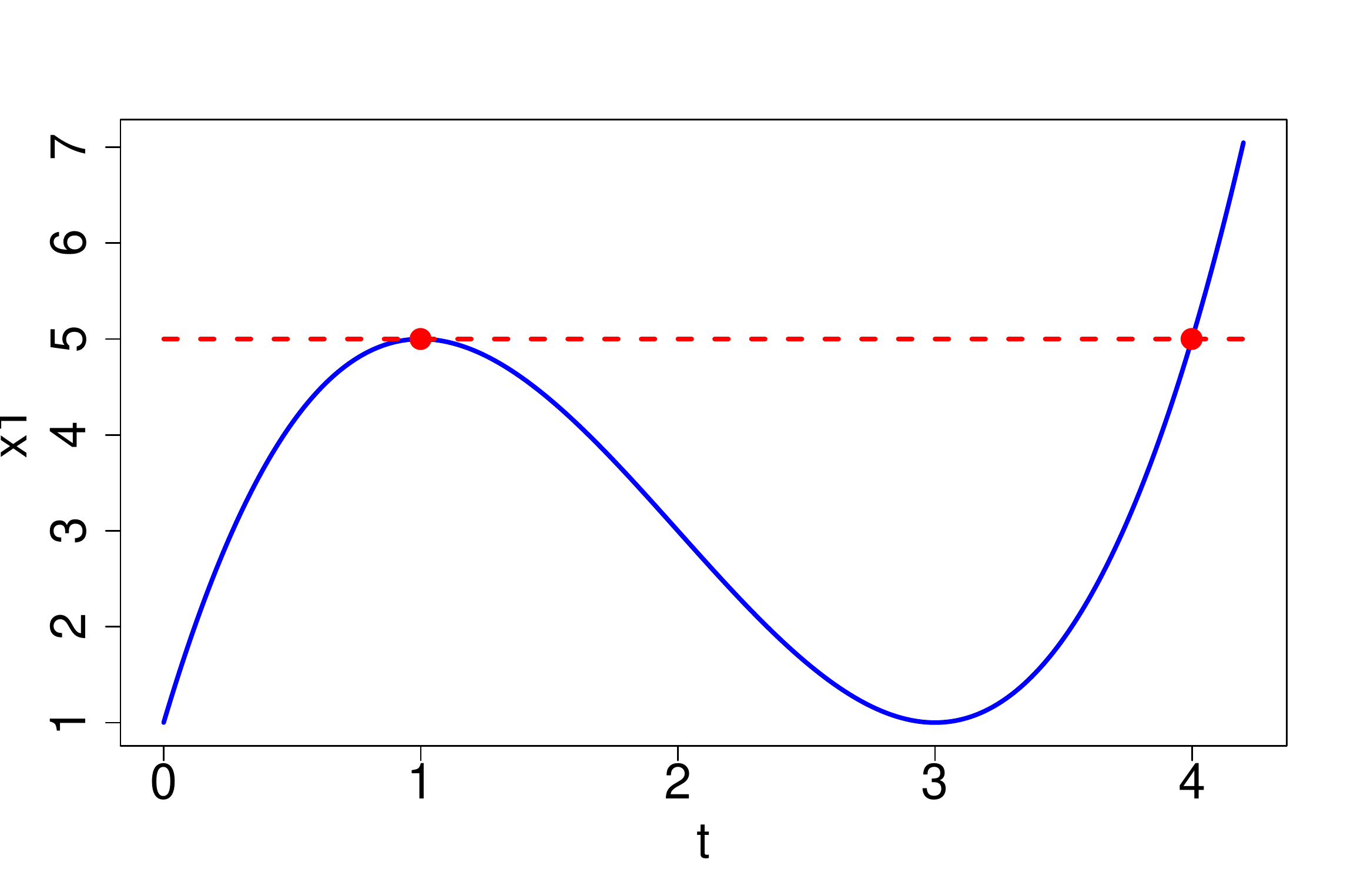} }
\subfigure[$\bbP\{Z\N(t)=1\}$] {\label{fig:tangentialZ}
\includegraphics[width=.47\textwidth]{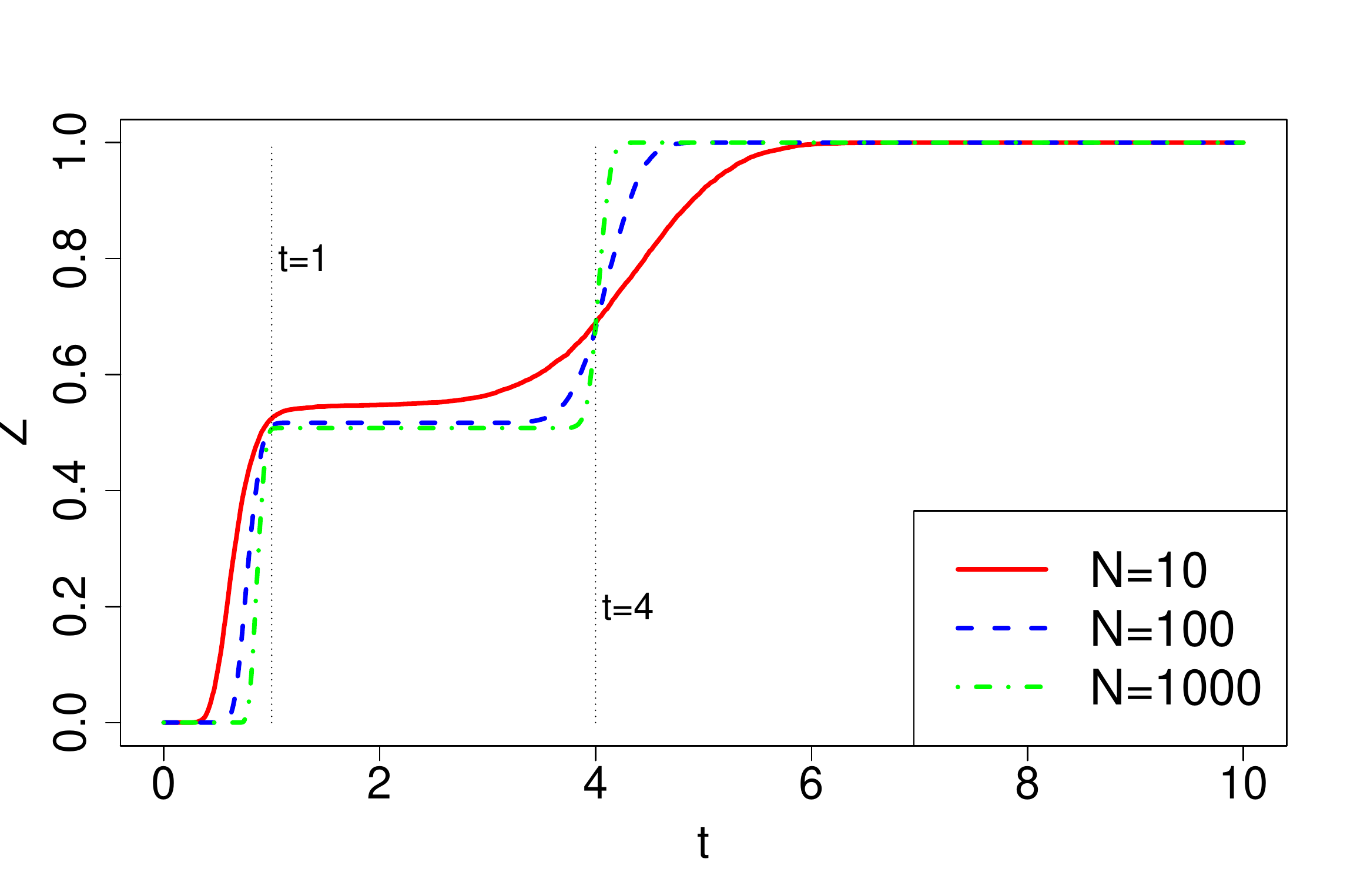} }
\end{center}

\caption{Left: limit trajectory of $\n{x}_1(t)$ for Example \ref{ex:tangential}. Right:  $\bbP\{Z\N(t)=1\}$ as a function of $t$, for different values of $N$. Bi-modality of the distribution around $t=1$ and $t=4$ is manifest. }
\end{figure}

Convergence of exit times does not hold in general.  
Focussing on a deterministic trait of the PDMP dynamics, the problem is created by trajectories of the vector field that activate a guard by touching \emph{tangentially} its boundary surface, as shown in Figure \ref{fig:tangential}. In fact, for $N$ large enough trajectories of the CTMC are contained in a small flow tube around this solution, hence some of them can cross the surface, while others may miss it.
Another class of trajectories that creates problems is that of trajectories remaining in the boundary of the activation region of the guard (i.e., the discontinuity surface of the guard predicate) for a non-negligible amount of time, say for the time interval $[t_1,t_2]$. Here, the problem is that a CTMC trajectory can activate the guard in any time instant between $t_1$ and $t_2$. 

However, convergence holds for  trajectories of the vector field which transversally cross  the discontinuity surface of a guard predicate, meaning that they intersect the surface at time $t$ and enter in the interior of the region in which the guard is true just after $t$. Fortunately, this is the situation we are more likely to find in practice. 

Now we prove a result about convergence of exit times that can be applied to the setting of Section \ref{sec:fluidLimit}. This result requires that almost surely only transversal crossings occur. We will then extend the hybrid limit theorem imposing this condition. We postpone discussion about how to check and/or enforce such a condition until later in the section.

We need some preliminary definitions. The first logical step that we need is to move from predicates to continuous functions in the definition of a guard. 

\begin{definition}
\label{def:activationFunction}
Let $\guardb{}(\xb{})$ be a guard predicate with closed activation region.  A function $h:\sspace{}\rightarrow \bbR$ is an \emph{activation function} or a  \emph{guard function} for $\guardb{}$ if it is a continuous function, and if the sets of points $\{\xb{}~|~h(\xb{})\geq 0\}$ defines the activation region of $\guardb{}$: $\guardb{}(\xb{})$ is true if and only if $h(\xb{})\geq 0$.\\
The function $h$ is a \emph{robust activation function} for $\guardb{}$ if and only if $\partial \{\xb{}~|~h(\xb{}) < 0\} = \partial \{\xb{}~|~h(\xb{}) > 0\} = \{\xb{}~|~h(\xb{}) = 0\}$.
\end{definition}

The notion of robust activation function essentially guarantees that the interior of the set in which $\guardb{}$ is true is $\{\xb{}~|~h(\xb{}) > 0\}$, so that it makes sense to define a transversal crossing of the guard $\guardb{}$ as a change of sign of the function $h$. The discontinuity surface of the guard is therefore $\h{} = \{\xb{}~|~h(\xb{}) = 0\}$. 

The notion of robust activation is a very reasonable assumption to make. In practical cases, the function $h$ will be a piecewise smooth function, usually piecewise linear, and the surface $\h{}$ will be a union of differentiable (or even analytic) manifolds of dimension $n-1$ or less. In particular, a function like $h(\xb{}) = x_1$ if $x_1 \leq 0$ and  $h(\xb{}) = 0$ if $x_1 >0$ is forbidden. This function is a bad activation function as its discontinuity surface $\h{}$ is the whole half-space $\{x_1\geq 0\}$, and furthermore the sequence of functions $h\N(\xb{}) = h(\xb{})-\frac{1}{N}$ converges uniformly to $h$ but their activation set is empty for all $N$. This justifies the use of the term robust: we are forbidding functions which under any small perturbation induce a discontinuous change in the activation set. From now on, we restrict our attention to robust activation functions. 

Coming back to PDMP derived from TDSHA, it is easy to see how to construct a guard function for the class of guards of instantaneous transitions. In fact, we are considering only positive boolean combinations of atoms of the form $h_i(\xb{}) \geq 0$, where $h_i$ is continuous. Then, we just need to combine the functions $h_i$ with maximum and minimum to take into account the structure of boolean combinators.
Furthermore, if in the TDSHA we have $k$ instantaneous transitions $\pi_1,\ldots,\pi_k$, with activation functions $h_1,\ldots,h_k$, we can combine them into a unique activation function $h$ by taking their maximum:
$h(\xb{}) = \max\{h_1(\xb{}),\ldots,h_k(\xb{})\}$. The function is a robust activation function which is greater than or equal to zero if and only if at least one guard is true. We call $h$ the \emph{activation function of the PDMP}.

Consider now a continuous trajectory $\xb{}:[0,\infty)\rightarrow \sspace{}$ and let $h:\sspace{}\rightarrow\bbR$ be a robust activation function, such that $h(\xb{}(0)) < 0$. 

\begin{definition}
\label{def:transversalActFunc} 
The robust activation function $h$ (or the corresponding activation surface $\h{}$) is \emph{transversal} for the trajectory $\xb{}(t)$ if and only if, letting $\zeta = \inf\{t~|~h(\xb{}(t)) \geq 0\}$,  there is a $\delta>0$ such that $h(t) > 0$ for $t\in (\zeta,\zeta+\delta]$. \\
\end{definition}

Suppose now $\Xb{}(t)$ is a stochastic process with almost surely continuous trajectories, like the fluid limit $\xb{}(t)$, with initial conditions (drawn from a distribution) $\xb{0}$. 

\begin{definition}
\label{def:rebustlyTransversal}
An activation function $h$ is  \emph{robustly transversal} to $\Xb{}(t)$ if and only if the set of trajectories for which it is transversal has probability 1.
\end{definition}

The notion of robustly transversal activation function can be lifted to PDMP, by requiring that all the guards of the PDMP are \emph{robustly transversal} in each continuous trait of the dynamics,\footnote{This means that, if the $i$-th jump of a PDMP trajectory $\xb{}(t)$, happening at time $T_{i}$,  corresponds to an instantaneous transition, then there is a $\delta>0$ such that by extending the continuous trajectory starting at $\xb{}(T_{i-1}^+=$ up to $T_{i} +\delta$, the crossing is transversal, i.e. $h(\xb{}(t)) > 0$ for $t\in(T_{i},T_{i} + \delta)$.} i.e. that instantaneous transitions are activated transversally:
\begin{definition}
\label{def:transversalPDMP}
A PDMP $\xb{}(t)$ is robustly transversal if and only if with probability one its trajectories are \emph{robustly transversal} in each continuous trait.
\end{definition}

Consider now  a sequence $\Xb{}\N(t)$ of normalized CTMC associated with a sequence $(\calA,\size{N})$ of population-\sCCP\ models, and assume that $\Xb{}\N(t)$ converges weakly to $\Xb{}(t)$ as $N\rightarrow\infty$, where $\Xb{}$ is a.s. continuous. Furthermore, let $h,h\N:\sspace{}\rightarrow\bbR$  be the activation functions for  $\Xb{}$ and $\Xb{}\N(t)$, respectively. Assume that $h\N$ converges to $h$ uniformly for each compact set $K\subseteq \sspace{}$ and call $\zeta\N  = \inf\{t~|~h\N(\Xb{}\N(t))\geq 0\}$. 

We can show the following lemma, whose proof is given in Appendix \ref{app:proofs}.

\begin{lemma}
\label{lemma:convergenceExitTimes}
Let $(\calA,\size{N})$ be a sequence of population-\sCCP\ models for increasing population size $\size{N}\rightarrow\infty$, as $N\rightarrow\infty$. 
Let $\Xb{}\N(t)$ be the associated normalized sequence of CTMC, and suppose  $\Xb{}\N\Rightarrow\Xb{}$, where $\Xb{}$ has a.s. continuous sample paths. 
Let $h\N$, $h$ be activation functions for $\Xb{}\N$ and $\Xb{}$, such that $h\N\rightarrow h$ uniformly, and suppose $h$ is \emph{robustly transversal} to $\Xb{}$. Then $\zeta\N\Rightarrow \zeta$. \qed
\end{lemma}

\begin{example}
\label{ex:tangential}
We discuss now a hand-crafted example to demonstrate the need for the request of transversal activation of a guard. We consider a population-\sCCP\ program $(\calA,\size{N})$ with three continuous variables $X_1$, $X_2$, and $X_3$ taking values in $\bbZ$, and one discrete variable $Z$. 
\begin{center}
{\tt
\begin{tabular}{lcl}
    agent1   & $\defeq$ & $[X_2\geq 0\rightarrow X_1 '= X_1 + 1 ]_{X_2}$.agent1\\ ù& + & $[X_2 < 0\rightarrow X_1 '= X_1 - 1]_{|X_2|}$.agent1\\   
    agent2   & $\defeq$ & $[*\rightarrow X_2 '= X_2 + 1 ]_{X_3}$.agent2\\
    & + & $[*\rightarrow X_2 '= X_2 - 1 ]_{12\size{N}}$.agent2\\
    agent3   & $\defeq$ & $[*\rightarrow X_3 '= X_3 + 1 ]_{6\size{N}}$.agent3\\

    doom & $\defeq$ & $[X_1\geq 5\size{N} \rightarrow Z = 1 ]_{\infty:1}$.\textbf{0}
\end{tabular}}
\end{center}
The initial network is \texttt{agent1} $\parallel$ \texttt{agent2} $\parallel$ \texttt{agent3} $\parallel$ \texttt{doom}, with initial value of variables $X_1(0) = \size{N}$,  $X_2(0) =  9\size{N}$,  $X_3(0) =  Z(0) = 0$. 

Normalizing the model, we observe that all non-instantaneous transitions satisfy the continuous scaling. If we compute the drift,  we obtain the following set of ODEs (as the guards in the transitions of \texttt{agent1} elicit with the modulus), with initial conditions $(1,9,0)$:
\[\left\{\begin{array}{l}
\dxdt{x_1}{t} = x_2 + 9\\
\dxdt{x_2}{t} = x_3 - 12\\
\dxdt{x_3}{t} = 6\\
\end{array} \right.  \] 
These equations can be integrated directly, obtaining $x_1(t) = t^3 - 6t^2 + 9t + 1$, whose trajectory can be seen in Figure \ref{fig:tangential}. Notice that, for $t=1$, the curve hits tangentially the line $x_1 = 5$, which is the activation surface associated with the robust activation function $h(\xb{}) = x_1-5$, while it transversally crosses such a line at $t=4$. In Figure \ref{fig:tangentialZ}, we show the hitting time distribution for the sequence of CTMC for increasing size $\size{N}$, by visualizing the passage-time distribution of the event $Z=1$, i.e. $\bbP\{Z(t)=1\}$ as a function of $t$. As we can see, the bimodal nature of the distribution persists also for large $N$, supporting the claim that tangential activation creates problems for convergence of exit times.
\end{example}

\subsection*{Convergence of Reset Kernels}

There is a second source of discontinuity induced by instantaneous transitions, namely in the reset kernel of the PDMP on the activation surface of the guards.
The problem lies in the fact that, if we have more than one instantaneous transition, a specific one will be active only in a subregion $\h{\pi}$ of the activation surface $\h{}$, where $\h{} = \{\xb{}~|~h(\xb{})=0\}$ and $\h{\pi} = \{\xb{}\in \h{}~|~h_\pi(\xb{})=0\}$. In particular, the reset kernel is not robust in the boundary $\partial_{\h{}}\h{\pi}$ of $\h{\pi}$ in $\h{}$. In fact, if a trajectory of the PDMP hits $\h{}$ in such a boundary, then a small perturbation can change the set of active guards (including or excluding $\pi$), and the fate of the system may be different. The same problem can manifest itself on the intersection between the activation surfaces of the guards of two instantaneous transitions $\pi_1$ and $\pi_2$: in any neighbourhood of (the boundary of) this region, we can find points in which only one of $\pi_1$ and $\pi_2$ is active.
This lack of robustness reflects itself in a loss of continuity of the reset kernel. Hence we can no longer rely on this property to prove the convergence of the state after the reset (a fact used in the proof of Theorem \ref{th:hybridBasic}).

Intuitively, convergence cannot hold for trajectories $\xb{}(t)$  of the PDMP that hit $\h{}$ in $\partial_{\h{}}\h{\pi}$. In fact, trajectories of the CTMC that converge to $\xb{}(t)$ can hit either $\h{\pi}$ or is complement in $\h{}$, implying that the CTMC can be reset to a different state from the PDMP. 
Furthermore, the probability of hitting $\h{\pi}$ or its complement in $\h{}$ will depend on the geometry of $\h{}$ around the boundary $\partial_{\h{}}\h{\pi}$, rather than on the priority functions governing the choice for the PDMP. We illustrate this point by the following simple example.

\begin{figure}
\subfigure[Example \ref{ex:rwInstTrans}] {\label{fig:rwHit}
\includegraphics[width=.47\textwidth]{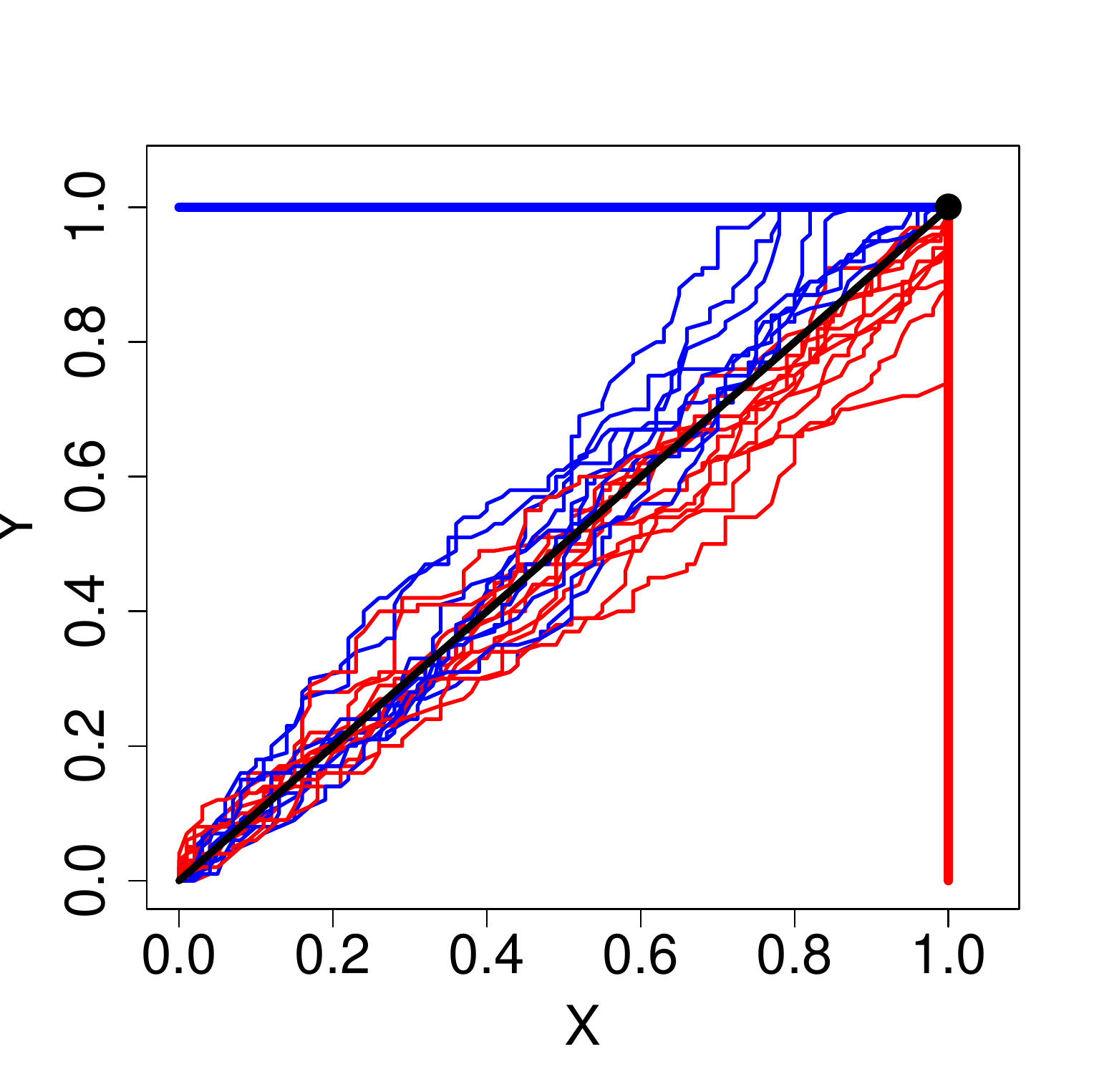} }
\subfigure[Example \ref{ex:rwNdep}] {\label{fig:rwN}
\includegraphics[width=.47\textwidth]{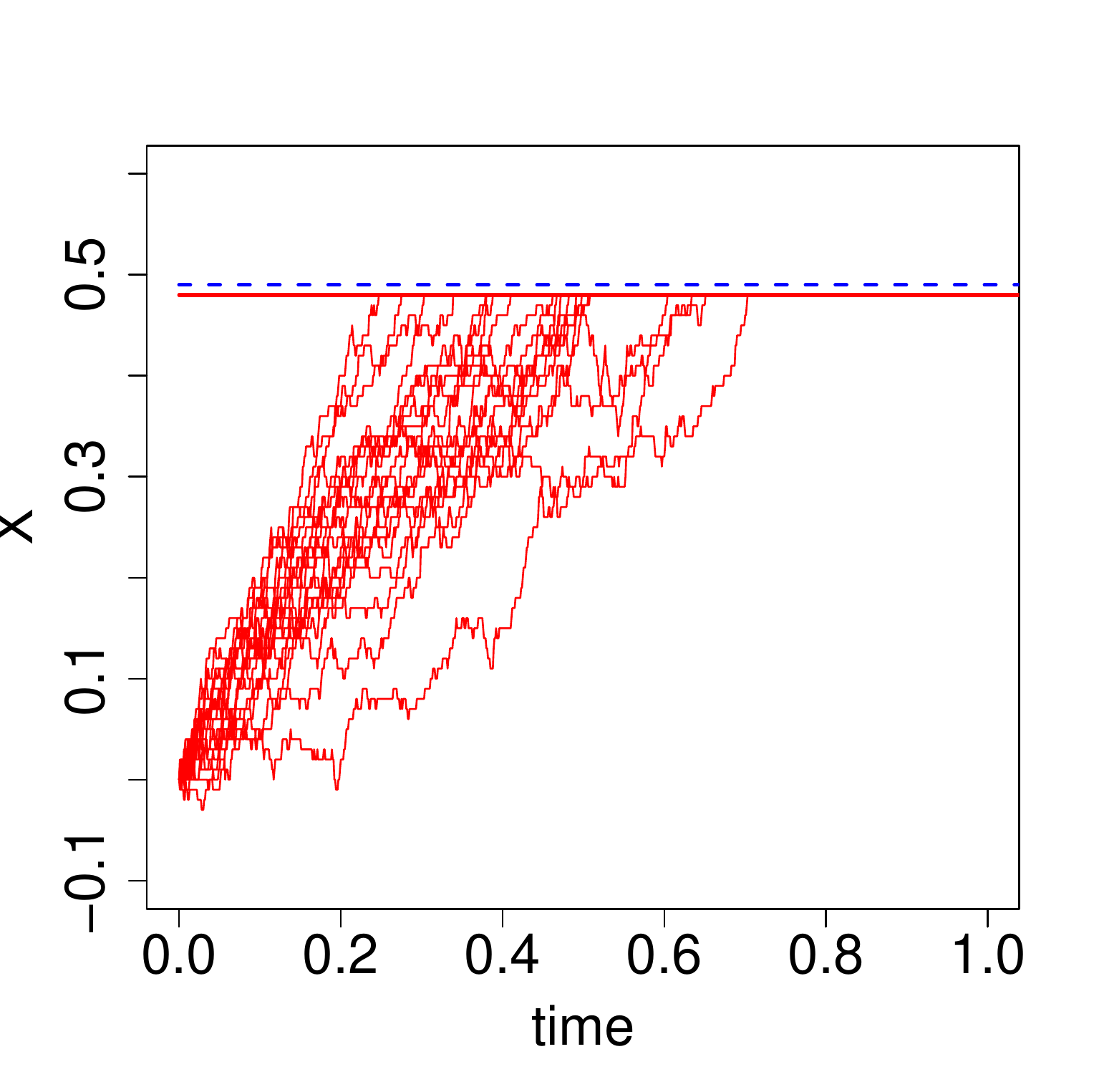} }

\caption{(left) Exemplification of the random walk model  of Example \ref{ex:rwInstTrans}. The activation surface of instantaneous transition \texttt{doom1} is shown in red, while the activation surface for \texttt{doom2} is visualized in blue. Trajectories are coloured according to the surface they hit. The black trajectory is the solution of the PDMP associated with the model.
(right) Exemplification of the random walk model of Example \ref{ex:rwNdep}. The trajectories coloured in red fire the \texttt{doom2} instantaneous transition, whose activation surface is also shown in red. The dotted blue line is the activation surface of \texttt{doom1}. Obviously, no trajectory is coloured in blue, as no trajectory can hit that surface. 
}
\end{figure}

\begin{example}
\label{ex:rwInstTrans}
We consider a model of a one-dimensional random walk in \sCCP. More specifically, we consider a population-\sCCP\ program $(\calA,\size{N})$ with two variables to be approximated continuously, $X$ and $Y$, and one variable $Z$ to be kept discrete. In particular, $X$ and $Y$ will count how many times we go up and  down, respectively. We have the following \sCCP\ code:
\begin{center}
{\tt
\begin{tabular}{lcl}
    up   & $\defeq$ & $[*\rightarrow X '= X + 1 ]_{\size{N}}$.up\\
    down   & $\defeq$ & $[*\rightarrow Y '= Y + 1 ]_{\size{N}}$.down\\
    doom1 & $\defeq$ & $[X\geq \size{N} \rightarrow Z = 1 ]_{\infty:99}$.\textbf{0}\\
    doom2 & $\defeq$ & $[Y\geq \size{N} \rightarrow Z = -1 ]_{\infty:1}$.\textbf{0}
\end{tabular}}
\end{center}
The initial network is \texttt{up} $\parallel$ \texttt{down} $\parallel$ \texttt{doom1} $\parallel$ \texttt{doom2}, with initial value of variables $X(0) = Y(0) = Z(0) = 0$. 
This system is easily seen a to be a one-dimensional random walk for the variable $W = X-Y$. When visualized in the plane $X,Y$, the trajectories of the random walk are  rotated by 45 degrees along the line defined by the equation $Y=X$ (see Figure \ref{fig:rwHit}).
In the normalized system, such trajectories will eventually hit one of the discontinuity surfaces at $\n{x}=1$ or $\n{y}=1$. The vector field of the PDMP is given by $F(x,y) = (1,1)$, hence the solution from the point $\n{x}(0) = \n{y}(0) = 0$ is $\n{x}(t) = \n{y}(t) = t$, which corresponds to the line  $\n{x} = \n{y}$. This line hits the activation surface in its corner point $(1,1)$, where both transitions are active, hence after the reset $Z=1$ with probability 0.99, and $Z=-1$ with probability 0.01. However, in the CTMC $\n{X}$ and $\n{Y}$ can only increment by $1/N$ asynchronously, meaning that each trajectory has to hit one of the two segments of the activation surface before the other. By a simple symmetry argument, we can see that for each $N$, the probability that $Z=1$ after the reset is 0.5 (see again Figure \ref{fig:rwHit}), hence convergence cannot hold for this model.
\end{example}

To have some hope to obtain convergence after a reset, we need to exclude trajectories of the PDMP $\xb{}(t)$ that are troublesome. Consider a \sCCP\ model with $m$ instantaneous transitions $\pi_1,\ldots,\pi_m$, and let $h_{\pi_1},\ldots,h_{\pi_m}$ be the corresponding activation functions, and $h=\max\{h_{\pi_1},\ldots,h_{\pi_m}\}$ be the activation function of the PDMP. 
Define the activation surface $\h{} = \{\xb{}~|~h(\xb{})=0\}$ and $\h{\pi_j} = \{\xb{}\in\h{}~|~h_{\pi_j}(\xb{})=0\}$. Let $D_{j} = \partial_{\h{}} \h{\pi_j}$ be the boundary of $\h{\pi_j}$ in $\h{}$ and $D = \bigcup_j D_j$ be the union of such boundaries, called the discontinuity region of $\h{}$. 

\begin{definition}
\label{def:robustActivationProperty}
A PDMP $\xb{}(t)$ obtained from a TDSHA $\calT$ has the \emph{robust activation property} if and only if the set of trajectories hitting the discontinuity region $D$ of $\h{}$ has probability zero.
\end{definition}

This property essentially tells us that we can ignore the situations in which the PDMP activates instantaneous transitions in non-robust points. 

However, this is not the only issue with reset kernels. There is another problem that can arise when we allow the activation functions $h_{\pi}$ of instantaneous transitions  to depend on the size $\size{N}$, i.e. when $h_\pi\N \neq h_\pi$. The problem, in particular, manifests itself if the activation surfaces of two or more guards with size-dependent activation functions overlap (robustly) in the limit model. 

\begin{example}
\label{ex:rwNdep}
We consider a simple random walk model, with one variable $X$ with values in $\bbZ$, that will be approximated continuously, and a variable $Z$ that will remain discrete.
\begin{center}
{\tt
\begin{tabular}{lcl}
    rw   & $\defeq$ & $[*\rightarrow X '= X + 1 ]_{\size{N}}$.rw\\
    & + & $[*\rightarrow X '= X - 1 ]_{\size{N}}$.rw\\
    doom1 & $\defeq$ & $[Z = 0 \wedge X\geq \size{N}k - 1  \rightarrow Z = 1 ]_{\infty:1}$.\textbf{0}\\
    doom2 & $\defeq$ & $[Z = 0 \wedge X\geq \size{N}k - 2 \rightarrow Z = -1 ]_{\infty:1}$.\textbf{0}
\end{tabular}}
\end{center}
The initial network is \texttt{rw}  $\parallel$ \texttt{doom1} $\parallel$ \texttt{doom2}, with initial value of variables $X(0) = Z(0) = 0$. 
The activation surface for $Z=0$ of \texttt{doom1} in the normalized model is the hyperplane $\n{X} = k - \frac{1}{\size{N}}$, while that of \texttt{doom2} is the hyperplane $\n{X} = k - \frac{2}{\size{N}}$. As $X$ is increased and decreased by one unit only (hence $\n{X}$ is modified by $\frac{1}{N}$ units), for any $N$ the system will always fire \texttt{doom2} (notice that the additional condition on $Z$ forbids  firing \texttt{doom1} once \texttt{doom2} has fired). Hence, $Z=-1$ eventually, for the CTMC models at any population level $N$ (see also Figure \ref{fig:rwN}, for $k=0.5$). However, in the limit model both activation surfaces converge to the limit hyperplane $\n{x} = k$, hence in the limit PDMP $Z$ takes value -1 only with probability 0.5. Convergence again fails.
\end{example}

This example suggest that, in order to avoid such problems, we should either forbid $N$-dependent guards in instantaneous transitions of population-\sCCP\ models, or try to forbid those situations in which more than one $N$-dependent guard can be robustly activated at the same time in the limit model. We state this in the following definition.

\begin{definition}
\label{def:NcompatibleGuards}
A set of activation functions of guards $h_1\N,\ldots,h_m\N$ of a population-\sCCP\ model is \emph{size-compatible} if and only if, for each $j$ such that $h_j\N$ is size-dependent (i.e. $h_j\N$ converges uniformly to $h_j$ in each compact set but $h_j\N\neq h_j$), then $int_{\h{}}(\h{j})\cap\h{i} = \emptyset$, for each $i\neq j$ (i.e. in any point in which the limit activation function $h_j$ is robustly zero in $\h{}$, no other $h_i$ function is zero).

The limit PDMP $\xb{}(t)$ obtained from a population-\sCCP\ model is \emph{size-compatible} if and only if the set of activation function of guards of instantaneous transitions is size compatible. 
\end{definition}

%
%

Technically, Definitions \ref{def:robustActivationProperty} and \ref{def:NcompatibleGuards}  are the key properties that allow us to extend a lemma on the convergence of continuous reset kernels (Lemma \ref{lemma:convergenceAfterReset}), into a more general result capable of dealing with discontinuous reset kernels of the form induced by instantaneous transitions. This will be formally discussed in Lemma \ref{lemma:convergenceAfterResetInstantaneousTransitions} in Appendix \ref{app:proofs}.

\subsection*{Hybrid Convergence Theorem}

The previous lemmas and hypothesis are the core argument for extending Theorem \ref{th:hybridBasic} in the presence of instantaneous transitions. Before proving it, we make explicit the scaling for instantaneous transitions.
\begin{scaling}[Discrete Scaling for Instantaneous Transitions]
\label{scaling:discreteInstantaneous}
A \emph{normalized} instantaneous \sCCP\ transition with \emph{random reset} $\norm{\pi} = (g\N(\Xb{}),\Xb{}' = \resetb{}\N(\Xb{}, \W{}\N(\Xb{}) ), \priorityb{\pi}\N(\Xb{}))$ of a population-\sCCP\ program $(\calA,\size{N})$ with variables partitioned into $\Xb{} = (\X{d},\Xb{c},\X{e})$, with $\Xb{}\in \sspace{}$, satisfies the \emph{discrete scaling} if and only if:
\begin{enumerate}
\item The activation function $h\N(\Xb{})$ of the guard $g\N(\Xb{})$ converge uniformly in  each compact $K \subset \sspace{}$ to a continuous function $h(\Xb{})$;
\item  $\priorityb{\pi}\N(\Xb{}) = O(1)$, $\priorityb{\pi}\N(\Xb{})$ is continuous and it  converges uniformly in each compact $K \subset \sspace{}$ to the continuous function $\priorityb{\pi}(\Xb{})$; 
\item Resets converge weakly (uniformly on compacts), i.e. for each $\xb{}\N\rightarrow\xb{}$ in $\sspace{}$, $\resetb\N(\xb{}\N,\W{}\N(\xb{}\N) \Rightarrow \resetb{}(\xb{},\W{}(\xb{}))$, as random elements on $\sspace{}$.  
\end{enumerate}
\end{scaling}

If an instantaneous \sCCP-transition of the form $\pi = [h_\pi\N(\Xb{})\geq 0 \linebreak\rightarrow \Xb{}' = \resetb{\pi}\N(\Xb{},\W{}\N(\Xb{}))]_{\infty:\priorityb{\pi}\N(\Xb{})}$ satisfies the previous scaling, then the corresponding transition in the limit TDSHA is given by $(h_\pi(\xb{}))\geq 0,\resetb{\pi}(\xb{},\W{}(\xb{})),\priorityb{\pi}(\xb{}))$. 
Consider now the limit  PDMP $\xb{}$ on $\sspace{}$, associated with the normalized TDSHA $\norm{\calT}(\calA)$ constructed from a sequence $(\calA,\size{N})$ of population-\sCCP\ models, in which all transitions satisfy Scalings \ref{scaling:continuousHybrid}, \ref{scaling:discreteStochastic}, or \ref{scaling:discreteInstantaneous}.

\begin{theorem}
\label{th:hybridInstantaneous}
Let $(\calA,\size{N})$ be a sequence of population-\sCCP\ models for increasing system size $\size{N}\rightarrow\infty$, as $N\rightarrow\infty$, with variables partitioned into $\X{} = (\X{d},\X{c},\X{e})$, with discrete stochastic actions satisfying scaling \ref{scaling:discreteStochastic}, instantaneous actions satisfying scaling \ref{scaling:discreteInstantaneous}, and continuous actions satisfying scaling \ref{scaling:continuousHybrid}.
Let $\Xb{}\N(t)$ be the associated sequence of normalized CTMC and $\xb{}(t)$ be the limit PDMP associated with the  normalized limit TDSHA $\norm{\calT}(\calA)$.

If $\xb{0}\N \Rightarrow \xb{0}$ (weakly) and the PDMP is \emph{non-Zeno}, \emph{robustly transversal}, has the \emph{robust activation property} and it is \emph{size-compatible}, then $\Xb{}\N(t)$ converges weakly to $\xb{}(t)$,  $\Xb{}\N \Rightarrow \xb{}$, as random elements in the space of cadlag function with the Skorohod metric. 
\end{theorem}

\proof The proof is only sketched here, see Appendix \ref{app:proofs} for further details. The idea is to reason as in Theorem \ref{th:hybridBasic}, just replacing the machinery about jump times by a more refined one taking into account also instantaneous jumps. Essentially, we have to take the minimum of the stochastic and instantaneous jump times, and choose which reset kernel to use according to which kind of event (stochastic or instantaneous) fires first. The weak convergence of these new jump times and reset kernels follows easily from the convergence of stochastic and instantaneous ones.\qed

%


\begin{remarkstar}
\label{rem:checkingInstantaneousConditions}
Theorem \ref{th:hybridInstantaneous} relies on three global properties of the PDMP associated with the population-\sCCP\ model, namely the robust transversal, the robust activation, and the size-compatibility property. 

The last requirement should be generally relatively easy to check, as it depends only on the activation functions of guards, and not on their interaction with the vector field. 
In fact, in most practical cases, guards are boolean combinations of linear predicates, hence if some of them depend on $N$, by computing the limit activation function (which should also be a combination of linear functions $h_{j,1},\ldots,h_{j,k_j}$), one can discover if there is a robust overlapping of guards by solving a linear system of equations for each pair $i,j$  of size-dependent guards
(say $h_{i,1}(\xb{}) = 0,\ldots,h_{i,k_i}(\xb{}) = 0,h_{j,1}(\xb{}) = 0,\ldots,h_{j,k_j}(\xb{}) = 0$) and checking if the solution has dimension $n-2$ or less in each mode. 
Here we are implicitly assuming that the PDMP and the sequence of CTMC can evolve in an open subset of $\sspace{}$ of dimension  $n$ in each mode (if this is not the case, due to conservation laws, we can just use these laws to reduce the dimensionality of the system).
If the activation functions are non-linear, then the previous approach can be still carried out, but checking the intrinsic dimensionality of a non-linear manifold is obviously more complicated. 

On the other hand, the robustness conditions on the PDMP are more complex to check. The robust transversal property requires that the PDMP transversally crosses an activation surface with probability one. 
If the PDMP is deterministic (i.e., there is no discrete and stochastic transition), then this check can be carried out along the single trajectory starting from the given initial state. In case the number of firings of instantaneous transitions is finite, or these events are ultimately periodic, then it may be possible to set up a semi-decision procedure for this task. 
The problem, also in this simple case, is that checking if a trajectory has a tangential crossing is the same as looking for a non-simple zero\footnote{A zero of a real valued differentiable function is non-simple if also the derivatives of the function up to order $k\geq 1$ are zero in the same point.} of the activation function. 
However, no root finding algorithm is able to properly deal with non-simple zeros, even for analytic functions, see e.g. \cite{COMP:Taylor:20120:RealLambdaCalculus}. In fact, we can only hope to compute a non-deterministic approximation of the trajectory, namely a flow tube around it, which for a tangential activation would intersect the surface but not cross it completely. 
This still does not prove that there is a tangential zero, just that we cannot ignore this possibility. Note that if a trajectory does not intersect the activation surface but a flow tube  of small radius around it does, then the behaviour of the sequence of CTMC can diverge from that of the PDMP due to small fluctuations around the limit trajectory, which can lead to completely different behaviours. In those cases, even if convergence will hold in the limit, the speed of convergence can be very slow.

If the PDMP is a proper stochastic process, then checking the robust transversal property can be even more challenging. In fact, the condition requires us to show that non-transversal activations happen with probability zero. 
One way to approach the problem is to exploit randomness to our advantage. Suppose  that, in a given mode $q$, the continuous state space $\sspace{q}$ has topological dimension $n$  and that we can show that the activation surfaces have (topological) dimension $n-1$ and set of points $B$ in the activation manifold corresponding to non-transversal crossing has (topological) dimension $n-2$ or less. 
Then, the set of points $\sspace{t}$ of $\sspace{q}$ such that $\xb{}(t) \in B$ if $\xb{0}\in \sspace{t}$ has dimension $n-2$ (it is the continuous image of $B$ under the flow of the vector field for $-t$ units of time) so that the subset $\sspace{B}\subseteq \sspace{q}$ of initial points for which $\xb{}(t)$ hits $B$ has dimension $n-1$. 
If we can further prove that the distribution at each time $t$ of the PDMP is absolutely continuous with respect to the Lebesgue measure (i.e. $\bbP(A)=0$ for each Borel set $A$ of Lebesgue measure 0), it necessarily follows that the probability that $\xb{}(t)\in B$ is zero (as $\sspace{B}$ has Lebesgue measure zero). 
This last property can be enforced by requiring that the initial conditions and the reset kernels are absolutely continuous probability distributions (e.g. $n$-dimensional multivariate Gaussian distributions).  
If the system satisfies some conservation law, so that we are interested in its dynamics in a manifold of dimension less than $n$, then we can reduce its dimensionality and analyse the reduced system in the way sketched above. 

Proving that the set $B$ has dimension $n-1$ or less, instead, is more challenging in general. If the activation function of guards are linear (or analytic), and  the vector field is analytic, then one may exploit properties of analytic manifolds for this task, studying the set of zeros of the scalar product of the normal vector to the activation surface with the vector field ($B$, in fact, is contained in this zero set). 
We do not pursue this direction any further in this paper, leaving its investigation for future work, with the goal of providing (semi-)automatic static analysis procedures to check for the applicability of the hybrid approximation method, at least for a practically relevant subclass of population-\sCCP\ models.  

The property of robust activation can be dealt with along the lines sketched above, looking at the dimension of the intersection of activation surfaces. In this case, the task should be considerably simplified if all guards are linear.  

\end{remarkstar}

\begin{example}
\label{ex:epidemicsOscillating}
We consider now a different scenario, in which we model the spreading of a worm epidemic in a computer network. The class of models used for this circumstance is usually drawn from the well developed field of epidemiology, and we make no exception to this rule. We will consider a simple SIR model \cite{STOC:AndersonBritton:2000:StochasticEpidemics}, in which each node of the network has three states: susceptible $X_s$, infected $X_i$ and recovered $X_r$. Here the size of the system $\size{N}$ coincides with the total population $N$ of nodes, which is assumed to be constant, i.e. $X_i + X_s + X_r = N$.  
We assume that infection happens by the malicious action of the worm in infected nodes, which try to send infected messages around the network. There is also a small chance that infection comes externally from the network. Recovery from an infection is obtained by patching an infected computer node. However, after some time new generations of worms appear, and we describe this by the loss of immunity of recovered nodes, that return to be susceptible. We assume that only infected nodes are patched. The \sCCP\ code for this model is as follows:
\begin{center}
{\tt
\begin{tabular}{lcl}
    infection   & $\defeq$ & $[*\rightarrow X_i '= X_i + 1 \wedge X_s '= X_s - 1]_{k_i X_s X_i/\size{N} }$.infection +\\
 & & $[* \rightarrow X_i '= X_i + 1 \wedge X_s '= X_s - 1]_{\size{N} k_e }$.infection\\
    loss\_immunity & $\defeq$ & $[*\rightarrow X_r '= X_r - 1 \wedge X_s '= X_s + 1]_{k_s X_r}$.loss\_immunity\\
    patching & $\defeq$ & $[*\rightarrow X_i '= X_i - 1 \wedge X_r '= X_r + 1]_{k_p X_i}$.patching
\end{tabular}}
\end{center}
The patching rate $k_p$ is the only controllable activity in the system, and we will use instantaneous transitions to model control policies. In particular, we consider here the following policy: if the fraction of infected computers is above a threshold $\alpha_1$, we increase the patch rate from $k_p^0$ to $k_p^1$. If the fraction of infected fall below the threshold $\alpha_0 < \alpha_1$, we switch back to the normal patching rate.
We model this in \sCCP\ by introducing a new variable $U$, taking values 0 or 1, modifying the agent \texttt{patching} as
 \begin{center}
{\tt
\begin{tabular}{lcl}
    patching & $\defeq$ & $[U = 0 \rightarrow X_i '= X_i - 1 \wedge X_r '= X_r + 1]_{k_p^0 X_i}$.patching\\
    & + & $[U = 1 \rightarrow X_i '= X_i - 1 \wedge X_r '= X_r + 1]_{k_p^1 X_i}$.patching
\end{tabular}}
\end{center}
and introducing the \texttt{control} agent
\begin{center}
{\tt
\begin{tabular}{lcl}
    control   & $\defeq$ & $[X_i/\size{N} > \alpha_1 \rightarrow U' = 1]_{\infty:1}$.control\\
 & + &  $[X_i/\size{N} < \alpha_0 \rightarrow U' = 0]_{\infty:1}$.control\\
\end{tabular}}
\end{center}

\begin{figure}
\subfigure[$k_p^1 = 2$, $\alpha_1 = 0.1$] {\label{fig:epidecmicsOscillatingNice}
\includegraphics[width=.47\textwidth]{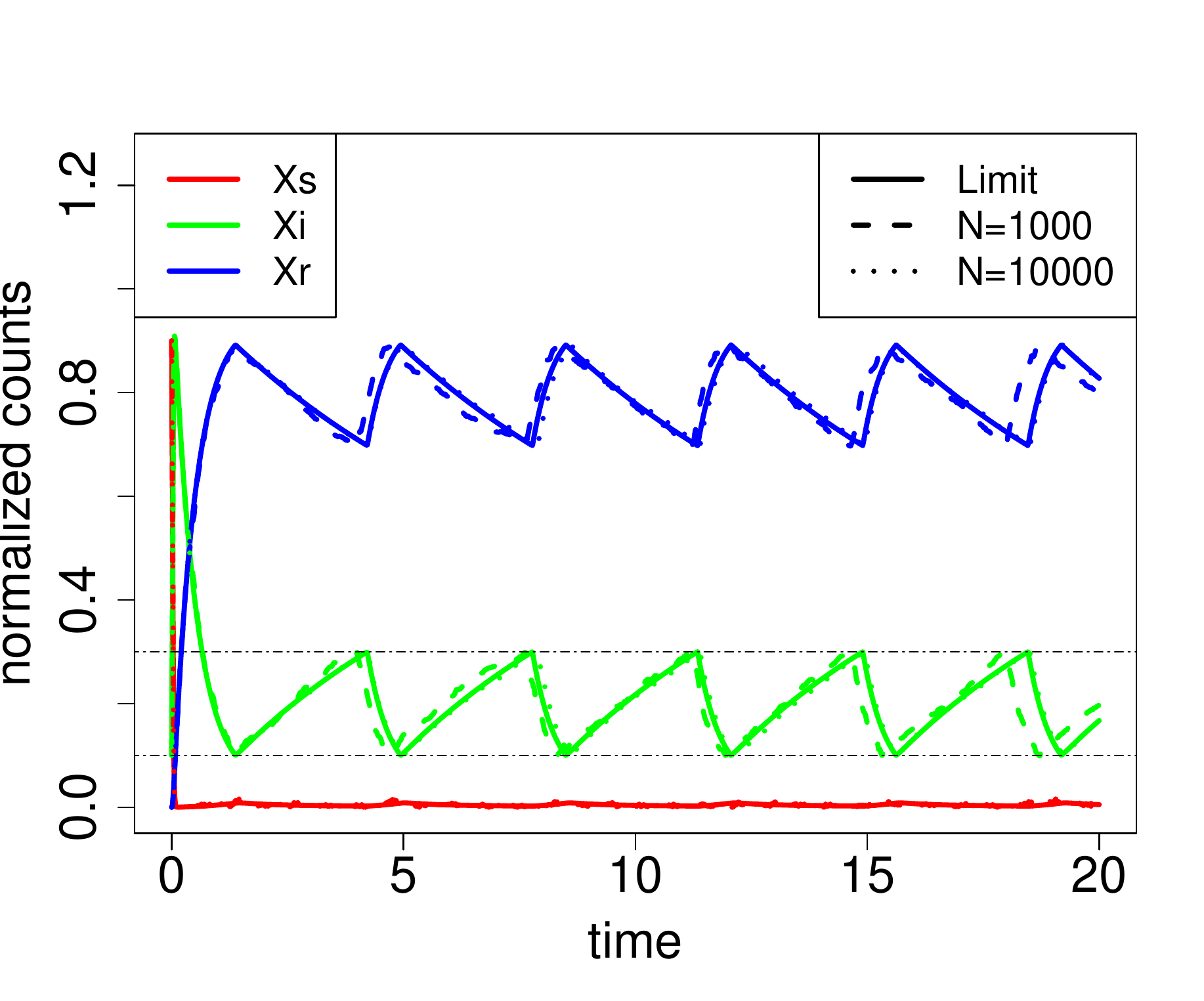} }
\subfigure[$k_p^1 = 1$, $\alpha_1 = 0.088$] {\label{fig:epidecmicsOscillatingBad}
\includegraphics[width=.47\textwidth]{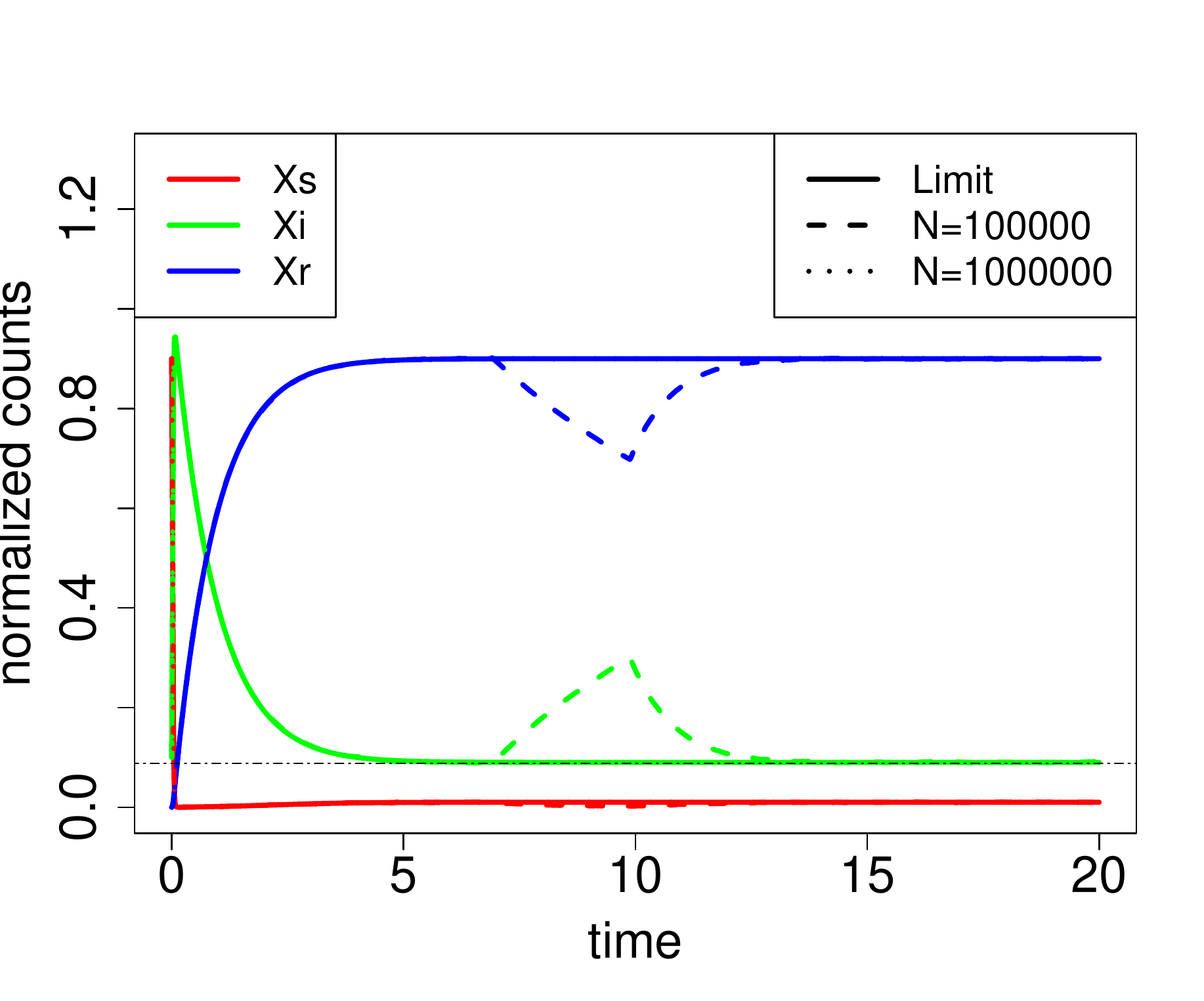} }
\caption{Comparison of hybrid and stochastic trajectories of the epidemics model of Example \ref{ex:epidemicsOscillating}, for parameters $k_i = 100$, $k_e = 0.001$, $k_s = 0.1$, $k_p^0 = 0.1$, $k_p^1 = 2.0$ (left) or $k_p^1 = 1.0$ (right). Control thresholds are $\alpha_0 = 0.3$ and $\alpha_1 = 0.1$ (left) or $\alpha_1 = 0.088$ (right). Initial conditions are $\n{x}_i(0) = 0.1$ and $\n{x}_s(0)=0.9$. Note that for $N$ large the stochastic trajectory is indistinguishable from the hybrid limit one. In the figure on the right, the threshold $\alpha_1$ is slightly smaller than the steady state of the ODE when $U=1$. However, the stochastic system can hit the threshold and change mode, even for $\size{N}$ large. In any fixed time horizon, this event becomes less and less likely as $N$ goes to infinity: here, we observe a spike for $N=100000$, but not for $N$ equal to one million. }
\label{fig:epidemicsOscillating}
\end{figure}

\noindent First, note that this model satisfies the scaling assumptions of Theorem \ref{th:hybridInstantaneous}, when all variables except $U$ and all stochastic \sCCP-transitions are considered as continuous. As there is no stochastic transition, the limit model is deterministic. Hence, if we start from a given initial state, we need to check that the single trajectory of the limit model satisfies the assumptions. For a given set of parameters, shown in Figure \ref{fig:epidemicsOscillating}, we can choose  $\alpha_1$ such that the steady state of the limit model with low patching rate is above $\alpha_1$ and $\alpha_0$ such that the steady state of the limit ODE model with high patching rate is below $\alpha_0$, inducing oscillations in the limit hybrid model. This is confirmed in Figure \ref{fig:epidecmicsOscillatingNice}, where we can also check visually that the crossing of the guard surfaces is always transversal. A formal proof can be given as well, by verifying that the projection of the vector field on the orthogonal direction to $\n{X}_i = \alpha_1$ or $\n{X}_i = \alpha_0$ is null in a single point, namely $(k_p^0/k_i,\alpha_1,1-k_p^0/k_i-\alpha_1)$ or $(k_p^1/k_i,\alpha_0,1-k_p^1/k_i-\alpha_0)$, respectively, and observing that the trajectory in Figure \ref{fig:epidemicsOscillating} never passes from these points. 
As the conditions of Theorem \ref{th:hybridInstantaneous} are satisfied, we can conclude that the sequence of CTMC models obtained by the \sCCP-program for increasing $\size{N}$ converges to the hybrid system.

In Figure \ref{fig:epidecmicsOscillatingBad}, we show the same model for a different high patching rate and a different threshold $\alpha_1$, such that the limit model in state $U=1$ converges to a steady state slightly greater than  $\alpha_1$, thus never activating the instantaneous transition. We can see that the stochastic model behaves in the same way, but for $N$ very large, because the proximity of $\alpha_1$ induces an activation of the transition in stochastic trajectories with small (but non null) probability for any $N$. In particular, this implies that if we leave enough time, almost surely a stochastic trajectory will eventually cross the surface $\n{x}_i = \alpha_1$, changing discrete mode. This shows that the notion of weak convergence is restricted to the transient behaviour, but does not bring in general information about the steady state. 
\end{example}

%
%
%
%
\subsection{Time-Dependent Guards}
\label{sec:timedEvents}

Guards depending on time on instantaneous transitions can be a valuable addition to the modelling language, as they allow us to describe global events that have a duration, which can be either deterministic or stochastic. 

More precisely, we consider the extension of \sCCP\ \cite{MyNETMAHIB2012} with a reserved keyword $time$, referring to simulation time, whose usage is confined to guards and update functions of instantaneous transition, and to update functions of stochastic transitions, which have to be kept discrete. 
Moreover, the special variable $time$ can never be updated. 
Specifically, we allow instantaneous transitions of the form $[\guard{\pi}(\X{},time) \rightarrow \X{}' = \reset{\pi}(\X{},\W{},time)]_{\infty:w}$. In particular, $\guard{\pi}(\X{},time)$ is required to be of the form $time \geq h_0(\X{}) \wedge \guard{\pi,1}(\X{})$, for some function $h_0$ and some standard guard predicate $\guard{\pi,1}(\X{})$ (with activation function $h_1(\Xb{})$).  We call a population-\sCCP\ model  $(\calA,\size{})$ with timed-guards a \emph{time-guarded population-\sCCP\ model}.

Translation of these transitions to TDSHA is straightforward and follows the same scheme as Section \ref{sec:sCCPtoTDHA}. The only difference is that in the TDSHA/PDMP setting it is more convenient to internalize the notion of time, by adding a dedicated clock variable keeping track of the global simulation time. This is done by adding a new continuous variable, $Time$, and a new automata, called \emph{time-monitor}, in the parallel composition of TDSHA, with a single continuous transition of the form $(\vr{1}_{Time},1)$, where $\vr{1}_{Time}$ is the vector equal to one in the position of the variable $Time$, and zero elsewhere.

In the following, we restrict our attention to time-guarded transitions that satisfy the following scaling assumption with respect to the population size $\size{N}$.
\begin{scaling}[Discrete Scaling for Time-Guarded Instantaneous Transitions]
\label{scaling:timedInstantaneous}
A \emph{normalized} time-guarded instantaneous \sCCP\ transition with \emph{random reset} $\norm{\pi} = (g\N(\Xb{},time),\Xb{}' = \resetb{}\N(\Xb{}, \W{}\N(\Yb{}), time),$ $\priorityb{\pi}\N(\Xb{}))$ of a population-\sCCP\ program $(\calA,\size{N})$ with  variables partitioned into $(\X{d},\X{c},\X{e})$, $\Xb{} \in \sspace{}$, has \emph{discrete scaling} if and only if:
\begin{enumerate}
\item The activation function of the guard $g\N(\Xb{})$, which is $\min\{time - h_0\N(\Xb{}),h_1\N(\Xb{})\}$, is such that $h_i\N(\Xb{})$ converges uniformly in each compact $K \subset \sspace{}$  to a continuous function $h_i(\Xb{})$, $i=0,1$. 
Furthermore, $h_0\N$ and $h_0$ \emph{do not} depend on the variables $\Xb{c}$ of $\Xb{}$ that are modified continuously;
\item  $\priorityb{\pi}\N$ satisfies the same conditions as in Scaling \ref{scaling:discreteInstantaneous}; 
\item Resets converge weakly (uniformly on compacts), i.e. for each $(\xb{}\N,t\N)\rightarrow(\xb{},t)$ in $\sspace{}\times \bbR_{\geq 0}$, $\resetb\N(\xb{}\N,\W{}\N(\xb{}\N),t\N) \Rightarrow \resetb{}(\xb{},\W{}(\xb{}),t)$, as random elements on $\sspace{}$.  
\end{enumerate}
\end{scaling}

Under the previous scaling, if we consider initial conditions (or the state after one jump) such that $\Xb{0}\N\Rightarrow\Xb{0}$, given  the independence of the activation function from variables modified continuously, we easily obtain $h_0\N(\Xb{0}\N)\Rightarrow h_0(\Xb{0})$ (reason as in Lemma \ref{lemma:convergenceExitTimes}). Recalling that in the limit PDMP, $Time$ is treated like a regular continuous variable, we can combine this observation with the discussion about exit times and reset kernels in the previous section to obtain convergence. Note, in particular, that the activation condition on $time$ has always a robustly transversal activation function (as $Time$ is monotonically increasing). Hence, the only problems for convergence of a timed-transition can come from the other component of the guard (i.e. from the activation function $h_1$).
Therefore, we obtain that the time $T\N$ in which $g\N$ becomes true converges weakly to the time $T$ in which $g$ becomes true. Then, a minor adaptation of the proof of Theorem \ref{th:hybridInstantaneous} gives the following

\begin{proposition}
\label{prop:hybridTimeGuards}
Let $(\calA,\size{N})$ be a sequence of time guarded population-\sCCP\ models for increasing systems size, 
$\size{N}\rightarrow\infty$, as $N\rightarrow\infty$, with variables partitioned into $\X{} = (\X{d},\X{c},\X{e})$, with discrete stochastic actions satisfying scaling \ref{scaling:discreteStochastic}, instantaneous actions satisfying scaling \ref{scaling:discreteInstantaneous}, time guarded actions satisfying scaling \ref{scaling:timedInstantaneous}, and continuous actions satisfying scaling \ref{scaling:continuousHybrid}.
Let $\Xb{}\N(t)$ be the associated sequence of normalized CTMC and $\xb{}(t)$ be the limit PDMP associated with the sequence of normalized TDSHA $\norm{\calT}(\calA,\size{N})$.

If $\Xb{0}\N \Rightarrow \xb{0}$ (weakly) and the PDMP is \emph{non-Zeno}, \emph{robustly transversal}, has the \emph{robust activation property} and it is \emph{size-compatible}, then $\Xb{}\N(t)$ converges weakly to $\xb{}(t)$,  $\Xb{}\N \Rightarrow \xb{}$, as random elements in the space of cadlag function with the Skorokhod metric. \qed 
\end{proposition}

\begin{figure}
\subfigure[Weibull Breakdown: $t=10000$] {\label{fig:csBreakWeibullHist}
\includegraphics[width=.48\textwidth]{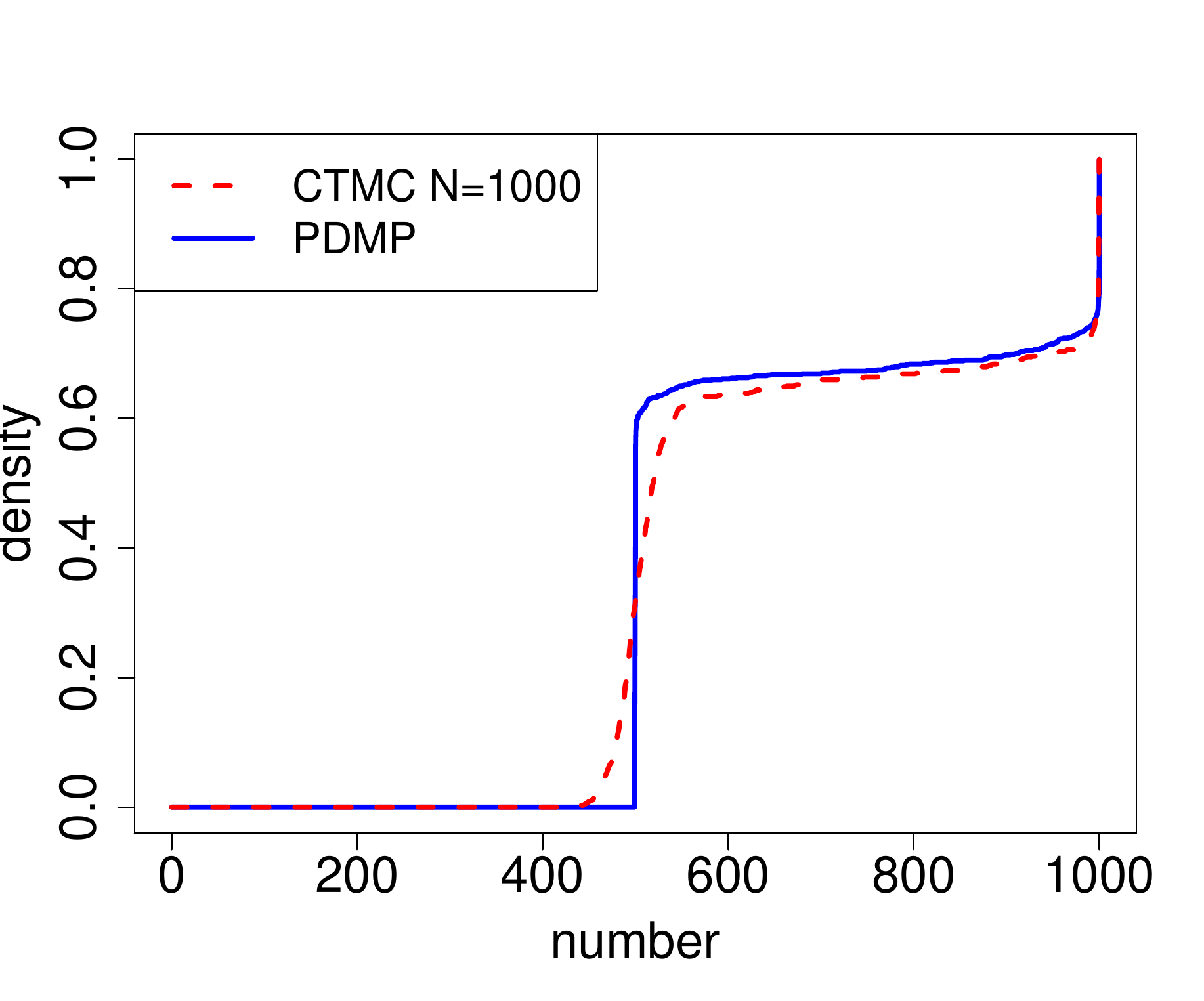}
}
\subfigure[Weibull Breakdown: average] {\label{fig:csBreakWeibullAverage}
\includegraphics[width=.48\textwidth]{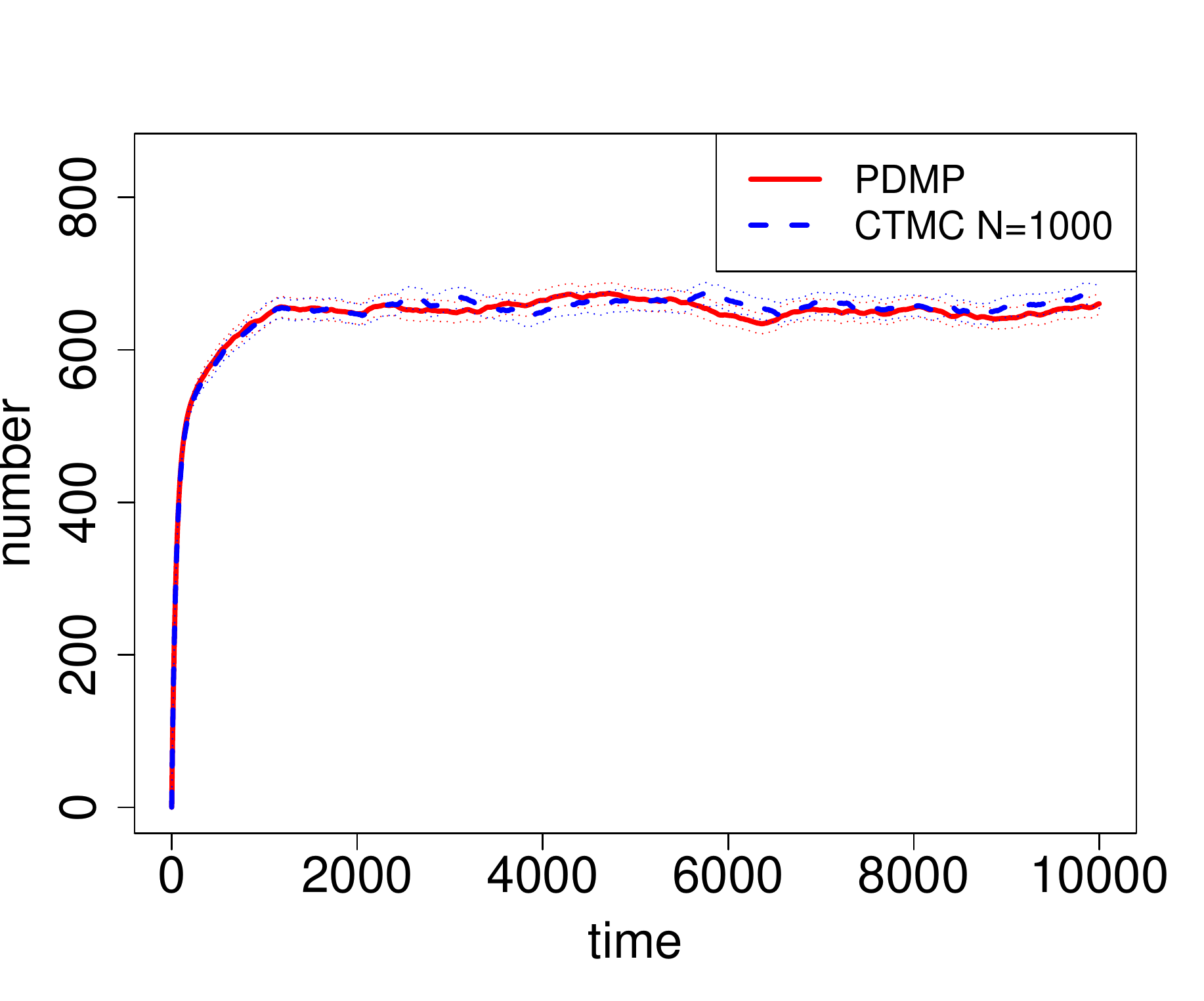} }

\caption{Empirical cumulative distribution of clients requesting service at time $t=10000$ and average number of clients request service for the client-server model of Example \ref{ex:csBreakDownGen}. The model has a fixing time sampled from a Weibull distribution with shape  1.5 and rate 1/1000. Other parameters are as in the caption of Figure \ref{fig:csH}. The bimodality of the distribution is captured, with less variability in the the hybrid model. The average is almost indistinguishable.}
\label{fig:csBreakdownWeibull}
\end{figure}

\begin{example}
\label{ex:csBreakDownGen}
As an example of time-dependent guards, we consider again the client-server model with breakdown, as in Example \ref{ex:csBreakdownVariable}. In that example, we used random resets to model a variable level of damage, reflecting in the time needed to repair the system. Here, instead, we consider a single damage level, but with a generally distributed repair time. 
In terms of \sCCP\ model, we need an environment variable, say $K$, representing the time in which the server repair will finish. It will be re-sampled from a given distribution  each time a breakdown occurs. More specifically, let $W$ be a random variable on the positive reals, with cumulative distribution function $F(t)$, independent of the current state. Each time the server breaks, we set $K$ to $time + W$. The \texttt{server} agent now becomes  
\begin{center}
{\tt
\begin{tabular}{lcl}
    server & $\defeq$ & $[*\rightarrow X_i '= X_i - 1 \wedge X_b '= X_b + 1 \wedge K' =  time + W]_{k_b X_i}$.server\\
    & + & $[time = K \rightarrow X_i '= X_i + 1 \wedge X_b '= X_b - 1]_{\infty:1}$.server
\end{tabular}}
\end{center}
It is easy to see that this modified model satisfies the scaling conditions of Proposition \ref{prop:hybridTimeGuards}, as the guard of the times transition is independent of $\size{N}$. It follows that convergence holds, as can be seen in Figure \ref{fig:csBreakdownWeibull}, where we consider a fixing time sampled according to a Weibull distribution.
\end{example}

%
%
%
\section{Dealing with Guards Depending on Continuous Variables}
\label{sec:guards}

In this section, we look at what happens if we allow guards depending on continuous variables in \sCCP\ transitions, which in the limit can either be approximated as continuous or be kept discrete and stochastic. 
This additional feature, which is straightforward from the point of view of the modelling language and which poses no problems in the definition of the CTMC semantics, has more complex consequences for what concerns the hybrid limits. 

We will first focus on guards on continuous transitions, as these are somehow more delicate to deal with. Guards on discrete stochastic transitions, which create problems that are, in a certain sense, analogous to those with instantaneous transitions, will be discussed later on in Section \ref{sec:guardsDiscreteStochastic}.

\subsection{Guards on Continuous Transitions}
\label{sec:guardsContinuousTrans}

Guards on continuous transitions introduce discontinuities in the vector field. In fact, the rate function $\srateb{\pi}(\xb{})$ of a continuous transition $\pi$ has to be multiplied by the indicator function of the guard predicate, which we assume to be of the form $h(\xb{})\geq 0$, obtaining the discontinuous function $\ratefb{\pi}(\xb{}) = \srateb{\pi}(\xb{})\cdot\ind{h(\xb{})\geq 0}$. In doing this operation, we leave the world of differential equations, entering into the more intricate realm of discontinuous or piecewise-smooth dynamical systems (PWSS) \cite{HA:Cortes:2008:DiscontinuousDynamicalSystems, HA:Filippov:1988:ODEdiscontinuousRHS} or, more generally, of differential inclusions \cite{THMAT:AubinCellina:1984:DiffInclusion}.

The problem, roughly speaking, is that existence and uniqueness of the solution of an ODE with discontinuous right-hand-side is not guaranteed even if all rate functions are regular (say Lipschitz continuous) and if guards are also described by smooth functions (say differentiable functions). Furthermore, solutions can exhibit strange behaviours, like \emph{sliding motion} (sliding on a discontinuity surface) or \emph{chattering} (Zeno behaviour in crossing discontinuity surfaces). 

The lack of uniqueness, in particular, is problematic in our context, as it is a fundamental condition in the definition of the class of PDMP we consider here. Indeed, more general frameworks can be considered, like PDMP based on differential inclusions, but we leave the investigation of this direction for future work.

In this paper, we will follow the treatment of \cite{MyQEST2011}, in which the author discusses mean field limits in presence of guards, when the limit is a PWSS. A more general approach is that of \cite{STOC:Gast:2010:DifferentialInclusionMeanField}, but we stick to the first one as we believe it is more intuitive.
In the next subsection, we will briefly give an introduction to PWSS, in which we will discuss conditions for existence and uniqueness of solutions.
Then, we will turn our attention to fluid approximation of those systems and plug these results into our framework.

\subsubsection{Piecewise-Smooth Dynamical Systems}
\label{PWSsystems}

Consider an ordinary differential equation $\frac{d\xb{}}{dt} = F(\xb{}{})$.
A solution in the classical sense is a (continuously) differentiable
function $\xb{}(t)$ such that $\frac{d}{dt} \xb{}(t) = F(\xb{}(t))$, and
$\xb{}(0) = \xb{0}$. A classical result is the Picard Lindel\"{o}f
theorem~\cite{THMAT:Rudin:1976:analysis}: if $F$ is (locally)
Lipschitz on a set $E\subseteq\bbR^n$ and $\xb{0}$ is in the interior of $E$, then there exists a unique global solution of the differential equation within $E$.

However, here we are interested in dynamical systems in
which the right hand side of the ODE can be a discontinuous
function, possibly undefined, on a set of points of measure
zero. This is the setting studied in the theory of ordinary
differential equations with discontinuous right-hand
side~\cite{HA:Filippov:1988:ODEdiscontinuousRHS}. In particular, we
will consider the so-called switching systems or piecewise smooth
(PWS) dynamical
systems~\cite{HA:Cortes:2008:DiscontinuousDynamicalSystems,HA:Dieci:2009:SlidingMotion}.
Let $F:E\rightarrow\bbR^n$, with $E\subseteq\bbR^n$, and suppose
there exist a finite set of domains $\calR_i$, $i=1,\ldots,s$, such
that $F$ is smooth (or at least Lipschitz) on $\bar{\calR}_i$, the closure of $\calR_i$, and
$\bigcup \bar{\calR}_i \supseteq \bar{E}$. Notice that $F$ can be
discontinuous only on the boundaries $\partial \calR_i$ of the
regions $\calR_i$, so that the discontinuous set is $\calH = \bigcup
\partial \calR_i$ and it has measure zero.

In the following, we will briefly sketch some basic notions of these
systems, which we will need in the following, starting from the
concept of a solution. In fact, given that the vector field is
discontinuous, we cannot look anymore for solutions which are
continuously differentiable functions. Therefore, we will look for
solutions among \emph{absolutely continuous} functions, i.e.
continuous functions which are equal to the integral of another
function~\cite{THMAT:Rudin:1973:functionalAnalysis} and are
henceforth differentiable almost everywhere.

\begin{figure*}[!t]

\begin{center}
\subfigure[Transversal motion]{\label{fig:transversalMotion}
  \includegraphics[width=5cm]{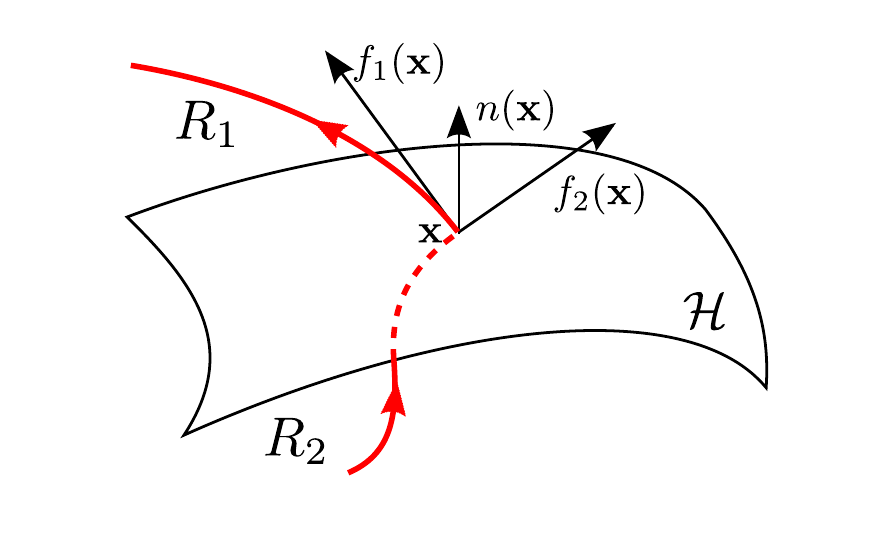}}
\subfigure[Sliding motion]{\label{fig:slidingMotion}
  \includegraphics[width=5cm]{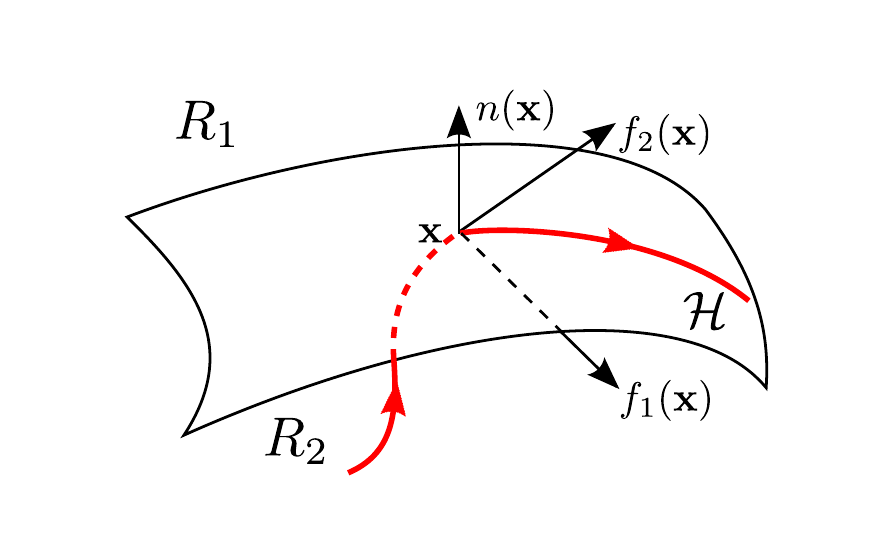}}
\subfigure[Sliding vector field]{\label{fig:slidingVectorField}
  \includegraphics[width=5cm]{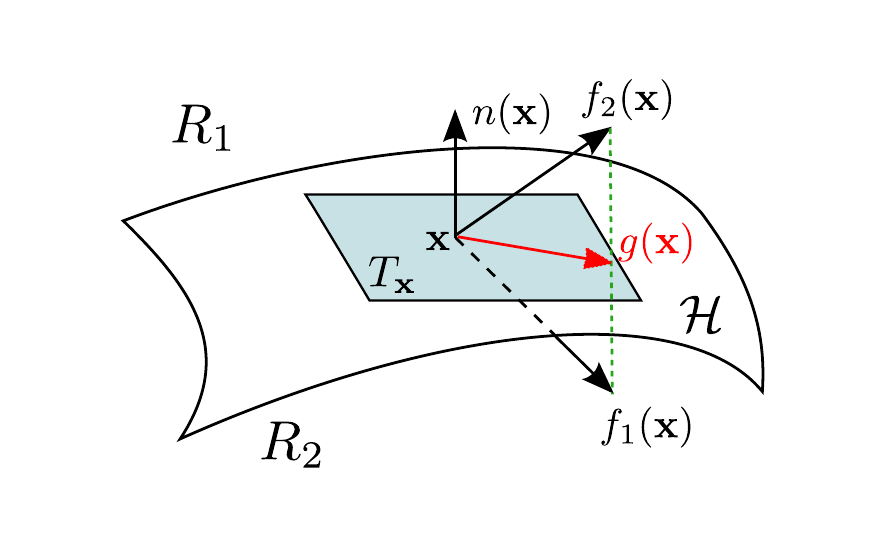}}
\end{center}
  \caption{Schematic representations of 
  transversal motion   across $\calH$, of sliding motion along $\calH$,
  and of the geometric construction of the sliding vector field $G(\xb{})$.}\label{fig:PWS}
\end{figure*}

In order to define such solutions, we will lift the function $F$ to
a set valued function $\bar{F}$, $\bar{F}(\xb{})\subseteq \bbR^n$,
known as the \emph{Filippov extension} of $F$. Then we define a
\emph{Filippov solution} as an absolutely continuous function
$\xb{}(t)$ such that $\xb{}(0) = \xb{0}$ and $\frac{d}{dt}\xb{}(t) \in
\bar{F}(\xb{}(t))$ almost everywhere. That is to say, we replace the
discontinuous differential equation by a \emph{differential
inclusion}~\cite{HA:Cortes:2008:DiscontinuousDynamicalSystems, THMAT:AubinCellina:1984:DiffInclusion}. More
specifically, we define $\bar{F}(\xb{})$ as
$\mathtt{co}\{\lim_{k\rightarrow\infty}
F(\xb{k})~|~\xb{k}\rightarrow\xb{},\ \xb{k}\not\in\calH\}$, where
$\mathtt{co}$ denotes the convex closure of a set. Notice that for
each continuity point $\xb{}$ of $F$, $\bar{F}(\xb{}) = \{F(\xb{})\}$, so
that we have a proper differential inclusion only in the
discontinuity region $\calH$.

For simplicity, consider a PWS system constituted by only two
regions $\calR_1$ and $\calR_2$, and let $\xb{}\in\calH$ be a point of
discontinuity of the vector field. Furthermore, suppose that $F$
equals the function $F_1$ on $\bar{\calR}_1$ and the function $F_2$
on $\bar{\calR}_2$, and that $F_{1,j}(\xb{}) < F_{2,j}(\xb{})$. In this setting,
$\bar{F_j}(\xb{}) = [F_{1,j}(\xb{}),F_{2,j}(\xb{})]$.

The existence of a solution, starting from a point $\vr{x_0}$, is
guaranteed under mild conditions on the Filippov extension $\bar{F}$
of $F$~\cite{HA:Filippov:1988:ODEdiscontinuousRHS}: $\bar{F}$ must
be \emph{(locally) bounded}\footnote{A set function $\bar{F}$ is
locally bounded at $\xb{}\in\bbR^n$ if there exists $\eps>0$ and
$M_{\xb{}}>0$ such that $||\zb{}||<M_{\xb{}}$ for each $\zb{}\in \bar{F}(\yb{})$
and $\yb{}\in B(\xb{},\eps)$.} and \emph{upper semicontinuous}\footnote{A
set function $\bar{F}$ is upper semicontinuous at $\xb{}\in\bbR^n$ is
for each $\eps>0$ there exists $\delta>0$ such that $\bar{F}(\yb{})
\subset\bar{F}(\xb{}) + B(\vr{0},\eps)$ for each $\yb{}\in
B(\xb{},\delta)$.}. Consider again a PWS system with two regions
$\calR_1$ and $\calR_2$, as above. Then, existence is guaranteed if
functions $F_1$ and $F_2$ are continuous on $\bar{\calR}_1$ and
$\bar{\calR}_2$.

In order to understand the behaviour of a PWS dynamical system on a
discontinuity point of the vector field, we restrict our attention
to the two regions system (which is a good local model, unless a
discontinuity point belongs to the boundary of more than two
regions), further assuming that $\calR_1$ and $\calR_2$ are
separated by a smooth surface $\calH$. In particular, $\calH$ is defined
as $\calH = \{\xb{}~|~h(\xb{}) = 0\}$, where $h$ is a function with
continuous second order derivatives, while $\calR_1 =
\{\xb{}~|~h(\xb{})>0\}$ and $\calR_2 = \{\xb{}~|~h(\xb{})<0\}$. We further require
that $\nabla h(\xb{}) \neq \vr{0}$ for each point $\xb{}\in\calH$, so that
the normal vector $n(\xb{})= \nabla h(\xb{})/||\nabla h(\xb{})||$ is always
defined for the surface $\calH$, and always points into $\calR_1$, see
Figure~\ref{fig:PWS}.

To understand the behaviour of a trajectory when it hits the surface
$\calH$, consider a situation in which the solution is in the interior
of $\calR_2$ and hits $\calH$ in $\xb{}$ at some time $t$. Then, two
things can happen, depending on the relative orientation of the
vectors $F_1(\xb{})$ and $F_2(\xb{})$ with respect to $\calH$. In particular,
as $\xb{}(t)$ hits $\calH$ from $\calR_2$, the vector $F_2(\xb{})$ must point
towards $\calR_1$. If also the vector $F_1(\xb{})$ points towards
$\calR_1$, then the trajectory $\xb{}(t)$ crosses the surface $\calH$,
possibly with a discontinuity in its derivative. This phenomenon is
called \emph{transversal motion}, see
Figure~\ref{fig:transversalMotion}. Alternatively, the vector
$F_1(\xb{})$ may point towards $\calR_2$. In this case, the trajectory
cannot enter $\calR_1$, as it will be pushed immediately back to
$\calH$, but, symmetrically, it cannot also remain in $\calR_2$.
Therefore, the motion is confined in the discontinuity surface $\calH$.
This kind of behaviour is known as \emph{sliding motion}, see
Figure~\ref{fig:slidingMotion}. In particular, the trajectory
$\xb{}(t)$ follows the solution of the vector field tangential to $\calH$
obtained by selecting the only vector in $\bar{F}(\xb{})$ tangential to
$\calH$, see Figure~\ref{fig:slidingVectorField}. More precisely, the
sliding motion is defined by the differential equation
$\frac{d}{dt}\xb{} = G(\xb{})$, where $G$ is the vector field $(\xb{}) =
\alpha(\xb{})f_1(\xb{}) + (1-\alpha(\xb{}))f_2(\xb{})$. The value of the
weight coefficient $\alpha(\xb{})$ is obtained by requiring that
$n^T(\xb{})G(\xb{}) = 0$ (i.e., that $G(\xb{})$ is tangential to $\calH$),
obtaining $$\alpha(\xb{}) = \frac{n^T(\xb{})F_2(\xb{})}{n^T(\xb{})F_2(\xb{}) -
n^T(\xb{})F_1(\xb{})},$$ where $n^T(\xb{})F_1(\xb{})$ is the projection of
$F_1(\xb{})$ along the normal vector $n(\xb{})$ of $\calH$ in $\xb{}$. Sliding
motion continues until one (and only one) of the two vectors fields,
say $F_1$, becomes tangential to $\calH$. In this case, the motion
continues in the region $\calR_1$. The condition that only one
vector out of $F_1(\xb{})$ and $F_2(\xb{})$ becomes tangential to $\calH$ is
known as the first order exit condition of the sliding
motion~\cite{HA:Dieci:2009:SlidingMotion}.  If both $F_1$ and $F_2$
become tangential, then the motion continues on a submanifold of
$\calH$, but we do not consider these situations in this paper, which
can, however, be treated similarly to the motion in the intersection
of the boundary between three or more
regions~\cite{HA:Dieci:2009:SlidingMotion}. Hence, from now on we
tacitly assume that sliding motion terminates with first order exit
conditions.

In general, if we are in a point $\xb{}$ of $\calH$, the behaviour of a
solution starting in $\xb{}$ depends on the values of $n^T(\xb{})F_1(\xb{})$
and $n^T(\xb{})F_2(\xb{})$:
\begin{itemize}
  \item If both $n^T(\xb{})F_1(\xb{})$ and  $n^T(\xb{})F_2(\xb{})$ are non-zero
  and have the same sign, then there is a \emph{transversal
  crossing} of the surface.
  \item If $n^T(\xb{})F_1(\xb{}) < 0$ and $n^T(\xb{})F_2(\xb{}) > 0$, we have a
  \emph{stable sliding motion} along $\calH$.
  \item If $n^T(\xb{})F_1(\xb{}) > 0$ and $n^T(\xb{})F_2(\xb{}) < 0$, we have an
  \emph{unstable sliding motion} along $\calH$.
  \item If only $n^T(\xb{})F_1(\xb{}) = 0$, then the trajectory continues
  in the region pointed by $F_2(\xb{})$, and similarly for
  $n^T(\xb{})F_2(\xb{}) = 0$ (\emph{tangential crossing}).
\end{itemize}

For the scope of this paper, the uniqueness result for PSW systems
plays a relevant role. More precisely,
Filippov~\cite{HA:Filippov:1988:ODEdiscontinuousRHS} proved that
there is a \emph{unique} solution starting in $\xb{}\in\calH$, provided
that at least one of $n^T(\xb{})F_1(\xb{}) < 0$ and $n^T(\xb{})F_2(\xb{}) > 0$
holds. Notice that this condition rules out unstable sliding motion. There is also a condition for existence and uniqueness expressed in terms of differential inclusions, which requires the set-valued function $\bar{F}$ to be one-sided Lipschitz continuous\footnote{ A set valued function $F$ is one-sided Lipschitz if and only if for each $\x{1},\x{2}\in \sspace{}$ and $\y{1}\in F(\x{1})$, $\y{2}\in F(\x{2})$, it holds that $(\x{1}-\x{2})^t\cdot (\y{1}-\y{2}) \leq L\|\x{1}-\x{2} \|$, for some $L>0$.}. We remark that the global existence and uniqueness of a solution allows us to define a semiflow $\flow{}(t,\xb{})$ for the discontinuous vector field $F$, which is the condition required in the definition of a PDMP adopted here. 

%
%
%

\subsection{Deterministic Approximation for PWS Limits}
\label{sec:KurtzPWS}

We will now present the limit result of \cite{MyQEST2011} in the framework of this paper, and then plug it in the proof of Theorem \ref{th:hybridBasic} in order to extend the hybrid convergence limit to this discontinuous setting. We start by expanding Scaling \ref{scaling:continuous} to deal with guards. 

\begin{scaling}[Continuous Scaling with Guards]
\label{scaling:continuousGuarded}
A normalized guarded \sCCP\ transition $\norm{\pi} = (\guardb{\pi}\N(\Xb{}),\Xb{}' = \Xb{} + \stoichb{\pi}\N, \srateb{\pi}\N(\Xb{}))$ of a population-\sCCP\ program $(\calA,\size{N})$, with $E$ the domain of normalised variables $\Xb{}$, assumed to be all continuous, has \emph{continuous scaling} if and only if:
\begin{enumerate}
\item The rate and update satisfy the conditions of Scaling \ref{scaling:continuous};
\item $\guardb{\pi}\N(\Xb{})$ is of the form $h_{\pi,1}(\Xb{})\geq 0 \wedge \ldots \wedge h_{\pi,k}(\Xb{})\geq 0$, where each $h_{\pi,j}$ is independent of $N$ and has continuous second order derivatives, with $\nabla h_{\pi,j}(\xb{})\neq 0$ for all $\xb{} \in \{\xb{}~|~h(\xb{})=0\}$. 
\end{enumerate}
\end{scaling}

Consider now a population-\sCCP\ program $(\calA,\size{N})$, with no discrete variables and no instantaneous transitions, and with stochastic actions satisfying either Scaling \ref{scaling:continuous} or \ref{scaling:continuousGuarded} (hence, all actions will be approximated continuously). Then, we can compute its drift according to equation \ref{eqn:fluidVFN}, which defines a piecewise-smooth system:
\[ F\N(\Xb{}) = \sum_{\pi}\bbE[\stoichb{\pi}\N] \ind{\guardb{\pi}(\Xb{})}\ratefb{\pi}\N(\Xb{})  \]
Notice, in particular that, as the guards are independent of $N$, it holds that $F\N\rightarrow F$ uniformly, where $F$ is defined by
\[F(\Xb{}) = \sum_{\pi}\bbE[\stoichb{\pi}] \vr{1}\{\guardb{\pi}(\Xb{})\}\ratefb{\pi}(\Xb{}). \]

Let us take a closer look to the PWS systems 
$\frac{d}{dt}\xb{} = F(\xb{})$. Each transition of the model having a
non-trivial guard, partitions the state space in several regions.
In fact, if the predicate $\guardb{\pi}$ is a conjunction of inequalities
defined by smooth functions $h_{\pi,j}$, then each such function
partitions the state space in two regions: $\calR_{\pi,j}^+$, where
$h_{\pi,j}$ is positive, and $\calR_{\pi,j}^-$, where $h_{\pi,j}$ is
negative. Therefore, in order to define the PWS system, we have to consider all possible intersections of regions $\calR_{j}^+$ and
$\calR_{j}^-$, for all distinct function $h_j$ appearing in guards
of transitions. If there are $m_0$ such functions, then we have
$2^{m_0}$ distinct regions. In practice, however, many transitions
usually have trivial guards, and there may be transitions sharing
the same functions $h_j$, so that this number should be reasonably
small. In the following, we indicate by $\calH_j$ the manifold defined
by $h_j$: $\calH_j = \{\xb{}\in\bbR^n~|~h_j(\xb{}) = 0\}$.

In the rest of the paper, we require that the PWSS defined by $F$ is globally regular, in the following sense:
\begin{enumerate}
  \item solutions exist globally in $E$, and are unique, so that the PWSS admits a semi-flow on $E$;
  \item sliding motion never happens on the intersection of more than one surface, and has first order exit condition;
  \item the PWSS has no Zeno trajectories, i.e. the number of transversal crossings and traits of sliding motion is finite in each compact time interval $[0,T]$, for each trajectory of the PWSS. 
\end{enumerate}
These conditions are essentially those introduced in \cite{MyQEST2011}, just extended to the whole domain $E$.\footnote{
This regularity condition can be simplified, if the problem is restated in terms of differential inclusions and we require that the set-values extension $\bar{F}$ of $F$ is one-sided Lipschitz. Under this milder assumptions, the following theorem still holds. However, we stick here to the formulation in terms of PWSS, which is more natural in the context of PDMP.} 
If the PWSS is regular, then each of its trajectories is regular and therefore the following theorem holds:

\begin{theorem}\label{th:kurtzPWSS}
Let $(\calA,\size{N})$ be a sequence of \sCCP\ models for increasing systems size, satisfying the conditions of Theorem \ref{th:Kurtz}, with all actions $\pi$ satisfying either Scaling \ref{scaling:continuous} or Scaling \ref{scaling:continuousGuarded}. 
Let $\Xb{}\N(t)$ be the associated sequence of normalized CTMC, and assume $\xb{}\N(0)\rightarrow \xb{0}$ (in probability/almost surely). 

Let $\xb{}(t)$ be the solution of the regular PWSS system $\frac{d}{dt} \xb{} = F(\xb{})$ starting in point $\xb{}(0) = \xb{0}\in E$. Fix a finite time horizon $T<\infty$, 
Then
$$\lim_{N\rightarrow\infty}\sup_{t\leq T} \left\|\Xb{}\N(t) - \xb{}(t)\right\|
= 0\ \  \mbox{in probability.}$$ \qed
\end{theorem}

As an immediate corollary, we get that if $\xb{0}\N\Rightarrow \xb{0}$, then $\Xb{}\N\Rightarrow\xb{}$ as random elements in the space of cadlag functions. 

The intuition behind the proof of the theorem is that each regular trajectory in $[0,T]$ of the PWSS can be sliced into a finite
number of pieces, such that each piece is either the solution of a standard ODE within a continuity region of the vector field, or it is a sliding motion along a discontinuity surface. The idea is to prove the convergence of the sequence $\Xb{}\N(t)$ to $\xb{}(t)$ in each piece separately, either using standard deterministic approximation
(Theorem~\ref{th:Kurtz}), or using a specialized version of such a
result for sliding motion (Theorem IV.2
in \cite{MyQEST2011}). Then, one simply proves convergence in $[0,T]$ by combining
the convergence in each piece, exploiting convergence of exit times. Sliding motion is the difficult case, because the trajectory of the PWSS evolves according to the
sliding vector field, which is different from the drift $F\N$ of the sequence of CTMC.

Differently from \cite{MyQEST2011}, we are not requiring that rate functions are globally bounded and Lipschitz on $E$, but they just satisfy these properties locally. However, the validity of the previous theorem depends only on a compact neighbourhood of the trajectory up to time $T$, and on the assumption of global existence of solutions.  Furthermore, we are also allowing random increments, which can be dealt with exactly as in Theorem \ref{th:Kurtz}.

We also note that we could have relaxed the scaling condition on guards by making guard functions $h_j$ depend on $N$, and assuming that they converge uniformly to a limit function, with the same properties. The previous theorem would still hold, but with some modification with respect to the proof of \cite{MyQEST2011}.\footnote{
The proof becomes more involved because if guards are varying with $N$, $F\N$ does not converge uniformly to $F$ any more. 
Essentially, one proves that the trajectories of the PWSS defined by the discontinuous vector field $F\N$ converge uniformly in each $[0,T]$ to the trajectories of $F$, and that $\Xb{}\N$ converges in probability, uniformly in $[0,T]$, to $\xb{}\N$, the solution of $d/dt\; \xb{}\N(t) = F\N(\xb{}\N(t))$. To show that $\xb{}\N$ is regular, one relies on the regularity of the limit trajectory $\xb{}$, and on the uniform convergence of activation functions of guards and of the components $\ratefb{\pi}\N$ of the vector field.  To show convergence of $\Xb{}\N$ to $\xb{}\N$, one either invokes an obvious modification of Theorem \ref{th:Kurtz}, or modifies the proof for sliding motion in \cite{MyQEST2011} using again the uniform convergence of guard's functions and of rates. Alternatively, one could work with differential inclusions, as in \cite{STOC:Gast:2010:DifferentialInclusionMeanField}.}




\subsection{Hybrid Limit with Guarded Continuous Transitions}
\label{sec:hybridPWSS}

The previous theorem can be easily plugged in the framework of Section \ref{sec:hybridLimits}, replacing Kurtz theorem in the proof device of Theorem \ref{th:hybridBasic}. This can be done under the \emph{regularity assumption} of the limit PWSS in each mode of the PDMP, i.e. for each possible combination of values of discrete variables. We call \emph{PWS regular} such a PDMP. 
The reason is that the proof of Theorem \ref{th:hybridBasic} relies only on the weak convergence implied by Kurtz theorem and on the continuity of limit trajectories, which are also satisfied by the subset of PWSS considered here. Furthermore, as the  treatment of instantaneous transitions or time-dependent guards in Section \ref{sec:InstantaneousTransitions} is also independent of the fine-grained details of the continuous dynamics, we can also include those kinds of transitions. Notice that the notion of non-Zeno PDMP and those of robustly transversal PDMP, robust activation property and size-compatible PDMP, extend automatically to this PWS setting, as they only depend on the existence of the semi-flow of the continuous dynamics of the PDMP. Before stating the theorem, we need to extend Scaling \ref{scaling:continuousGuarded} to the hybrid setting, as done for scaling \ref{scaling:continuous}. 

\begin{scaling}[Hybrid Continuous Scaling with Guards]
\label{scaling:continuousGuardedHybrid}
A normalized guarded \sCCP\ transition $\norm{\pi} = (\guardb{\pi}\N(\Xb{}),$ $\Xb{}' = \Xb{} + \stoichb{\pi}\N, \srateb{\pi}\N(\Xb{}))$ of a population-\sCCP\ program $(\calA,\size{N})$, with variables $\Xb{}\in\sspace{}$ partitioned into $(\X{d},\Xb{c},\X{e})$, has \emph{hybrid continuous scaling} if and only if:
\begin{enumerate}
\item The rate and update satisfy the conditions of Scaling \ref{scaling:continuousGuarded};
\item $\guardb{\pi}\N(\Xb{}) =\guardb{\pi,d}\N(\X{d},\X{e})\wedge \guardb{\pi,c}(\Xb{c})$, where $\guardb{\pi,c}(\Xb{c})$ satisfies the condition of Scaling \ref{scaling:continuousGuarded}.
\end{enumerate}
\end{scaling}

Therefore, we have the following:

\begin{proposition}
\label{prop:hybridPWSS}
Let $(\calA,\size{N})$ be a sequence of time-guarded population-\sCCP\ models for increasing systems size $\size{N}\rightarrow\infty$, as $N\rightarrow\infty$, with variables partitioned into $\X{} = (\X{d},\X{c},\X{e})$, with discrete stochastic actions satisfying scaling \ref{scaling:discreteStochastic}, instantaneous actions satisfying scaling \ref{scaling:discreteInstantaneous}, time guarded actions satisfying scaling \ref{scaling:timedInstantaneous}, and continuous actions satisfying either scaling \ref{scaling:continuousHybrid} or scaling \ref{scaling:continuousGuardedHybrid}.
Let $\Xb{}\N(t)$ be the associated sequence of normalized CTMC and $\xb{}(t)$ be the limit PDMP associated with the limit normalized TDSHA $\norm{\calT}(\calA)$.

If $\xb{0}\N \Rightarrow \xb{0}$ (weakly) and the PDMP is \emph{non-Zeno},  \emph{robustly transversal}, has the \emph{robust activation property}, is \emph{size-compatible} and \emph{PWS regular}, then $\Xb{}\N(t)$ converges weakly to $\xb{}(t)$,  $\Xb{}\N \Rightarrow \xb{}$, as random elements in the space of cadlag function with the Skorohod metric. \qed 
\end{proposition} 

\begin{figure}
\subfigure[Example \ref{ex:epidemicsPWS}] {\label{fig:epiPWS}
\includegraphics[width=.48\textwidth]{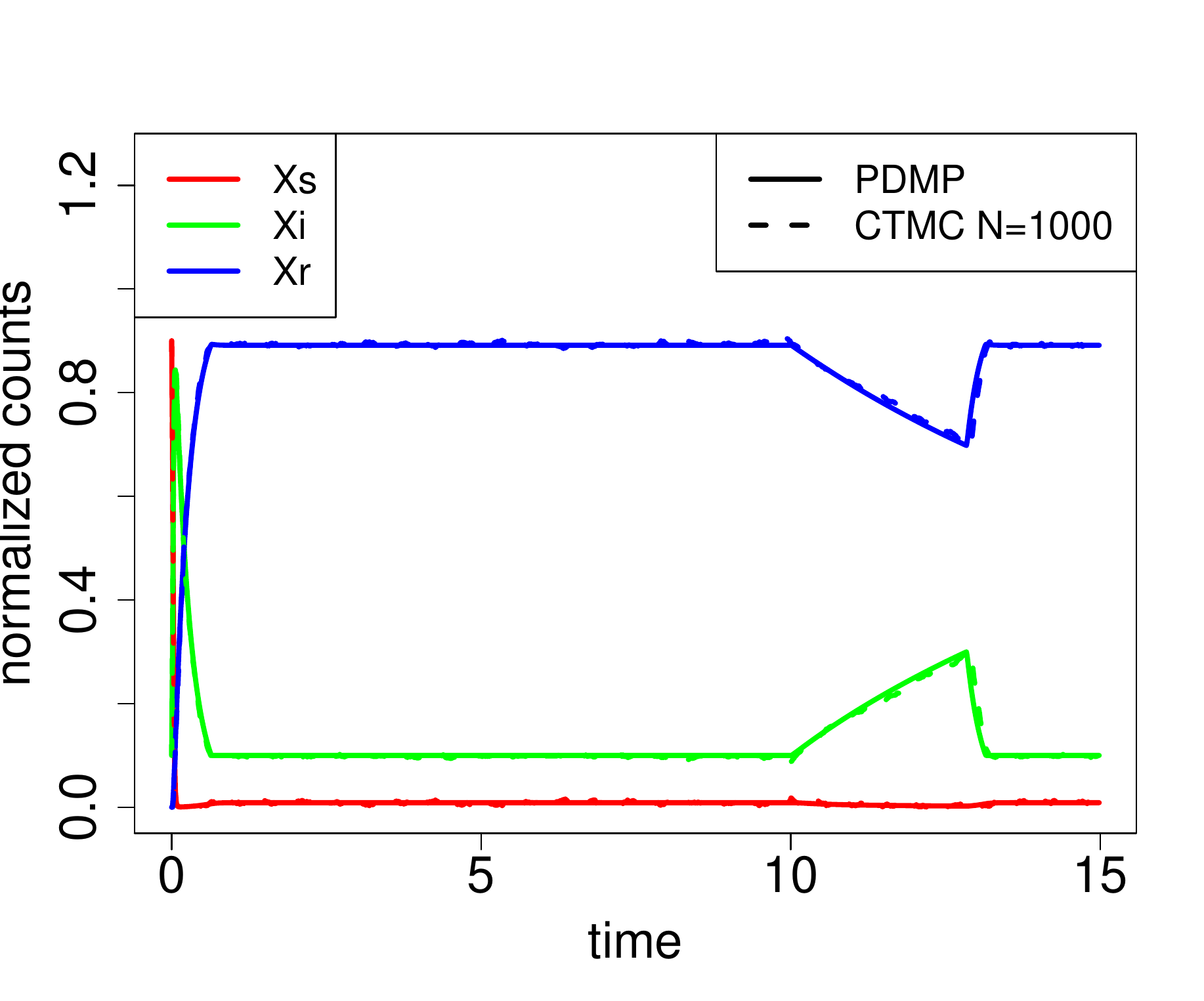} }
\subfigure[Example \ref{ex:epidemicsStochGuard}] {\label{fig:epiStocG}
\includegraphics[width=.48\textwidth]{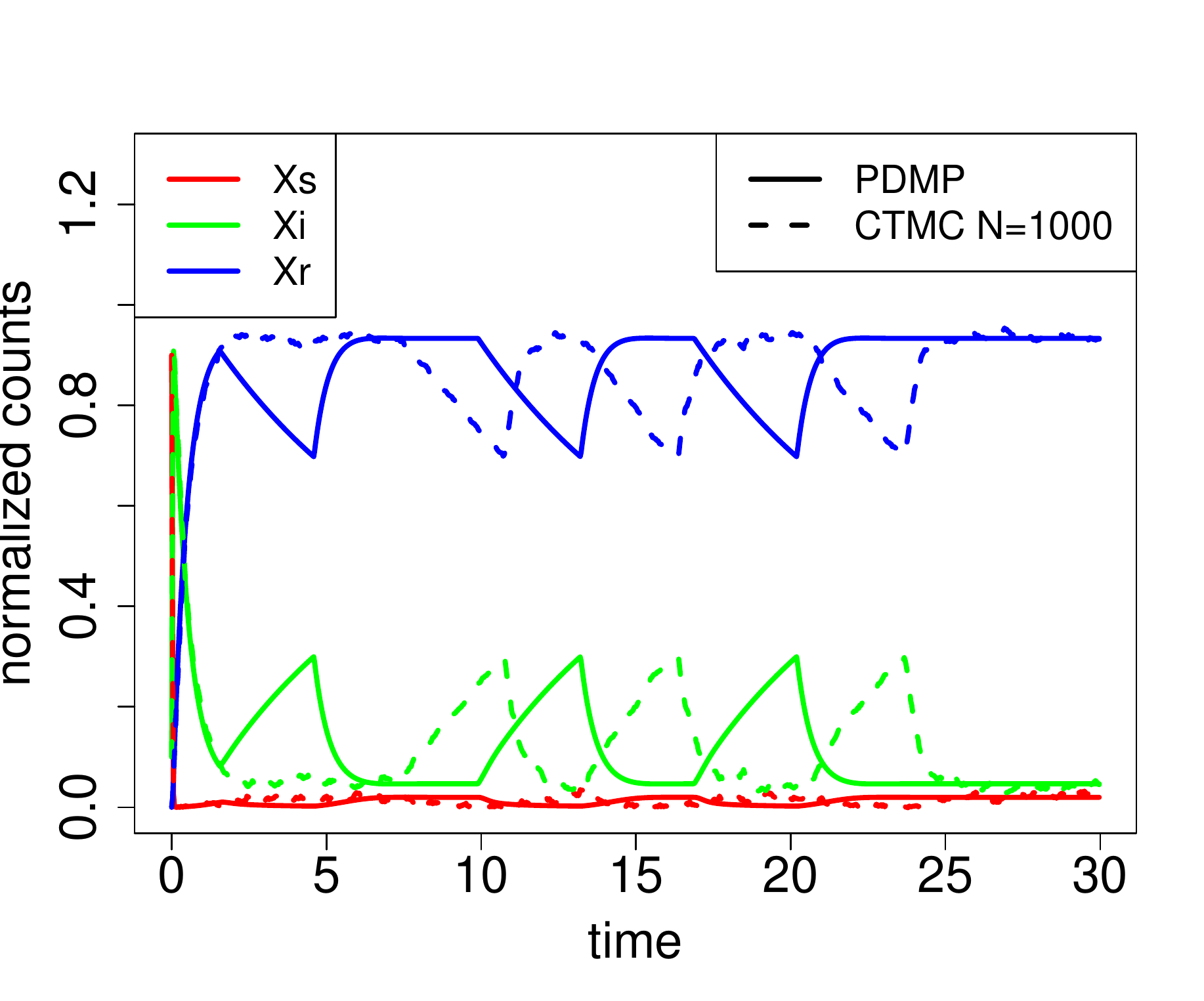} }

\caption{Comparison of single trajectories of the limit PDMP and the CTMC model for system size 1000, for the control policies of the network epidemic models of Example \ref{ex:epidemicsPWS} (left) and Example \ref{ex:epidemicsStochGuard} (right). The behaviour is essentially the same (modulo randomness of switch times in the figure on the right). Parameters are as in the caption of Figure \ref{fig:epidemicsOscillating}. Additionally, in the model of Example \ref{ex:epidemicsPWS}, we have $k_p^1 = 0.05$, $k_p^2 = 4.0$, $k_p^3 = 0.5$, $\alpha = 0.1$, $\beta = 0.3$, and the duration of the emergency policy is 10 units of time. the rate of switch from emergency to normal patching policy in the model of Example  \ref{ex:epidemicsStochGuard} is 0.1.}
\label{fig:epiDisc}
\end{figure}

\begin{example}
\label{ex:epidemicsPWS}
We consider a variant of the computer network epidemic model of Example \ref{ex:epidemicsOscillating}. In particular, we consider the following normal and emergency patching policies. 
Under the normal policy,  patches are applied at constant rate $k_p^1$. Under the emergency policy, instead, computers in the network are patched with a rate $k_p^2 > k_p^1$ if the fraction $\norm{X}_{i}$ of infected nodes is above a threshold $\alpha$, and at rate $k_p^3 < k_p^2$, $k_p^3 > k_p^1$, if the fraction of infected nodes is below $\alpha$. 
The emergency policy is initiated as soon as the fraction of infected nodes becomes greater than a threshold $\beta>\alpha$, and is executed for $w\in\bbR^+$ units of time. 
We will use an environmental variable $K$ to remember the next firing time of such a delayed transition. When the emergency policy is aborted, the normal policy is restored. 
We can model this policy in \sCCP\ by suitably modifying the code of Example \ref{ex:epidemicsOscillating}, with particular regard to the \texttt{patching} and the \texttt{control} agents. The variable $U$ is the discrete variable modelling the patching policy: $U=1$ indicates the normal policy, while $U=2$ indicates the emergency one.
 \begin{center}
{\tt
\begin{tabular}{lcl}
    patching & $\defeq$ & $[U = 1\rightarrow X_i '= X_i - 1 \wedge X_r '= X_r + 1]_{k_p^1 X_i}$.patching\\
    & + & $[U = 2 \wedge \n{X}_i \geq \alpha \rightarrow X_i '= X_i - 1 \wedge X_r '= X_r + 1]_{k_p^2 X_i}$.patching\\
    & + & $[U =  2 \wedge \n{X}_i < \alpha \rightarrow X_i '= X_i - 1 \wedge X_r '= X_r + 1]_{k_p^3 X_i}$.patching\\
    control   & $\defeq$ & $[\n{X}_i > \beta \rightarrow U' = 2 \wedge K' = time + w]_{\infty:1}$.control\\
 & + &  $[time \geq K  \rightarrow U' = 1]_{\infty:1}$.control\\
\end{tabular}}
\end{center}

The limit model is a PWSS when $U=2$. In Figure \ref{fig:epiPWS}, we show a trajectory of the PWSS exhibiting sliding motion on the plane $\n{X}_i = \alpha$. It is easy to check that the PWSS has a unique solution from any initial state. In fact, taking the scalar product of the two vector fields $F_1$, $F_2$ with the normal to $\n{X}_i = \alpha$ on the two sides of the plane $\n{X}_i = \alpha$, we obtain $k_i \alpha \n{X}_s - k_p^2\alpha$ and $k_i\alpha \n{X}_s - k_p^3\alpha$. Now, $k_i\alpha \n{X}_s - k_p^3\alpha > 0$ for $\n{X}_s > k_p^3/k_i$. But as $k_p^2 > k_p^3$, for $\n{X}_s \leq  k_p^3/k_i$, we have that $k_i \alpha \n{X}_s - k_p^2\alpha \leq \alpha k_p^3 - k_p^2\alpha < 0$, which shows that the uniqueness condition is verified. 
Also in this case, the hybrid limit model is deterministic, and we can see its trajectory from a fixed set of initial conditions in Figure \ref{fig:epiDisc}. Inspecting this trajectory, we can easily convince ourselves that the crossing of activation surfaces is always transversal, so that we can  apply Proposition \ref{prop:hybridPWSS}.

\end{example}

%

\subsection{Guards on discrete stochastic transitions}
\label{sec:guardsDiscreteStochastic}

In this section, we consider a relaxation of Scaling \ref{scaling:discreteStochastic}, in which we allow guards on discrete stochastic transitions to depend on continuous variables. Similarly to continuous transitions, this extension introduces discontinuities in the rate functions of the PDMP. Intuitively, as the jump time distribution is obtained by the cumulative rate, i.e. by integrating the rate function, these discontinuities should not create problems, as far as the trajectories of the PDMP do not remain in a discontinuity surface of a rate for too long. 
Essentially, problems emerge if a continuous trajectory of the PDMP slides on the discontinuity surface of a guard for some time interval $[t_1,t_2]$. Suppose that on the surface the guard is false, hence the transition of the PDMP cannot fire. In this case, even if trajectories of the CTMC converge to the one of the PDMP, they may remain on the ``wrong'' side of the discontinuity surface, i.e. on the side in which the guard is true and the transition active, so that the event can fire in the sequence of CTMC. If this transition determines the fate of the system, than convergence to the PDMP can fail quite dramatically. 

To explain better the problem, we consider the following example. 
\begin{example}
\label{sec:guardStochRateFate}
Consider a simple \sCCP\ model of a random walk in 1 dimension, for variable $X$, initially set to zero. Variable $Z$ instead, can take values 0 and 1, and it is the fate variable. Initially it is set to zero, and it may become 1 by the firing of a stochastic transition with rate 1, but active only if $X>0$. The \sCCP\ program has initial configuration  \texttt{random\_walk} $\parallel$ \texttt{doom}, where
\begin{center}
{\tt
\begin{tabular}{lcl}
    random\_walk   & $\defeq$ & $[*\rightarrow X '= X + 1 ]_{\size{N}}$.random\_walk\\
 & + & $[*\rightarrow X '= X - 1 ]_{\size{N}}$.random\_walk\\
    doom & $\defeq$ & $[X > 0\rightarrow Z = 1 ]_{1}$.\textbf{0}
\end{tabular}}
\end{center}

Notice that the rate of the transitions of the \texttt{random\_walk} agent grow with $\size{N}$, hence they can be approximated continuously, and, once normalized, induce the drift $F\N(\Xb{}) = F(\Xb{}) = 0$. 
Hence, the limit PDMP model has quite boring continuous dynamics, in fact a constant one on the discontinuity surface $\n{x}=0$. It follows that the discrete stochastic transition will never fire in the PDMP model, and $Z$ will remain 0. However, for any $N$, the CTMC model will spend half of its time on the subspace $\n{X}>0$, meaning that the \texttt{doom} agent will eventually fire its transition, on average in 2 time units.  Hence $Z$ will be equal to 1 with probability going to 1 as $T$ increases. Hence convergence does not hold for this model. Notice, however, that if we set the initial value of $\n{x}$ to $-\eps$, for $\eps>0$, then convergence will hold. In particular, by the Kurtz theorem, for any time $T<\infty$, the CTMC $\Xb{}\N(t)$ will be smaller than $-\eps/2$, in $[0,T]$, for $N$ large enough, hence the \texttt{doom} transition will not fire in $[0,T]$. 
However, it will eventually fire for any $N$, as with probability one, $\Xb{}\N(t)$ will get eventually above zero and remain there for a long enough time. In addition, the time when this event will happen is pushed further and further into the future as $N$ grows. This does not create problems for weak convergence, as the Skorohod metric on which weak convergence is based discounts the future and will give a smaller and smaller weight to the difference of $Z$-values as $N$ grows (see Appendix \ref{app:background} for details on the Skorohod metric).   
\end{example}   
  
From the previous discussion, it should be clear that the problems in introducing guards for discrete stochastic transitions are somehow related to the way the flow of the vector field interacts with the discontinuity surface of the guards. This suggests the following definition: 
\begin{definition}
\label{def:robustCompatibilityGuardStochasticTransitions}
A cadlag function $\xb{}(t)$ taking values in $\sspace{}$ is \emph{robustly compatible} with the activation function $h(\xb{})$ of a guard predicate $\guardb{}(\xb{})$  if and only if the set $\{t\geq 0~|~h(\xb{}(t))=0\}$ has Lebesgue measure zero.\\
A PDMP is \emph{robustly compatible} with a guard $\guardb{\pi}(\xb{})$ if almost surely its trajectories are robustly compatible with the activation function of $\guardb{\pi}(\xb{})$.\\ 
A PDMP derived from a TDSHA $\calT$ is \emph{robustly compatible} if and only if it is \emph{robustly compatible} with all guards of discrete stochastic transitions of $\calT$.
\end{definition}

We can now introduce the scaling condition for guarded discrete stochastic transitions.

\begin{scaling}[Discrete Scaling for Guarded Stochastic Transitions]
\label{scaling:DiscreteStochasticGuarded}
A \emph{guarded} \emph{normalized} \sCCP\ transition with \emph{random reset}, $\norm{\pi} = (\guardb\N{\pi}(\Xb{}),\Xb{}' = \resetb{}\N(\Xb{}, \W{}\N(\Xb{}) ), \srateb{\pi}\N(\Xb{}))$ of a population-\sCCP\ program $(\calA,\size{N})$ with variables $\Xb{}\in\sspace{}$ partitioned into $(\X{d},\Xb{c},\X{e})$, has \emph{discrete scaling} if and only if:
\begin{enumerate}
\item $\guardb{\pi}\N(\Xb{})$ has activation function $h\N(\Xb{})$, converging uniformly in every compact subset $K$ of $\sspace{}$ to the continuous  function $h(\Xb{})$;
\item  $\srateb{\pi}\N$ and $\resetb{\pi}\N$ satisfy the same conditions as Scaling \ref{scaling:discreteStochastic}.
\end{enumerate}
\end{scaling}

Technically, the condition of robust compatibility of a PDMP is needed to prove the convergence of the stochastic jump times and of the states after the reset. The problem here lies in the fact that, on the surface $\{h(\xb{}) = 0\}$, the reset kernel $R$ of the PDMP is discontinuous, and a sequence $\xb{}\N$ of points approaching $\xb{}\in \{h(\xb{}) = 0\}$ may activate a different subset of guards for each $R\N$.  This may lead to radically different behaviours and compromise convergence. This problem is essentially the same as we had with the discontinuity of the reset kernel for instantaneous transitions. The robust compatibility condition permits us to ignore such points, as there is probability zero of jumping from them.
With these assumptions in force, we get the following proposition, whose proof can be found in Appendix \ref{app:proofs}.

\begin{proposition}
\label{prop:hybridDiscontinuousJumps}
Let $(\calA,\size{N})$ be a sequence of population-\sCCP\ models for increasing systems size $\size{N}\rightarrow\infty$, as $N\rightarrow\infty$, with variables partitioned into $\X{} = (\X{d},\X{c},\X{e})$, with discrete stochastic actions satisfying either scaling \ref{scaling:discreteStochastic} or scaling \ref{scaling:DiscreteStochasticGuarded}, no instantaneous actions, and continuous actions satisfying scaling \ref{scaling:continuousHybrid}.
Let $\Xb{}\N(t)$ be the associated sequence of normalized CTMC and $\xb{}(t)$ be the limit PDMP associated with the  normalized limit TDSHA $\norm{\calT}(\calA)$.

If $\xb{0}\N \Rightarrow \xb{0}$ (weakly) and the PDMP is \emph{non-Zeno} and \emph{robustly compatible}, then $\Xb{}\N(t)$ converges weakly to $\xb{}(t)$,  $\Xb{}\N \Rightarrow \xb{}$, as random elements in the space of cadlag function with the Skorohod metric. \qed 
\end{proposition}

\begin{example}
\label{ex:epidemicsStochGuard}
We consider again the computer network epidemics model of Examples \ref{ex:epidemicsOscillating} and \ref{ex:epidemicsPWS}. In particular, we modify the control policy with respect to  Example \ref{ex:epidemicsPWS} by assuming that the emergency policy is dropped in favour of the normal one in an exponentially distributed time, with rate $k_d$, provided $\n{X}_i$ is below $\beta_2<\beta$. 
The control agent then becomes
 \begin{center}
{\tt
\begin{tabular}{lcl}
    control   & $\defeq$ & $[U=1 \wedge \n{X}_i > \beta \rightarrow U' = 2]_{\infty:1}$.control\\
 & + &  $[U = 2 \wedge \n{X}_i < \beta_2 \rightarrow U' = 1]_{k_d}$.control\\
\end{tabular}}
\end{center}
Furthermore, we assume that there is a single emergency patch rate $k_p^2 > k_p^1$, so that the patch agent is 
 \begin{center}
{\tt
\begin{tabular}{lcl}
    patching & $\defeq$ & $[U = 1\rightarrow X_i '= X_i - 1 \wedge X_r '= X_r + 1]_{k_p^1 X_i}$.patching\\
    & + & $[U = 2 \rightarrow X_i '= X_i - 1 \wedge X_r '= X_r + 1]_{k_p^2 X_i}$.patching
\end{tabular}}
\end{center}
If we take the scalar product between the vector field and the normal $\vr{1}_i$ to the plane $\n{X}_i - \beta_2 = 0$ and set it to zero in the plane, we get the equation $k_i \n{X}_s - k_p^2 = 0$, which has only one solution $(k_p^2/k_i,\beta_2,1-k_p^2/k_i-\beta_2)$ if $k_p^2/k_i+\beta_2\leq 1$, and no solution otherwise. In particular, it follows that the trajectories of the vector field do not slide on $\n{X}_i=\beta_2$, hence the PDMP is robustly compatible. Thus, the model satisfies the hypothesis of Proposition \ref{prop:hybridDiscontinuousJumps}, and convergence to the hybrid limit holds.
\end{example}

\begin{remarkstar}
\label{rem:checkingRobustCompatibility}
In order to check the robust compatibility of a PDMP with respect to a guard of a stochastic transition, we can proceed similarly to Remark \ref{rem:checkingInstantaneousConditions}, by applying the randomization trick. In particular, the property holds if we can show that the set of trajectories of the vector field sliding on the discontinuous surface has dimension $n-1$ or less,\footnote{This holds, for instance, if the set of zeros of the scalar product of the normal to the discontinuity surface of the guard with the vector field has dimension $n-2$ or less.} where $n$ is the number of continuous variables, and if initial conditions and resets are absolutely continuous with respect to the Lebesgue measure.
If guards are linear, this check should be relatively easy to carry out, see Example \ref{ex:epidemicsStochGuard}.

%

\end{remarkstar}

\subsection{Collecting all results together.}
In this subsection, we collect in a unique statement all the approximation results spread throughout the paper.

\begin{theorem}
\label{th:final}
Let $(\calA,\size{N})$ be a sequence of time-guarded population-\sCCP\ models for increasing system size $\size{N}\rightarrow\infty$, as $N\rightarrow\infty$, with variables partitioned into $\X{} = (\X{d},\X{c},\X{e})$, satisfying the following scaling conditions:
\begin{enumerate}
\item discrete stochastic actions with guards \emph{not} depending on continuous variables satisfy scaling \ref{scaling:discreteStochastic};
\item discrete stochastic actions with guards depending on continuous variables satisfy scaling \ref{scaling:DiscreteStochasticGuarded};
\item instantaneous actions satisfy scaling \ref{scaling:discreteInstantaneous};
\item time-guarded actions satisfy scaling \ref{scaling:timedInstantaneous};
\item continuous actions with guards \emph{not} depending on continuous variables satisfy  scaling \ref{scaling:continuousHybrid};
\item continuous actions with guards depending on continuous variables satisfy  scaling  \ref{scaling:continuousGuardedHybrid};
\end{enumerate}

Let $\Xb{}\N(t)$ be the associated sequence of normalized CTMC and $\xb{}(t)$ be the limit PDMP associated with the limit normalized TDSHA $\norm{\calT}(\calA)$. 
If
\begin{enumerate}
\item the PDMP is \emph{non-Zeno};
\item the PDMP is \emph{robustly transversal}, has the \emph{robust activation property}, and is \emph{size-compatible} (for instantaneous transitions);
\item the PDMP is \emph{robustly compatible} with all guards of discrete stochastic actions;
\item the PDMP is \emph{PWS regular} (if continuous transitions guarded by continuous variables are present);
\item $\xb{0}\N \Rightarrow \xb{0}$ (weakly)
\end{enumerate}
then $\Xb{}\N(t)$ converges weakly to $\xb{}(t)$,  $\Xb{} \Rightarrow \xb{}$, as random elements in the space of cadlag function with the Skorohod metric. \qed 
\end{theorem}

\section{Conclusions}\label{sec:conc}

In this paper, we discussed hybrid limits of Markov population processes, described as stochastic concurrent constraint programs. We considered the limit behaviour when only a subset of system variables, corresponding to populations of growing size, is approximated continuously, while the other variables remain discrete. 
We proved that the sequence of CTMC for increasing population size converges to a stochastic hybrid system, described as a Piecewise-Deterministic Markov Process. 

We first considered the simplest case, in which continuous transitions, i.e. those becoming fluxes of the limit vector field, satisfy the standard scaling \emph{\`a la Kurtz}, while discrete transitions are stochastic and have continuous rates and reset kernels. 

We then extended these results by including several sources of discontinuity in the evolution of the system: instantaneous transitions in the \sCCP\ program, which induce forced transitions at the PDMP level, and guards depending on continuous variables in continuous and stochastic transitions. In all these cases, discontinuities create potential problems in the interactions between the deterministic vector field in each mode of the PDMP and the discontinuity surfaces of the involved functions. Essentially, the reset kernels become non-Feller (i.e.\ non-continuous), and one has to impose additional conditions to enforce convergence of the times at which transitions fire and the states of the system after a jump. 
In general, the conditions required can be quite difficult to check. However, in practical cases the geometry of discontinuous surfaces should be reasonably simple (mainly linear hyperplanes), hence checking the required conditions should not be too hard. 

Nevertheless, it is unlikely that one can find  general algorithmic procedures to check automatically if a sequence of models is amenable of hybrid approximation, unless there is no source of discontinuity and rate functions and resets satisfy further constraints, see Remark \ref{rem:checkingContinuousStochasticScaling}. 

The moral is that one should avoid introducing too many discontinuities in a model, if approximation results are needed to perform more efficient analysis. 

In this direction, we are investigating some relaxation techniques. In most of the cases, boolean conditions may be replaced by smooth counterparts without altering significantly the model behaviour. For instance, instantaneous events may be replaced by stochastic ones with a rate that changes continuously from zero to a very large value in the proximity of the activation surface, as done in \cite{HA:Abate:2008:RemovingForcedJumps}. Moreover, guards in discrete and stochastic transitions may be replaced by sharp sigmoid functions modulating the rates. However, this operation is not always possible without introducing spurious behaviours. In this case, one has to verify additional regularity conditions before using hybrid approximation. 

Another strategy to simplify the conditions required for the hybrid limits, similar in spirit to the previous one, is to introduce randomness in the continuous evolution. In particular, we could replace the vector field of the PDMP by a stochastic differential equation, obtaining a Stochastic Hybrid System in the sense of \cite{HA:BujorianuLygeros:2004:GSHA}. The simplest possibility is to perturb the trajectories of the vector field with Gaussian noise, obtaining the so-called \emph{central-limit approximation}, for which a limit result analogous to Theorem \ref{th:Kurtz} exists (see \cite{STOC:Kurtz:1986:MarkovProcesses}, Chapter 11). This fact guarantees that convergence proofs presented in this paper extend straightforwardly to this new setting. Furthermore, in doing this, we get the advantage of removing the bad behaviours happening at the discontinuity boundary. Intuitively, in the central-limit regime, the probability of a trajectory to slide on a surface of dimension $n-1$ is zero. Similarly, the probability that a trajectory tangentially hits  a surface is zero. It follows that in this setting, most of the additional conditions on PDMP required for convergence hold almost automatically. On the downside, simulating a SDE is more expensive from a computational point of view, although the regularity of Gaussian Processes may be exploited to improve efficiency. We are currently investigating this direction. 

We also plan to investigate the definition of algorithms to check the conditions for convergence and to suggest a partition of variables into discrete and continuous (also for models in which the dependency on system size is not explicit).

An important question related to approximation theorems is if the weak convergence result, which are limited to the transient dynamics, can be extended to steady state. In the deterministic case, this can be done only in a limited number of cases. Essentially, convergence of steady state depends on the phase space properties of the limit ODE, and it is guaranteed only in presence of a unique globally attracting steady state, see e.g. \cite{PA:LeBoudec:2008:MeanFieldContinuousTime, STOC:Benaim:2003:MeanFieldGames}. In the future, we would like to investigate if  similar results can be found for the hybrid limit case. The situation in this case is more complex. On the one hand, if the limit process is deterministic, then we can exploit recent results \cite{STOC:BenaimLeBoudec:2011:stationaryConvergence}, provided we can characterise invariant measures for deterministic hybrid systems. On the other hand, if the limit process is stochastic, one has to prove that it does indeed have a steady state, and extend the result on invariant measures from the deterministic case to the stochastic one. In this setting, computing the invariant measure of the PDMP can be quite challenging in itself \cite{STOC:Costa:2008:ErgodicityPDMP}.

Finally, we want to understand what happens if we include non-determinism in the framework, especially in terms of uncertainty on parameters. This would require consideration of stochastic hybrid systems combining differential inclusions \cite{THMAT:AubinCellina:1984:DiffInclusion} with imprecise probabilities \cite{STAT:Walley:1991:ImpreciseProbabilities}. 

\paragraph{Acknowledgements.}
Work partially supported by ``FRA-UniTS'' grant. 

%
%
%
%


\clearpage

\appendix

\section{Notation}
\label{app:notation}

Here are some notational conventions followed throughout the paper. 
\begin{itemize}
\item We denote the closure of a set $A$ by $\bar{A}$ and its boundary by $\partial A$.
\item We denote with $\bbR$ the reals, $\bbZ$ the integers, $\bbN$ the natural numbers, and $\bbR_{\geq 0}$ the  non negative reals. 
\item $Q$ indicates a countable subset of $\bbR^k$, the set of modes. $\sspace{}$ is the state space of the fluid limit (in this case $\sspace{}\subseteq \bbR^\cdim$) or of the hybrid limit, and in this case $\sspace{}\subseteq  Q\times \bbR^\cdim$. 
\item $\cdim$ is the dimension of continuous variables, $\ddim$ is the dimension of discrete variables, $\fdim$ is the dimension of the full vector of variables.
\item $\X{}$ denotes the non-normalized vector of variables. We assume an ordering of variables, so we can interchange sets and vectors of variables freely. $\Y{}$ or $\X{c}$ are vectors of non-normalized continuous variables, while   $\Z{}$ or $\X{d}$ are vectors of discrete variables. 
\item  $\x{}$, $\y{}$, $\z{}$ denote points in the non-normalized space $\bbR^\fdim$.
\item $\Xb{}$ denotes the normalized vector of variables, taking values in  $\sspace{}$. $\Yb{}$ or $\Xb{c}$ are vectors of normalized continuous variables.
\item  $\xb{}$, $\yb{}$, $\zb{}$ denote points in the normalized space $\sspace{}$.
\item Given a vector function $f:\bbR^n\rightarrow\bbR^n$, with $f(\X{})$ we indicate the variables it depends to, while with $f[X](\X{})$ we indicate the function corresponding to variable $X$. 
\item $\size{}$ and $\size{N}$ denote the system size.
\item $(\calA,\size{N})$ is a population-\sCCP\ program. $\calT(\calA)$ is the TDSHA associated with \sCCP\ program $\calA$. $\n{\calT}(\calA,\size{N})$ is the TDSHA associated with the normalized population-\sCCP\ program  $(\calA,\size{N})$ and $\n{\calT}(\calA)$ is the limit normalized TDSHA.
\item $\pi$ denotes a \sCCP\ action, while $\eta$ denotes a TDSHA transition.
\item $\X{}\N(t)$ denotes a CTMC associated with a population \sCCP\ model with system size  $\size{N}$, while $\Xb{}\N(t)$ denotes the corresponding normalised CTMC. 
\item $\xb{}(t)$ denotes either the (normalised) limit PDMP or the fluid limit.
\item With $B_\eps(\xb{})$ we indicate the ball of radius $\eps$ centred in $\xb{}$, while $B_\eps(\xb{}([0,T]) =  \bigcup_{t\in[0,T]}B_\eps(\xb{}(t))$. 
\item $\T{i}$ is the jump time of the $i$-th stochastic event in the PDMP. $\zeta_i$ is the jump time of the $i$-th instantaneous event. $T_i$ is the jump time of the $i$-th event. 

\end{itemize}

\section{Additional material on \sCCP}
\label{app:sccp}

In this appendix, we provide a few additional details about \sCCP. We start by formalising the construction transforming a generic \sCCP\ program $\calA$ into a flat \sCCP model $flat(\calA)$. 

\begin{definition}
\label{def:flatSCCP}
Let  $\calA = (A,\Def,\vr{X},\calD,\vr{x_0})$ be a \sCCP\ program. Its \emph{flattened} version $flat(\calA) = (A^+,\Def^+,\vr{X}^+,\calD^+,\vr{x_0}^+)$ is constructed as follows:
\begin{itemize}
\item We add a new variable for each component $C\in\Def$: $\vr{X}^+ = \vr{X} \cup \vr{P}$, with $\vr{P} = \{P_C~|~C\in\Def\}$ taking values in $\bbN$.
\item Each component $C$ is replaced by a component $C^+$. If $C = \pi.A + M$, with $\pi = [g(\vr{X}),u(\vr{X},\vr{X'},\mu)]_{\lambda(\vr{X}}$, then $C^+ = \pi^+.C^+ + M^+$, with  $\pi^+ = [g^+(\vr{X}),u^+(\vr{X},\vr{X'},\mu)]_{\lambda^+(\vr{X})}$.
\item The guard of $\pi^+$ is $g^+(\vr{X}) = g(\vr{X})\wedge P_C > 0$.
\item The update of $\pi^+$ is  $u^+(\vr{X},\vr{X'},\mu) = u(\vr{X},\vr{X'},\mu)\wedge P_C' = P_C - 1 + \#(C,A) \wedge \bigwedge_{C_1 \neq C} P_{C_1}' = P_{C_1}  + \#(C_1,A)$, where $\#(C,A)$ is the number of occurrences of component $C$ in the parallel composition $A$.
\item The rate of $\pi^+$ is  $\lambda^+(\vr{X}) = P_C\cdot \lambda(\vr{X})$.
\item If $\pi$ is an instantaneous action, then $\infty:p(\vr{x})$ becomes $\infty:P_C \cdot p(\vr{x})$ in $\pi^+$.
\item The initial value of variables $\vr{x_0}^+$ equals $\vr{x_0}$ for variables in $\vr{X}$ and $\#(C,A)$ for each variable $P_C$. 
\item the initial network of $flat(\calA)$ is  $A^+ = \parallel_{C^+\in\Def^+} C^+$. 
\end{itemize}
\end{definition}

The notion of flattening has been previously defined in \cite{My2009TCSBjournalSCCPandODE}, but was called the extended version of a \sCCP\ program. In \cite{My2011LogicsCSjournalHybridSCCPLattice} we also showed that a \sCCP\ program and its flattened version have isomorphic labelled transitions systems, hence they are stochastically equivalent.

\begin{example}
\label{ex:bacteria}
We illustrate this transformation with a simple example. 
Consider a simple model  of a population of bacteria, in which each bacterium can consume a source of food and reproduce, or die. Both actions happen after some exponentially delayed time. We can model this in \sCCP\ by using a single integer variable $F$, representing the available food, initially set to $f_0$, and by describing each bacterium as an agent as follows:\\
\begin{center}
{\tt
\begin{tabular}{lcl}
    bacterium   & $\defeq$ & $[F > 0 \rightarrow F' = F - 1]_{k_r}$.(bacterium $\parallel$ bacterium) +     $[* \rightarrow  *]_{k_d}$.$\mathbf{0}$\\
\end{tabular}}\\
\end{center}
The initial network  {\tt bacterium $\parallel\ldots\parallel$ bacterium} consists of $m$ copies of the agent. This model can be flattened by introducing a new variable, $B$, counting the number of bacteria in the system, initially set to $m$, and replacing the agent {\tt bacterium} by \\
\begin{center}
{\tt
\begin{tabular}{lcl}
    bacterium\_flat   & $\defeq$ & $[F > 0 \wedge B > 0 \rightarrow F' = F - 1 \wedge B = B+1]_{k_r\cdot B}$.bacterium\_flat\\ & + &     $[B>0 \rightarrow  B' = B-1]_{k_d\cdot B}$.bacterium\_flat\\
\end{tabular}}\\
\end{center}
The new initial network will contain only the agent {\tt  bacterium\_flat}. Notice how the rates are updated to take into account the shift of perspective from the single agent to the population view. 
\end{example}

\section{Background Material}
\label{app:background}

In this section we briefly recall some notions that are needed in the proofs.

\paragraph{Hybrid state space.}

Let $Q$ be a countable subset of $\bbR^\ddim$, and consider $Q\times \bbR^\cdim$, the hybrid space. A point  $\x{}\in Q\times \bbR^n$ is  a pair $\x{}=(q,\y{})$, $\y{}\in\bbR^\cdim$.

In $Q\times \bbR^\cdim$, we define a metric $\bar{d}$ for which $Q\times \bbR^\cdim$ is a complete and separable metric space. This metric is derived from the euclidean metric $d$ in $\bbR^n$ by 
$$\bar{d}(\x{1},\x{2}) = \left\{\begin{array}{ll}
d(\y{1},\y{2})/(1+d(\y{1},\y{2})) & if\ \x{i} = (q_i,\y{i})\ and\ q_1= q_2,\\
1& if\ \x{i} = (q_i,\y{i})\ and\ q_1\neq q_2
\end{array} \right.$$
In particular,  $\bar{d}(\x{1},\x{2})<1$ if and only if $\x{1}$ and $\x{2}$ have the same $Q$-coordinate. Hence, a sequence converges with respect to $\bar{d}$, $\x{N}\rightarrow \x{}$, if and only if $\x{} = (q,y{})$, $\x{N} = (q,\y{N})$ for $N\geq N_0$, and $\y{N}\rightarrow \y{}$ in $\bbR^\cdim$.

Each subset $A$ of  $Q\times \bbR^n$ is of the form $A = \bigcup_{q\in Q}\{q\}\times A_q$, $A_q\subset\bbR^n$. 
A sub-base for the topology of $Q\times \bbR^n$ is given by the open balls of the form $\{q\}\times B_\eps(\y{})$. The boundary of a set $A$ is denoted by $\partial A$ and the closure by $\overline{A}$. 
Borel sets $\ssborel{}$ for $Q\times \bbR^n$ are defined from the Borel sets $\ssborel{q}$ of $\bbR^n$ as $\ssborel{} = \bigcup_{q\in Q}\{q\}\times \ssborel{q}$.  See \cite{STOC:Davis:1993:PDMP} for further details.

\paragraph{Skorohod metric.}

Continuous Time Markov Chains and Piecewise Deterministic Markov Processes considered in this paper can be seen as random variables on the space of cadlag functions $D([0,\infty),\sspace{})$ with values in $\sspace{}\subseteq Q\times\bbR^\cdim$. A cadlag function $f:[0,\infty)\rightarrow E$ is right continuous and has left limits for any $t\in[0,\infty)$.

The space $D([0,\infty),\sspace{})$ is given the structure of a  metric space by the Skorohod metric. 
The Skorohod metric is first defined on compact time intervals $[0,T]$ and then extended over the whole positive time axis $[0,\infty)$.

Consider the uniform metric on the space $D([0,T],\sspace{})$, i.e.\ $\sup_{0\leq t \leq T} \|\x{}\N(t) - \x{}(t) \|$. If we have a sequence $\x{}\N$ of cadlag functions, then they will converge to $\x{}$ in the uniform norm if and only if the discontinuous jumps of $\x{}\N$ happen precisely at the same times as those of $\x{}$ (for $N\geq N_0$).
The idea behind the Skorohod metric is to allow a small difference in these jump times by re-synchronizing them. That is to say, if the uniform metric allows one to wiggle space a bit, the Skorohod metric allows us also to wiggle time.
To formalize this statement, let $\omega(t):[0,T]\rightarrow [0,T]$ be a  time-wiggle function, i.e a strictly increasing continuous function. Call $\calI_T$ the set of such functions. Then, the Skorohod distance between $\x{},\y{}\in D([0,T],\sspace{})$ is
\[ d_T(\x{},\y{}) =\inf_{\omega\in \calI_T}\max\{\sup_{t\in[0,T]}\|\omega(t)-t\|,\sup_{t\in[0,T]}\|\x{}(t)-\y{}(\omega(t))\|\}.\]

The metric $d_T$ is extended to a metric on $D([0,\infty),\sspace{})$ by discounting large times as follows:
\[d(\x{},\y{}) = \sum_{K\in\bbN} 2^{-K}\min\{1,d_K(\x{},\y{})\}.\]

The Skorohod metric defines a topology for which $D([0,\infty),\sspace{})$ is complete and separable, i.e. it is a Polish space. See \cite{STOC:Pollard:1984:convStocProc, STOC:Billingsley:1999:convergenceProbabilities} for a detailed introduction to the metric and its properties.

We note here that a sequence of functions $\x{}\N\in D([0,\infty),\sspace{})$ converges to $\x{}\in D([0,\infty),\sspace{})$ if and only if for each $T>0$ there is a sequence of time-wiggle functions $\omega\N\in \calI_T$ satisfying $\sup_{t\in[0,T]}\|\omega\N(t)-t\|\rightarrow 0$ and $\sup_{t \in[0,T]}\|\x{}\N(t)-\x{}(\omega\N(t)) \|\rightarrow 0$.

\paragraph{Weak Convergence.}
 
The notion of weak convergence of a sequence of random variables $\X{}\N$ on a Polish space $\sspace{}$ to a random variable $\X{}$ is essentially the convergence of the induced probability measures on $\sspace{}$. Weak convergence of probability measures is defined as convergence in the weak* topology \cite{THMAT:Rudin:1973:functionalAnalysis}. More specifically, denote by $\calC_b(\sspace{})$ the set of bounded continuous functions $f:\sspace{}\rightarrow\bbR$ (note that we can have $\sspace{} = D([0,\infty),\sspace{0})$; in this case $f$ is usually called a functional), and let $P,P\N$ be probability measures on $\sspace{}$. We refer the reader to  \cite{STOC:Pollard:1984:convStocProc, STOC:Billingsley:1999:convergenceProbabilities} for an introduction to the subject.

\begin{definition}
\label{def:weakConvergence}
$P\N$ \emph{converges weakly} to $P$, $P\N\Rightarrow P$, if and only if, for each $f\in \calC(\sspace{})$,
\[\lim_{N\rightarrow\infty}\int_{\sspace{}}f(\x{})P\N(\x{})d\x{} = \int_{\sspace{}}f(\x{})P(\x{})d\x{}.\]
\end{definition} 
\noindent In case we have random variables $\X{},\X{}\N$, then the previous condition can be written as \[\lim_{N\rightarrow\infty}\bbE[f(\X{}\N)]=\bbE[f(\X{})].\] In this case, we write $\X{}\N \Rightarrow\X{}$.

The Portmanteau theorem provides a set of equivalent conditions for weak convergence of $\X{}\N$ to $\X{}$:
\begin{enumerate}
\item $\X{}\N\Rightarrow \X{}$; 
\item $\lim_{N\rightarrow\infty}\bbE[f(\X{}\N)]=\bbE[f(\X{})]$ for all bounded, uniformly continuous functions $f:\sspace{}\rightarrow\bbR$;
\item $\limsup_{N\rightarrow\infty}\bbP\{\X{}\N\in F\} \leq \bbP\{\X{}\in F\}$ for all closed sets $F$;
\item $\liminf_{N\rightarrow\infty}\bbP\{\X{}\N\in G\} \geq \bbP\{\X{}\in G\}$ for all open sets $G$;
\item $\lim_{N\rightarrow\infty} \bbP\{\X{}\N\in A\} = \bbP\{\X{}\in A\}$ for all $\X{}$-continuity sets $A$ (i.e., such that $\bbP\{\X{}\in\partial A\}=0$).
\end{enumerate} 

Recall that there are other modes of convergence of random variables, among which \emph{almost sure convergence} and \emph{convergence in probability}. 
These two notions, differently from weak convergence, require to have fixed the sample space in which random variables are  defined. 
More precisely, let $\X{},\X{}\N$ be random variables on $\sspace{}$, defined on the sample space $\Omega$, with $\sigma$-algebra $\calA$ and probability measure $\bbP$. (i.e. $\X{}:\Omega\rightarrow\sspace{}$ is a $\calA,\ssborel{}$ measurable function). 
Then $\X{}\N$ converges to $\X{}$ almost surely if and only if $\bbP\{\lim_{N\rightarrow\infty} \X{}\N(\omega) = \X{}(\omega)\}=1$, while it converges in probability if and only if, for each $\delta>0$, $\lim_{N\rightarrow\infty} \bbP\{\|\X{}\N-\X{}\|>\delta\}=0$. 

These three notions are linked in several ways. Almost sure convergence implies convergence in probability, which in turn implies weak convergence. Furthermore, the Skorohod representation theorem states that, if $\X{}\N\Rightarrow\X{}$, then there is a sample space $(\Omega,\calA,\bbP)$, and realizations $\asrep{\X{}},\asrep{\X{}}\N$ of $\X{},\X{}\N$ on $\Omega$ (i.e. $\asrep{\X{}}:\Omega\rightarrow\sspace{}$ induces the same probability on $\sspace{}$ as $\X{}$) such that $\asrep{\X{}}\N$ converges to $\asrep{\X{}}$ almost surely. 
Furthermore $\Omega$ can be taken as the unit interval with the Lebesgue measure.
In particular, for real-valued random variables $X\N,X$ on $\sspace{}\subseteq\bbR$, the Skorohod representation can be constructed using the quantile function $F^{\leftarrow}$, i.e. the pseudo-inverse of the cumulative distribution function $F(t) = \bbP\{X \leq t\}$: $\asrep{X\N} = (F\N)^{\leftarrow}(U)$ and  $\asrep{X} = F^{\leftarrow}(U)$, with $U$ uniform in $[0,1]$ \cite{STOC:Pollard:1984:convStocProc}.

Finally, weak convergence to a deterministic limit, i.e. a random variable concentrating all the probability mass to a point of $\sspace{}$, implies convergence in probability.

In the following, we also need the notion of \emph{tight} probability measure.
\begin{definition}
\label{def:TightProb}
A probability measure $P$ on $\sspace{}$ is \emph{tight} if and only if, for each $\eps>0$, there is a compact set $K_\eps\subset \sspace{}$ such that $P(K_\eps)>1-\eps$. 

A sequence $P\N$ of probability measure s on $\sspace{}$ is \emph{uniformly tight} if and only if for each $\eps>0$, there is a compact set $K_\eps\subset \sspace{}$ such that $P\N(K_\eps)>1-\eps$ for each $N\geq 0$. 
\end{definition}

If the space $\sspace{}$ is Polish, i.e. a complete and separable metric space, then each probability measure on $E$ is tight. Furthermore, if $P\N\Rightarrow P$, then $P\N$ is uniformly tight. 

Tightness is the right notion to characterize weak convergence in the space $D([0,\infty),\sspace{})$ equipped with the Skorohod topology. 
Let $\pi_{t_1,\ldots,t_k}:D([0,\infty),\sspace{}) \rightarrow\sspace{}^k$ be the projection at fixed times $t_1,\ldots,t_k\in[0,\infty)$. If $\X{}$ is a random variable in $D([0,\infty),\sspace{})$, then $\pi_{t_1,\ldots,t_k}(\X{})$ is called a finite dimensional distribution. 
Now, if $\X{},\X{}\N$ are random variables in $D([0,\infty),\sspace{})$, $\X{}\N$ is uniformly tight and $ \pi_{t_1,\ldots,t_k}(\X{}\N)\Rightarrow\pi_{t_1,\ldots,t_k}(\X{})$, for $t_j$ taken from a subset $\Gamma\subseteq [0,\infty)$ whose complement is at most countable (convergence of finite dimensional distributions), then $\X{}\N\Rightarrow\X{}$.   
Uniform tightness of $\X{}\N$ can be checked using some criteria based on the modulus of continuity, see \cite{STOC:Pollard:1984:convStocProc, STOC:Billingsley:1999:convergenceProbabilities} for further details. 

Finally, we need the \emph{continuous mapping theorem}: Let $\X{}\N,\X{}$ be random variables on $\sspace{}$ and let $h:\sspace{}\rightarrow\sspace{1}$. If $h$ is $\X{}$-almost surely continuous (i.e., $\bbP\{\X{}\in C_h\}=1$, where $C_h\subseteq \sspace{}$ is the set of continuity points of $h$), and $\X{}\N\Rightarrow\X{}$, then $h(\X{}\N)\Rightarrow h(\X{})$.


\paragraph{Markov Kernels.}

A Markov kernel or Markov transition kernel on $\sspace{}$, with $\sigma$-algebra $\ssborel{}$, is a function 
$R:\sspace{}\times\ssborel{}\rightarrow [0,1]$ such that
\begin{enumerate}
\item $R(\cdot,A)$ is a measurable function of for each  $A\in\ssborel{}$.  
\item $R(\y{},\cdot)$ is a probability measure for each $\y{}\in \sspace{}$. 
\end{enumerate} 

We now prove a Lemma that will be used in many proofs in next section, that allows us to propagate weak convergence by Markov kernels, under a suitable notion of continuity of the kernel. 

\begin{lemma}
\label{lemma:convergenceAfterReset}
Let $R\N(\yb{})=R\N(\yb{},\cdot)$ and $R(\yb{})=R(\yb{},\cdot)$ be Markov transition kernels on some Polish space $E$ such that $R\N(\yb{}\N) \Rightarrow R(\yb{})$, whenever $\yb{}\N\rightarrow \yb{}$.
If $\Yb{}\N \Rightarrow \Yb{}$, where $\Yb{}\N$, $\Yb{}$ are random elements in $E$, then $R\N(\Yb{}\N)\Rightarrow R(\Yb{})$ and $(\Yb{}\N,R\N(\Yb{}\N))\Rightarrow(\Yb{},R(\Yb{}))$.
\end{lemma}

\proof The proof is essentially the same as that of the core argument of Theorem 1 in \cite{STOC:Karr:1975:WeakConvergenceDTMP}. We reproduce it here just for completeness. Fix a bounded and uniformly continuous function $g:E\rightarrow\bbR$. We need to show that $|\bbE[g(R\N(\Yb{}\N))] - \bbE[g(R(\Yb{}))]|\rightarrow 0$. 
For simplicity, call $R\N g(\yb{}) = \integral{E}{}{g(\xb{})R\N(\yb{},\xb{})}{\xb{}}$ and similarly $Rg(\yb{})$, and further let $P\N(\yb{}) = \bbP\{\Yb{}\N = \yb{}\}$, and similarly for $P(\yb{})$. Then
\begin{eqnarray*}
|\bbE[g(R\N(\Yb{}\N))] - \bbE[g(R(\Yb{}))]|  & = &
\left|\integral{E}{}{R\N g(\yb{})P\N(\yb{})}{\yb{}} - \integral{E}{}{Rg(\yb{})P(\yb{})}{\yb{}}\right| \\
& \leq & \underbrace{\integral{E}{}{|R\N g(\yb{}) - Rg(\yb{})|P\N(\yb{})}{\yb{}}}_{(a)}\\ 
& + & \underbrace{\left|\integral{E}{}{R g(\yb{})P\N(\yb{})}{\yb{}} - \integral{E}{}{Rg(\yb{})P(\yb{})}{\yb{}}\right|.}_{(b)}
\end{eqnarray*}
The term (b) in the previous inequality goes to zero as the hypothesis imply that $Rg(\yb{})$ is a continuous function (see \cite{STOC:Karr:1975:WeakConvergenceDTMP}), and so we just need to focus on (a). 
As $E$ is a Polish space, i.e. a separable complete metric space, it follows from $P\N\Rightarrow P$ and $P$ tight that $P\N$ is uniformly tight. Then, we find a compact set $E_\eps$ such that $P\N(E_\eps) > 1-\eps/2\|g\|_\infty$, and so 
\begin{eqnarray*}
\integral{E}{}{|R\N g(\yb{}) - Rg(\yb{})|P\N(\yb{})}{\yb{}} & \leq & \integral{E_\eps}{}{|R\N g(\yb{}) - Rg(\yb{})|P\N(\yb{})}{\yb{}}+\eps\\ & \leq & \sup_{\yb{}\in E_\eps} |R\N g(\yb{}) - Rg(\yb{})| + \eps \rightarrow \eps,
\end{eqnarray*}
and therefore
$$\limsup_{N\rightarrow\infty} \integral{E}{}{|R\N g(\yb{}) - Rg(\yb{})|P\N(\yb{})}{\yb{}}  \leq \eps,$$
as the hypothesis of the lemma imply that $R\N g$ converges to $Rg$ uniformly on compact sets \cite{STOC:Karr:1975:WeakConvergenceDTMP}. As the previous inequality holds for each $\eps>0$, then $R\N(\Yb{}\N)\Rightarrow R(\Yb{})$.
To prove the second part of the theorem, observe that in the previous computation we can always restrict the expectation to a closed set $F_1\subset E$, showing that $\bbE[g(R\N(\Yb{}\N))\vr{1}_{F_1}(\Yb{}\N)] \rightarrow \bbE[g(R(\Yb{}))\vr{1}_{F_1}(\Yb{})]$. Now, if we fix another closed set $F_2\subset E$, and choose bounded uniformly continuous functions $g_\rho\downarrow \vr{1}_{F_2}$, approximating from above the indicator function of $F_2$, we then have
$$
\bbP\{(\Yb{}\N,R\N(\Yb{}\N))\in F_1\times F_2\} \leq 
\bbE[g_\rho(R\N(\Yb{}\N))\vr{1}_{F_1}(\Yb{}\N)], 
$$
from which, fixing $\rho$,
$$
\limsup_{N\rightarrow\infty} \bbP\{(\Yb{}\N,R\N(\Yb{}\N))\in F_1\times F_2\} \leq 
\bbE[g_\rho(R(\Yb{}))\vr{1}_{F_1}(\Yb{})], 
$$
by letting $\rho\rightarrow 0$ and invoking the bounded convergence theorem, as $g_\rho$ converges to $\vr{1}_{F_2}$, we have
$$
\limsup_{N\rightarrow\infty} \bbP\{(\Yb{}\N,R\N(\Yb{}\N))\in F_1\times F_2\} \leq 
\bbP\{(\Yb{},R(\Yb{}))\in F_1\times F_2\},$$
which by the Portmanteau theorem \cite{STOC:Billingsley:1999:convergenceProbabilities}, implies that $(\Yb{}\N,R\N(\Yb{}\N))\Rightarrow(\Yb{},R(\Yb{}))$.
\qed

\paragraph{Functional Analysis.}
In the following, we will also need the famous Gronwall inequality, which we recall here for convenience.
\begin{proposition}
\label{prop:app:Gronwall}
For any real valued integrable function $f$ on the interval $[0,T]$, if 
$$f(t) \leq C + D\integral{0}{t}{f(s)}{s},$$
then $$f(T) \leq Ce^{DT}.$$
\end{proposition}

\section{Proof of Main Lemmas and Theorems}
\label{app:proofs}

We will start by providing a quick proof of Theorem \ref{th:Kurtz}, the classic fluid theorem. This will be helpful in proving subsequent lemmas and theorems. 

\begin{apptheorem} [\ref{th:Kurtz}]
Let $(\calA,\size{N})$ be a sequence of population-\sCCP\ models for increasing system size $\size{N}\rightarrow\infty$, satisfying the conditions of this section, and with all \sCCP-actions $\pi$ satisfying the continuous scaling condition. Let $\Xb{}\N(t)$ be the sequence of normalized CTMC associated with the \sCCP-program and $\xb{}(t)$ be the solution of the fluid ODE.
\\
If $\xb{0}\N \rightarrow \xb{0}$ almost surely, then 
for any $T<\infty$, $\sup_{t\leq T}\|\Xb{}\N(t) - \xb{}(t) \| \rightarrow 0$ as $N\rightarrow\infty$, almost surely.
\end{apptheorem}

\proof 

Intuitively, the result of the theorem is a limit result, hence it depends only on what happens in a neighbourhood $B_\eps(\xb{}([0,T])$ of the solution of the ODE.
Thus, by restricting our attention to a compact set $K \subset E$ containing $B_\eps(\xb{}([0,T]) \cap E$ for some $\eps$, we can assume that all functions $g_\eta$ defining the rate of normalized transitions are bounded, say by $B_\eta$, and Lipschitz, say with Lipschitz constant $L_\eta$. This assumption is not restrictive, as  we can always extend the functions $g_\eta$ on the whole $E$ so that they are globally bounded and Lipschitz continuous. Clearly, $\xb{}(t)$ will remain unchanged by this operation, as it depends only on the value of $g_\eta$ in $K$.

We consider now the representation of the CTMC $\Xb{}\N(t)$ in terms of Poisson processes \cite{STOC:Kurtz:1986:MarkovProcesses}. We will use one Poisson process for each transition $\eta$ and each possible value of $\stoich{\eta}$ (which is a random element with bounded first and second moments). In the following, we indicate with $p_\eta(\w{})$ the probability that $\stoich{\eta} = \w{}$.
\begin{equation}
\label{eqn:app:CTMCintegral}
\Xb{}\N(t) = \Xb{0}\N + \sum_{\eta\in\calT}\sum_{\w{}\in \bbZ^n} \frac{\w{}}{N} \poisson{\eta}{Np_\eta(\w{}) \integral{0}{t}{ g\N_\eta(\Xb{}\N(s))}{s}}
\end{equation}
Furthermore, $\xb{}(t)$ in integral form is:
\begin{equation}
\label{eqn:app:OdeIntegral}
\xb{}(t) = \xb{0} + \integral{0}{t}{\sum_{\eta\in\calT} \bbE[\stoich{\eta}]g\N_\eta(\xb{}(s))}{s}
\end{equation}

In the following, we need the notion of centred Poisson process \cite{STOC:Kurtz:1986:MarkovProcesses}, defined by $\cpoisson{}{\lambda t} = \poisson{}{\lambda t} - \lambda t$, for which the following law of large numbers holds: $\sup_{t\leq T} \frac{1}{N}\cpoisson{}{N \lambda t} \rightarrow 0$ almost surely.

Now, we define 
\begin{eqnarray*}
\eps\N(t) & = & \Xb{}\N(t) - \Xb{}\N(0) - \integral{0}{t}{\sum_{\eta\in\calT} \bbE[\stoich{\eta}]g\N_\eta(\xb{}(s))}{s}\\
& = & \sum_{\eta\in\calT}\sum_{\w{}\in \bbZ^n} \frac{\|\w{}\|}{N} \cpoisson{\eta}{N p_\eta(\w{}) \integral{0}{t}{ g\N_\eta(\Xb{}\N(s))}{s}},
\end{eqnarray*}
so that $$\|\eps\N(t)\| \leq \sum_{\eta\in\calT}\sum_{\w{}\in \bbZ^n} \frac{\w{}}{N} \cpoisson{\eta}{N p_\eta(\w{}) B_\eta t}.$$
By the finiteness of second order moments for $\stoich{\eta}$, the previous equation is summable and we can further exchange limit and summation over $\w{}$ \cite{STOC:Kurtz:1986:MarkovProcesses}, to conclude that $\sup_{t\leq T}\|\eps\N(t)\| \rightarrow 0$ by the law of large numbers for centred Poisson processes.
Therefore, we have that
\begin{eqnarray*}
\sup_{t\leq T}\|\Xb{}\N(t)-\xb{}(t)\| & \leq&  \underbrace{\|\Xb{0}\N-\xb{0}\| + \sup_{t\leq T}\|\eps\N(t)\| + \sup_{t\leq T}\|\vfield{}\N(\Xb{}\N(t))-\vfield{}(\Xb{}\N(t))\|}_{=\delta\N(T)\rightarrow 0\ a.s.}\\ &+& \integral{0}{t}{\|\vfield{}(\Xb{}\N(t)) - \vfield{}(\xb{}(t))\|}{s}.
\end{eqnarray*}

Calling $\beta\N(T) = \sup_{t\leq T}\|\Xb{}\N(t)-\xb{}(t)\|$ and applying Lipschitz condition to the last term, we have that 
$$\beta\N(T) \leq \delta\N(T) + L\integral{0}{T}{\beta\N(t)}{t}.$$
By applying Gronwall's inequality (see Proposition \ref{prop:app:Gronwall}), we finally obtain 
$$\beta\N(T) \leq \delta\N(T)e^{L T} \rightarrow 0\ \ \mbox{almost surely.}$$\qed


We will turn now to prove Theorem \ref{th:hybridBasic}. We will need some auxiliary lemmas. The first one is a straightforward generalization of the Kurtz theorem, to the case in which the initial condition of the limit process is sampled from a distribution on the space $E$.

\begin{lemma}
\label{lemma:app:KurtzInitialDistribution}
Let $\Xb{}(t)$ and $\xb{}(t)$ be as in Theorem \ref{th:Kurtz}. Furthermore, assume that
$\|\Xb{0} - \xb{0} \|$ converges to zero almost surely. 
Then,  for any $T<\infty$, $\sup_{t\leq T}\|\Xb{}\N(t) - \xb{}(t) \| \rightarrow 0$ as $N\rightarrow\infty$ almost surely. 

\end{lemma}

\proof 
To begin with, suppose $\xb{0}$ is supported on a compact set $K_0$. Then the proof proceeds as in Theorem \ref{th:Kurtz}, with the only caveat that we need to consider a compact set $K$ containing an $\eps$-neighbourhood of all trajectories starting in $K_0$ up to time $T$ (which is a compact set, by continuity of the ODE flow). In fact, the argument of Theorem \ref{th:Kurtz} does not require that the initial condition of ODE is deterministic, but just  the convergence in probability of $\Xb{0}$ to $\xb{0}$.

Now, as $\xb{0}$ is \emph{tight}, for each $\eps>0$ there is a compact set $K_\eps$ such that $\bbP\{\xb{0}\not\in K_\eps\} < \eps$.
Conditional on $\xb{0}\in K_\eps$, the convergence of $\Xb{}\N(t)$ to $\xb{}(t)$ is then almost sure. Hence, fix a sequence $\eps_k\downarrow 0$, such that the corresponding $K_{\eps_k} \uparrow E$, and with $\sum_k \eps_k< \infty$. By discarding a set of measure 0, we can assume that convergence in each $K_{\eps_k}$ is sure. Then, any other point $u$ of the probability space $(\Omega,\calA,\bbP)$ on which processes are defined that makes convergence fail has to belong to the complement of $K_{\eps_k}$ infinitely often. By the Borel-Cantelli lemma \cite{STOC:Billingsley:1979:ProbabilityTheory}, the set of all such $u$ has probability zero. \qed

\bigskip

We will now turn to consider convergence of stochastic jump times, focussing attention on a single jump time, given convergent initial conditions of the stochastic and the piecewise-deterministic system. As we consider the first stochastic jump time, we can focus our attention on deterministic systems (with random initial conditions). In order to do this, we combine simple properties of the space of cadlag functions with the Skorohod representation theorem (see Appendix \ref{app:background}).

\begin{proposition}
\label{prop:convergenceIntegrals}
Let $\vr{x}\N$, $\vr{x}$ be elements of $D([0,\infty),\sspace{})$, such that $\x{}\N\rightarrow\x{}$ (with respect to the Skorohod metrics). Then, for each $T>0$, $\integral{0}{T}{\x{}\N(s)}{s} \rightarrow \integral{0}{T}{\x{}(s)}{s}$.
\end{proposition}

\proof By the definition of the Skorohod metrics, let $\omega\N(t)$ be a sequence of time-wiggle functions such that $\sup_{t\leq T}\|\omega\N(t)-t\|\rightarrow 0$, and $\sup_{t\leq T}\|\x{}\N(t)-\x{}(\omega\N(t)) \|\rightarrow 0$. Now, 
\[ \integral{0}{T}{\|\x{}\N(s)-\x{}(s)\|}{s} \leq \underbrace{\integral{0}{T}{\|\x{}\N(s)-\x{}(\omega\N(s))\|}{s}}_{(a)} + \underbrace{\integral{0}{T}{\|\x{}(\omega\N(s))-\x{}(s)\|}{s}}_{(b)}. \]
The term (a) goes to zero by the uniform convergence of $\x{}\N(t)$ to $\x{}(\omega\N(t))$, while for (b), observe that the function $g\N(t) = \|\x{}(\omega\N(t))-\x{}(t)\|$ goes to zero in every continuity point of $\x{}$, thus almost everywhere. Furthermore, in $[0,T]$ the function $\x{}$ is bounded by a compactness argument (see \cite{STOC:Billingsley:1999:convergenceProbabilities}), and so is $g\N$, so that we can apply the bounded convergence theorem \cite{STOC:Billingsley:1979:ProbabilityTheory} to conclude that (b) converges to zero. \qed


\paragraph{Jump times.}
In order to show convergence of jump times of discrete stochastic transitions, we need the notion of \emph{cumulative rate} for the PDMP $\Lambda(t)$ and for the CTMCs.  

Consider discrete stochastic actions $\pi\in\disc{\calA}$, and the rate of firing of discrete actions in the PDMP, $\srateb{}(\xb{}) = \sum_{\pi\in\disc{\calA}}\srateb{\pi}(\xb{})$, and in the CTMC at level $N$, $\srateb{}\N(\xb{}) = \sum_{\pi\in\disc{\calA}}\srateb{\pi}\N(\xb{})$.
Then, the \emph{cumulative rate} for the PDMP $\xb{}$ is 
\begin{equation}
\label{eqn:app:cumulativeRatePDMP}
\Lambda(t) = \integral{0}{t}{\srateb{}(\xb{}(s))}{s}, 
\end{equation}
while for the CTMC $\Xb{}\N(t)$ is
\begin{equation}
\label{eqn:app:cumulativeRatePDMP}
\Lambda\N(t) = \integral{0}{t}{\srateb{}\N(\Xb{}\N(s))}{s}. 
\end{equation}

These are the cumulative rates of non-homogeneous Poisson processes. We are interested in the first firing time, whose cumulative distribution function is given by $1-e^{-\Lambda(t)}$. By a standard inversion method \cite{STOC:Billingsley:1979:ProbabilityTheory, SB:Wilkinson:2006:StochasticModellingSB},
the first firing time for the PDMP is given by $\tau = \inf\{t\geq 0~|~1-e^{-\Lambda(t)} \geq U \} = \inf\{t\geq 0~|~\Lambda(t) \geq \xi \}$, where $U$ is a uniform random variable and $\xi$ is an exponentially distributed random variable with rate $1$ (it holds that $\xi = -\log U$). We assume $\inf\emptyset = \infty$. Similarly, we can define $\tau\N = \inf\{t\geq 0~|~\Lambda\N(t) \geq \xi \}$, the first firing time of a discrete transition $\pi\in\disc{\calA}$ in $\Xb{}\N(t)$.  
By the Skorohod representation theorem for unidimensional random variables (see Section \ref{app:background} or, for instance, \cite{STOC:Pollard:1984:convStocProc}), if the pointwise convergence of $\Lambda\N(t)$ to $\Lambda(t)$ holds, then $\tau\N\rightarrow \tau$ almost surely.
We can combine these facts with Proposition \ref{prop:convergenceIntegrals}, to prove the following lemma.

\begin{lemma}
\label{lemma:convergenceJumpTimes}
Let $\Xb{}\N(t)\Rightarrow \xb{}(t)$, and $\tau\N$, $\tau$ be defined as above. If $\srateb{}$ is continuous and $\srateb{}\N\rightarrow\srateb{}$ uniformly in each compact set $K\subseteq \sspace{}$, 
then $\tau\N \Rightarrow \tau$, as $N\rightarrow\infty$.
\end{lemma}
\proof The first step of the proof is to use the Skorohod representation theorem to construct realizations $\tilde{\X{}}\N$ of $\Xb{}\N$ and $\tilde{\X{}}$ of $\Xb{}$ on some probability space $\calP$ such that $\tilde{\X{}}\N \rightarrow \tilde{\X{}}$ almost surely, as random elements in the space of cadlag functions. \\
It then follows that $\srateb{}\N(\tilde{\X{}}\N) \rightarrow \srateb{}(\tilde{\X{}})$ almost surely. In fact, consider the time-wiggle functions $\omega\N$ (depending also on the sample space $(\Omega,\calA,\bbP)$, i.e. $\omega\N = \omega\N(u,t)$, for $u\in\Omega$) such that $\omega\N\rightarrow id$ and $\tilde{\Y{}}\N = \tilde{\X{}}\N \circ \omega\N \rightarrow \tilde{\X{}}$ uniformly in $[0,T]$. Then
\[\begin{split} \sup_{t\leq T} \| \srateb{}\N(\tilde{\X{}}\N(\omega\N(t))) - \srateb{}(\tilde{\X{}}(t)) \| & \leq  \underbrace{\sup_{t\leq T} \| \srateb{}\N(\tilde{\Y{}}\N(t)) - \srateb{}(\tilde{\Y{}}\N(t)) \|}_{(a)}\\ 
& + \underbrace{\sup_{t\leq T} \| \srateb{}(\tilde{\Y{}}\N(t)) - \srateb{}(\tilde{\X{}}(t)) \|}_{(b)}.
\end{split}\]
Term (a) goes to zero by uniform convergence of $\srateb{}\N$ to $\srateb{}$ (as in $[0,T]$ $\asrep{\X{}}\N$ and $\asrep{\X{}}$ are contained in a compact set), while term $(b)$ goes to zero due to uniform convergence of $\tilde{\Y{}}\N$ to $\tilde{\X{}}$ in $[0,T]$ and uniform continuity of $\srateb{}$ in $[0,T]$. \\
Now, we can apply Proposition \ref{prop:convergenceIntegrals} to $\srateb{}\N(\tilde{\X{}}\N) \rightarrow \srateb{}(\tilde{\X{}})$, to conclude that $\Lambda\N(T)\rightarrow \Lambda(T)$ almost surely for each $T>0$. Combining this with the Skorohod representation theorem for real random variables,\footnote{We are effectively coupling $\Xb{}\N$, $\xb{}$, $\T{}\N$ and $\T{}$ on the probability space $\Omega\times [0,1]$. Note in particular that we allow $\tau$ and $\tau\N$ to take the value $\infty$. This can happen with non-null probability if and only if  $\Lambda(T)$ does not diverge as $T\rightarrow\infty$. } we get $\tilde{\tau}\N\rightarrow \tilde{\tau}$ almost surely, where $\tilde{\tau}\N$ and $\tilde{\tau}\N$ are the jump times obtained from the realizations of the original processes. It then follows that $\tau\N\Rightarrow \tau$. \qed

\bigskip

We finally consider the convergence of states at times $\tau\N$ and $\tau$.

\begin{proposition}
\label{prop:app:convergenceValuesAtJumpTimes}
Let $\Xb{}\N\Rightarrow\xb{}$ and let $\tau\N, \tau$ be stopping times satisfy conditions of the previous lemma.
Then, conditional on $\tau < \infty$, $\Xb{}\N(\tau\N) \Rightarrow \xb{}(\tau)$.
\end{proposition}

\proof In fact, let $t\N\rightarrow t<\infty$ (here we use implicitly the fact that $\tau<\infty$). Then, use Skorohod representation theorem and take representations $\tilde{\X{}}\N$ and $\tilde{\X{}}$ of $\Xb{}\N$ and $\Xb{}$ such that $\tilde{\X{}}\N\rightarrow\tilde{\X{}}$ almost surely. By continuity of $\tilde{\X{}}$ and uniform convergence of $\tilde{\X{}}\N$ and $\tilde{\X{}}$ in $[0,T]$ (the Skorohod metrics and the uniform metrics on compact sets are the same when the limit function is continuous), it follows that $\tilde{\X{}}\N(t\N)\rightarrow\tilde{\X{}}(t)$ almost surely, hence $\Xb{}\N(t\N)\Rightarrow\xb{}(t)$. Hence, by seeing $\Xb{}\N$ and $\Xb{}$ as Markov kernels, we can apply Lemma \ref{lemma:convergenceAfterReset} to conclude. \qed

In order to prove Theorem \ref{th:hybridBasic}, we will use an inductive argument whose core is the following corollary, which combines the previous results in the light of weak convergence. 

\begin{corollary} 
\label{cor:weakKurtz}
Let $(\calA,\size{N})$ be a sequence of \sCCP\ models for increasing systems size, satisfying the conditions of this section, and with all actions $\pi$ satisfying the continuous scaling condition. Let $\Xb{}\N(t)$ be the associated sequence of normalized CTMC and $\xb{}(t)$ be the solution of the fluid ODE.

If $\Xb{0}\N \Rightarrow \xb{0}$, then 
\begin{enumerate}
\item $\Xb{}\N \Rightarrow \xb{}$, as random elements in the space of cadlag function on $E$, with the Skorohod metrics.
\item If $\tau\N$, $\tau$, are the jump times of a stochastic event with rate $\lambda\N$ and $\lambda$, respectively, then $\tau\N\Rightarrow \tau$;
\item $\Xb{}\N(\tau\N) \Rightarrow \xb{}(\tau)$;
\item If $R\N(\yb{})$ and $R(\yb{})$ are reset kernels satisfying $R\N(\yb{}\N) \Rightarrow R(\yb{})$ whenever $\yb{}\N\rightarrow\yb{}$, then 
$R\N(\Xb{}\N(\tau\N))\Rightarrow R(\xb{}(\tau))$;
\item Under the previous conditions,\\ 
$(\Xb{0}\N,\Xb{}\N,\T{}\N,\Xb{}\N(\T{}\N),R\N(\Xb{}\N(\T{}\N)))\Rightarrow (\xb{0},\xb{},\T{},\xb{}(\tau),R(\xb{}(\tau)))$
\end{enumerate}
\end{corollary}

\proof The proof works simply by constructing an a.s.\ convergent realization of the initial conditions. 
Then, by Lemma \ref{lemma:app:KurtzInitialDistribution}, we obtain point 1. Point 2 follows from Lemma \ref{lemma:convergenceJumpTimes} and point 1, while point 3 from Proposition \ref{prop:app:convergenceValuesAtJumpTimes} and point 2.  Point 4 follows from Lemma \ref{lemma:convergenceAfterReset} and point 3. Point 5, instead, follows again from Lemma \ref{lemma:convergenceAfterReset}, observing that each element on the vector is defined conditionally on the previous one (e.g. $\Xb{}\N$ depends conditionally on $\Xb{0}\N$, $\tau\N$ depends conditionally on $\Xb{}\N$, and so on), and this dependence satisfies the assumptions of the Lemma. For instance, if $\yb{}\N\rightarrow\yb{}$, then $(\Xb{}\N|\Xb{0}\N = \yb{}\N) \Rightarrow (\xb{}|\xb{0} = \yb{})$, and the dependency is measurable (in fact, continuous) on $\yb{}\N$, $\yb{}$. Similar observations hold for the other elements of the vector. Then an iterated application of Lemma \ref{lemma:convergenceAfterReset} is enough to conclude. \qed

We can now prove Theorem \ref{th:hybridBasic}.

\begin{apptheorem}[\ref{th:hybridBasic}]
Let $(\calA,\size{N})$ be a sequence of population-\sCCP\ models for increasing system size $\size{N}\rightarrow\infty$, satisfying the conditions of this section, with variables partitioned into discrete $\X{d}$, continuous $\X{c}$, and environment ones $\X{e}$. 
Assume that discrete actions  satisfy scaling \ref{scaling:discreteStochastic} and continuous actions satisfy scaling \ref{scaling:continuousHybrid}.
Let $\Xb{}\N(t)$ be the sequence of normalized CTMC associated with the \sCCP\ program  and $\xb{}(t)$ be the PDMP associated with the limit normalized TDSHA $\norm{\calT}(\calA)$.

If $\xb{0}\N \Rightarrow \xb{0}$ (weakly) and the PDMP is non-Zeno, then $\Xb{}(t)$ converges weakly to $\xb{}(t)$,  $\Xb{} \Rightarrow \xb{}$, as random elements in the space of cadlag function with the Skorohod metric.
\end{apptheorem}

\proof

The basic idea of the proof to show weak convergence is to apply inductively the previous corollary, to show weak convergence of $\Xb{m}\N(t) = \Xb{}\N(t\wedge\T{m+1}\N)$ to $\xb{m}(t) = \xb{}(t\wedge\T{m+1})$, i.e. to processes stopped after $m+1$ jumps, and then lift this to the full weak convergence.  

\paragraph*{Step 1: weak convergence conditional on $m$ jumps or less.}

Consider the sequence $\T{1},\T{2},\ldots$ of jump times of discrete stochastic transitions in the PDMP $\xb{}(t)$ and the sequence  $\T{1}\N,\T{2}\N,\ldots$ of jump times of discrete transitions $\pi\in\disc{\calA}$ in $\Xb{}\N(t)$. In the following, we need to take care also of the fact  that $\T{m}$ may be infinite with probability greater than zero. Note that, conditional on $\T{m}$ being infinite, all $\T{m+j}$ will be infinite, too.

In order to be more concise, let us introduce some additional local notation. First, denote $\Zb{m}\N = \Xb{}\N(\T{m}\Np)$ and $\zb{m} = \xb{}(\T{m}^+)$ the states of the CTMC at level $N$ and of the PDMP after the $m$-th discrete jump. If $\T{m}$ or $\T{m}\N$ are infinite, we assume $\zb{m}$ or $\Zb{m}\N$ be equal to a special value $(q_\Delta,\vr{0})$, where $q_\Delta$ is a special state of $Q$ where nothing happens: the vector field and the jump rate are null (i.e.\ it is a cemetery point). Note that $(q_\Delta,\vr{0})$ has  distance 1 from any point $(q,\vr{x})$ in $E$. 
\\
Let also $\Zb{0}\N = \Xb{0}\N$, $\zb{0} = \xb{0}$, $\T{0}\N = \T{0} = 0$. 
We  define $\Yb{m}\N(t)$ to be the CTMC starting from $\Zb{m}\N$ with no discrete jumps, and 
$\yb{m}(t)$ the PDMP starting in $\zb{m}$ with no discrete jumps (in fact, an ODE with random initial conditions). Notice that, if $\T{m}\N$ (resp. $\T{m}$) is finite, then $\Yb{m}\N(t)$ (resp. $\yb{m}(t)$) coincides with $\Xb{}\N(\T{m}\N+t)$ (resp. $\xb{}(\T{m}+t)$) for $\T{m}\N\leq t<\T{m+1}\N$ (resp. $\T{m}\leq t<\T{m+1}$), by the strong Markov property of CTMC \cite{STOC:Norris:1997:MarkovChains} and of PDMP \cite{STOC:Davis:1993:PDMP}. 

We will now prove that, for each $m>0$, conditional on $\T{m}<\infty$, 
$\Yb{m}\N \Rightarrow \yb{m}$ and $\T{m+1}\N\Rightarrow \T{m+1}$. Moreover, if $\T{m+1}<\infty$, then also $\Zb{m+1}\Rightarrow \zb{m+1}$. 
Finally, we will also show that, conditional on $\T{m+1}<\infty$, $(\Zb{0}\N,\Yb{0}\N,\T{1}\N,\Zb{1}\N,\ldots,\Yb{m}\N,\T{m+1}\N,\Zb{m+1}\N) \Rightarrow (\zb{0},\yb{0},\T{1},\zb{1},\ldots,\yb{m},\T{m+1},\zb{m+1})$. 
The argument is a simple induction. In particular, the induction hypothesis is that $(\Zb{0}\N,\Yb{0}\N,\T{1}\N,\Zb{1}\N,\ldots,\Zb{m}\N) \Rightarrow (\zb{0},\yb{0},\T{1},\zb{1},\ldots,\zb{m})$ conditional on $\T{m}<\infty$.\footnote{In particular, this implies that $\zb{m}\neq (q_\Delta,\vr{0})$, and as $\Zb{m}\N\Rightarrow\zb{m}$, ultimately also $\Zb{m}\N\neq (q_\Delta,\vr{0})$.} From this, $\Yb{m}\N \Rightarrow \yb{m}$  is immediate from Lemma \ref{lemma:app:KurtzInitialDistribution}, and $\T{m+1}\N\Rightarrow \T{m+1}$  follows from Lemma \ref{lemma:convergenceJumpTimes}. Now, conditional on $\T{m+1}<\infty$, we can apply Proposition \ref{prop:app:convergenceValuesAtJumpTimes} to conclude that $\Zb{m+1}\Rightarrow \zb{m+1}$. As $\T{m+1}<\infty$ implies $\T{m}<\infty$, reasoning as in Corollary \ref{cor:weakKurtz} (using the same argument there to extend inductively its length), we obtain the weak convergence of vectors of the random elements.

Consider now $\Xb{m}\N(t) = \Xb{}\N(t\wedge\T{m+1}\N)$ and $\xb{m}(t) = \xb{}(t\wedge\T{m+1})$. We can write $\Xb{m}\N(t) = \sum_{i=0}^{m} \Yb{i}\N(t-\T{i}\N)\ind{\T{i}\N\leq t < \T{i+1}\N}  + \ind{\T{m+1}\N\leq t}\Zb{m+1}\N$ and $\xb{m}(t) = \sum_{i=0}^{m} \yb{i}(t-\T{i})\ind{\T{i}\leq t < \T{i+1}}  + \ind{\T{m+1}\leq t}\zb{m+1}$. Now, the functional that associates with $T$ the cadlag element $\ind{t\leq T}$ is continuous with respect to Skorohod metrics, and if we consider the previous definitions  of $\Xb{m}\N$ and $\xb{m}$ as a function of $(\Zb{0}\N,\Yb{0}\N,\T{1}\N,\Zb{1}\N,\ldots,\Zb{m+1}\N)$ and $(\zb{0},\yb{0},\T{1},\zb{1},\ldots,\zb{m+1})$, respectively, then this function is continuous. Hence, conditional on $\T{m+1} < \infty$,  $\Xb{m}\N \Rightarrow \xb{m}$ by the continuous mapping theorem. The same property holds also when $\T{m+1} = \infty$. 
In fact, conditioning on $\T{j} < \infty$ and $\T{j+1} = \infty$, for $j\leq m$, we observe that the process $\w{j}(t)  = \sum_{i=0}^{j-1} \yb{i}(t-\T{i})\ind{\T{i}\leq t < \T{i+1}} + \ind{t \geq \T{j}}\yb{j}(t-\T{j})$ coincides with $\xb{m}(t)$, and by the same argument above, applied to vectors $(\Zb{0}\N,\Yb{0}\N,\T{1}\N,\Zb{1}\N,\ldots,\T{j+1}\N)$ and $(\zb{0},\yb{0},\T{1},\zb{1},\ldots,\T{j+1})$, the processes $\W{j}\N(t)  = \sum_{i=0}^{j-1} \Yb{i}\N(t-\T{i}\N)\ind{\T{i}\N\leq t < \T{i+1}\N} + \ind{t \geq \T{j}\N}\Yb{j}\N(t-\T{j}\N)$  converge weakly to $\w{j}$. 
Now, the processes $\W{j}\N(t)$ and $\Xb{m}\N(t)$ are the same up to time $\T{j+1}\N$, which is a divergent sequence  under the event $\{\T{j} < \infty, \T{j+1} = \infty\}$\footnote{In fact, $\T{j+1}\N\Rightarrow\T{j+1}$ conditional on $\T{j}<\infty$. Moreover, if $\T{j+1}=\infty$, it also holds that $\T{j+k}\N\Rightarrow\T{j+k}$ for any $k>0$, as $\T{j+k} = \infty$ and $\T{j+k}\N \geq \T{j+1} \rightarrow\infty$. This means that by induction we can conclude $\T{j}\N\Rightarrow\T{j}$ for any $j$.}. This implies that $\Xb{m}\N$ converges weakly to $\W{j}\N$ (in fact, their Skorokhod distance converges weakly to zero, in fact a.s. under any a.s. realisation of $\T{j+1}\N\rightarrow \infty$), and hence, by uniqueness of the limit, to  $\W{j} = \xb{m}$. Now, as the events $\{\T{j} < \infty, \T{j+1} = \infty\}$, for $j=0,\ldots,m$ and $\{\T{m+1}<\infty\}$ are disjoint and their union has probability one,  we can remove the conditioning and conclude $\Xb{m}\N \Rightarrow \xb{m}$.\footnote{
To see this more precisely, couple all processes $\W{j}\N$, $\w{j}$, $\xb{m}$, and $\Xb{m}\N$, and the exit times $\T{j}$ on a common space $\Omega$, and let $\Omega_j$ the subset corresponding to the event $\{\T{j} = \infty$, $\T{j-1} < \infty\}$, for $j = 1,\ldots,m$, and $\Omega_0$ be the subset corresponding to the event $\{\T{m}<\infty\}$. 
Clearly $\Omega_j$, for $j = 0,\ldots,m$, form a partition of $\Omega$. Let $\bbP$ be the probability measure on $\Omega$,  let $\mu_j$ the push-forward measure on the space of cadlag functions on $E$ of $\w{j}$, i.e. of $\xb{m}$ conditioned on $\Omega_j$, and let $\mu_j\N$ be the push-forward measure of $\Xb{m}\N$ conditioned on $\Omega_j$, for $j$ such that $p_j=\bbP(\Omega_j) > 0$ (call $J$ such a set of indices). 
We know $\mu_j\N\Rightarrow \mu_j$. Then $\mu$, the push-forward measure of $\xb{m}$, coincides with $\sum_{j\in J} p_j \mu_j$, and similarly $\mu\N$, the push-forward measure of $\Xb{m}\N$, is  $\sum_{j\in J} p_j \mu_j\N$. Now, let $F$ be a closed set of the cadlag space $D([0,\infty),E$. Then 
\[\limsup_N \mu_N(F) \leq \sum_{j\in J} p_j \limsup_N \mu_j\N(F) \leq \sum_{j\in J} p_j  \mu_j(F) = \mu(F),\] 
which implies $\mu\N\Rightarrow\mu$, and hence $\Xb{m}\N\Rightarrow\xb{m}$, by the Portmanteau theorem. 
}

\paragraph*{Step 2: Weak convergence.}
We now lift the weak convergence $\Xb{m}\N \Rightarrow \xb{m}$ to weak convergence of the full processes $\Xb{}\N \Rightarrow \xb{}$. 
Consider a bounded uniformly continuous function $f$ from the space $D = D([0,\infty),E)$ of cadlag functions with values in $E$ to $\bbR$. 
By the definition of the Skorohod metric $d$ in $D$, for each $\rho>0$, there is a $T > 0$ such that $d(x,y) < d_T(x,y) + \rho$, where $d_T$ is the metric restricted to the compact time interval $[0,T]$ (see Section \ref{app:background}).  
By the uniform continuity of $f$, given $\eps>0$, we fix a $\rho>0$ such that $|f(x)-f(y)|<\eps/4$ whenever $d(x,y)<\rho$.

Now, fix $T>0$ according to the previous condition on $\rho$, and choose $m$ such that $\bbP\{\T{m+1}>T\}>1-\eps/(16\|f\|)$, which can be found since the expected number of discrete transitions fired by the PDMP at time $T$ is finite. As $\T{m+1}\N\Rightarrow\T{m+1}$, we can also find an $N_0$ such that, for all $N\geq N_0$, $\bbP\{\T{m+1}\N>T\}>1-\eps/(8\|f\|)$ (using the liminf condition in the Portmanteau theorem). 

Now, conditioning on $\T{m+1} > T$, we have that $\xb{}(t\wedge T) = \xb{m}(t\wedge T)$, and so $d(\xb{},\xb{m})\leq \rho$. Similarly, conditioning on  $\T{m+1}\N > T$, we have $\Xb{}\N(t\wedge T) = \Xb{m}\N(t\wedge T)$ and $d(\Xb{}\N,\Xb{m}\N)\leq \rho$. 
Now
\begin{eqnarray*}
\left|\bbE[f(\Xb{}\N)] - \bbE[f(\xb{})]\right| & \leq & \underbrace{\bbE[|f(\Xb{}\N) - f(\Xb{m}\N)|]}_{(a)} + \underbrace{\bbE[|f(\xb{}) - f(\xb{m})|]}_{(b)}\\
& + & \underbrace{\left|\bbE[f(\Xb{m}\N)] - \bbE[f(\xb{m})]\right|}_{(c)}
\end{eqnarray*}
Now, term (c) goes to zero as $\Xb{m}\N \Rightarrow \xb{m}$. To bound (b), instead, using properties of conditional expectation, we have 
\begin{eqnarray*}
\bbE[|f(\xb{}) - f(\xb{m})|] & = & \bbE[\bbE[|f(\xb{}) - f(\xb{m})| ~\big|~\ind{\T{m+1}>T}]]\\
& \leq &  \bbE[|f(\xb{}) - f(\xb{m})| ~\big|~\ind{\T{m+1}>T}=1]\cdot\bbP\{\T{m+1}>T\} + 2\|f\|\bbP\{\T{m+1}\leq T\}\\
&\leq &  \bbE[|f(\xb{}) - f(\xb{m})| ~\big|~\ind{\T{m+1}>T}=1] + \eps/4 \leq \eps/2, 
\end{eqnarray*}
where the last inequality follows from the choice of $\rho$ and the fact that $d(\xb{},\xb{m})\leq \rho$.
A similar argument can be used for term (a), allowing us to conclude that 
$$\limsup_{N\rightarrow\infty} \bbE[|f(\Xb{}\N) - f(\Xb{m}\N)|] \leq \eps/2,$$
from which we have
$$\limsup_{N\rightarrow\infty} \left|\bbE[f(\Xb{}\N)] - \bbE[f(\xb{})]\right| \leq \eps.$$
By the arbitrariness of $\eps>0$, we can finally conclude that $\left|\bbE[f(\Xb{}\N)] - \bbE[f(\xb{})]\right| \rightarrow 0$. \qed


\subsection{Instantaneous transitions}

In this subsection, we give the proofs of results contained in Section \ref{sec:InstantaneousTransitions} of the paper.

\begin{applemma}[\ref{lemma:convergenceExitTimes}]

Let $(\calA,\size{N})$ be a sequence of population-\sCCP\ models for increasing population size.
Let $\Xb{}\N(t)$ be the associated sequence of normalized CTMC, and suppose  $\Xb{}\N\Rightarrow\Xb{}$, where $\Xb{}$ has a.s.\ continuous sample paths. 
Let $h\N$, $h$ be activation functions for $\Xb{}\N$ and $\Xb{}$, such that $h\N\rightarrow h$ uniformly, and suppose $h$ is transversal to $\Xb{}$. Then $\zeta\N\Rightarrow \zeta$.
\end{applemma}

\newcommand{\az}{\asrep{\zeta}}

\proof First of all, use the Skorohod representation theorem to construct representations $\Xt{}\N$ of $\Xb{}\N$ and $\Xt{}$ of $\Xb{}$ such that $\Xt{}\N\rightarrow\Xt{}$ almost surely. Now fix sample trajectories $\xt{}\N\rightarrow \xt{}$ in the Skorohod metrics. As $\Xt{}$ is almost surely continuous, we can assume $\xt{}$ continuous. In this case, the Skorohod metric is the same as the uniform metric on compact sets $[0,T]$.  
In particular, we can take $T$ larger than $\az+\delta$, for any $\delta >0$, where $\az$ is the exit time for $\xt{}$. 
Now, since $h$ is transversal for $\Xb{}$, there is a $\delta>0$ such that $h(\xt{}(t))>0$ for $t\in(\az,\az+\delta]$. Let $\bar{h} = \min\{ \max\{-h(\xt{}(t))~|~t\in[\az-\delta,\az]\},\max\{h(\xt{}(t))~|~t\in[\az,\az+\delta]\}\}$. 
Fix $\eps>0$, $\eps < \bar{h}$, and let $\az_\eps^- = \sup\{t\leq \az~|~ h(\xt{}(t)) \leq -\eps  \}$ and $\az_\eps^+ = \inf\{t\geq \az~|~  h(\xt{}(t)) \geq \eps \}$. 
By continuity of $\xt{}$, it follows that $\|h(\xt{}(t)) - h(\xt{}(\az))\| < \eps$ for any $t\in(\az_\eps^-,\az_\eps^+)$, and that $\az_\eps^-,\az_\eps^+\rightarrow \az$ as $\eps\rightarrow 0$. 

Now, choose a compact set $K$ in $E$ that contains the $\eps$-neighbourhood of $\xt{}$ in $[0,T]$, for $T>\az_\eps^+$. As $h$ is uniformly continuous in $K$, pick a $\rho>0$ such that $\|h(\xb{1}) - h(\xb{2})\|<\eps/4$ whenever $\|\xb{1}-\xb{2}\| < \rho$, and fix $N_0>0$ such that $\xt{}\N(t)$ is $\rho$-close to $\xt{}(t)$ for $N\geq N_0$, uniformly in $[0,T]$. Furthermore, find $N_1$ such that, for $N\geq N_1$, $\sup_{\xb{}\in K}\|h\N(\xb{})-h(\xb{})\| < \eps/4$. Let $\bar{N} = \max\{N_0,N_1\}$.
It follows that, for $N\geq \bar{N}$, $\|h\N(\xt{}\N(t)) - h(\xt{}(t))\| \leq \|h\N(\xt{}\N(t)) - h(\xt{}\N(t))\| + \|h(\xt{}\N(t)) - h(\xt{}(t))\| < \eps/2$, and so 
$h\N(\xt{}\N(t)) < 0$ for $t\in[0,\az_\eps^-]$, hence $\az\N > \az_\eps^-$. Furthermore, $h\N(\xt{}\N(\az_\eps^+)) > 0$, and so $\az\N < \az_\eps^+$. It follows $\az\N\rightarrow\az$ a.s., and therefore $\zeta\N\Rightarrow\zeta$. \qed

\bigskip

We now turn the attention to resets of instantaneous guards, proving a version of Lemma \ref{lemma:convergenceAfterReset} dealing with the discontinuities in the reset kernels under some regularity assumptions. We first recall some notation.
Let $\priorityb{i}\N$, $\priorityb{i}$ be the weight functions, with $\priorityb{i}$ continuous and $\priorityb{i}\N$ uniformly convergent to $\priorityb{i}$ on each compact set $K\subseteq \sspace{}$. Furthermore, let $\priorityb{}(\xb{}) = \sum_{i} \priorityb{i}(\xb{})$, and similarly $\priorityb{}\N(\xb{}) = \sum_{i} \priorityb{i}\N(\xb{})$. Let $R_i\N$ and $R_i$ be the reset kernels associated with the instantaneous transitions satisfying $R_i\N(\xb{}\N)\Rightarrow R_i(\xb{})$ whenever $\xb{}\N\rightarrow\xb{}$. Finally, let $h_i\N$, $h_i$ be the activation functions of guards, with $h_i$ continuous and $h_i\N$ converging uniformly to  $h_i$. The further properties that are required for the activation functions are the following:
\begin{itemize}
\item Each $h_i$ is a robust activation function, according to Definition \ref{def:activationFunction}.
\item The PDMP is robustly transversal, see Definition \ref{def:rebustlyTransversal}.
\item The set of activation functions $h_i$ enjoys the size-compatibility property, see Definition \ref{def:NcompatibleGuards}.
\item The PDMP has the robust activation property, as stated in Definition \ref{def:robustActivationProperty}.
\end{itemize}

Consider now the activation function for the PDMP, defined by $h(\xb{}) = \max\{h_1(\xb{}),\ldots,\linebreak h_m(\xb{})\}$, and let $\h{} = \{\xb{}~|~h(\xb{})=0\}$ be the activation surface of instantaneous transitions. $\h{}\N$ is defined similarly. Furthermore, let $\h{i} = \h{}\cap \{\xb{}~|~h_i(\xb{})=0\}$ be the portion of $\h{}$ in which transition $i$ is active. Call $D_i = \partial_{\h{}}\h{i}$ the boundary of $\h{i}$ in $\h{}$ and $D = \bigcup_{i=1}^m D_i$. The robust activation property implies that the probability of jumping from $D$ is zero. Furthermore, let $I_{dep}$ be the index of size-dependent activation functions, i.e. such that $h\N\neq h$, and $I_{ind}$ the index set of size-independent activation functions. The size-compatibility condition states that, for each $i\in I_{dep}$ and $\xb{}\in int_{\h{}}(\h{i})$, $h_j(\xb{}) \neq 0$ for $j\neq i$, i.e. only $h_i$ is zero. 

Finally, recall the definition of the reset kernels on $\h{}\N$ and $\h{}$:  
\[R\N(\xb{},\cdot) = \sum_{i=1}^m \vr{1}\{h_i\N(\xb{})\geq 0\}(\priorityb{i}\N(\xb{})/\priorityb{}\N(\xb{}))R_i\N(\xb{},\cdot),\] and 
\[R(\xb{},\cdot) = \sum_{i=1}^m \vr{1}\{h_i(\xb{})\geq 0\}(\priorityb{i}(\xb{})/\priorityb{}(\xb{}))R_i(\xb{},\cdot).\]

Under the previous hypothesis, we can prove the following lemma.
\begin{lemma}
\label{lemma:convergenceAfterResetInstantaneousTransitions}
Let $R\N(\yb{})$ and $R(\yb{})$ defined as before and 
let $\Yb{}\N \Rightarrow \Yb{}$, where $\Yb{}\N$, $\Yb{}$ are random elements with support in $\h{}\N$ and $\h{}$, respectively, such that $\bbP\{\Yb{}\in D\} = 0$. Then $R\N(\Yb{}\N)\Rightarrow R(\Yb{})$ and $(\Yb{}\N,R\N(\Yb{}\N))\Rightarrow(\Yb{},R(\Yb{}))$.
\end{lemma}

\proof 
Fix a bounded and uniformly continuous function $g:\sspace{}\rightarrow\bbR$. We need to prove that $|\bbE[g(R\N(\Yb{}\N))] - \bbE[g(R(\Yb{}))]|\rightarrow 0$ as $N\rightarrow\infty$. We use the same notation as in Lemma \ref{lemma:convergenceAfterReset}. The idea is to split the integrals $\integral{\sspace{}}{}{R\N g(\yb{}) P\N(\yb{})}{\yb{}}$ and $\integral{\sspace{}}{}{R g(\yb{}) P(\yb{})}{\yb{}}$ into several regions, surrounding the discontinuity points by a small probability region in such a way that the probability mass is concentrated on a continuity region, in which we can apply Lemma \ref{lemma:convergenceAfterReset}. There are some technical details that we have to work out, as $\Yb{}\N$ and $\Yb{}$ are concentrated on a manifold of  $\sspace{}$. 

Let $\delta>0$ (to be fixed afterwards) and $K\subseteq \sspace{}$ be a compact set. Call $D_\delta = \bigcup_{\xb{}\in D}B_\delta(\xb{})$ the $\delta$-neighbourhood of $D$. Clearly, $D_\delta\downarrow D$ as $\delta\downarrow 0$, and therefore $P(D_\delta)\downarrow 0$. The same holds for the closure $\overline{D_\delta}$:  $P(\overline{D_\delta})\downarrow 0$

Consider now a size-dependent activation function, $i\in I_{dep}$, and let $\h{i,\delta} = \h{i}\cap D_\delta^c$. By the size-compatibility condition, it follows that $|h_j(\xb{})| > 0$ for  each $\xb{}\in\h{i,\delta}$ and each $j\neq i$. 
In particular, $d(\xb{},\h{j})>0$ for each  $\xb{}\in\h{i,\delta}$, where the distance between a point and a set is defined in the usual way as $d(\xb{},A) = \inf_{\yb{}\in A}d(\xb{},\yb{})$. 
Now, notice that $\h{i,\delta}$ is closed and so $\h{i\,\delta}\cap K$ is compact. 
Therefore, by continuity of the distance $d$, there is a $\rho_{i,j} >0 $ such that $d(\xb{},\h{j}) > \rho_{i,j}$ 
for each $\xb{}\in \h{i\,\delta}\cap K$.\footnote{
Let $f:K\rightarrow \bbR$ be a continuous function on a compact set $K$ such that $|f(\xb{})| > 0$ for each $\xb{}\in K$. 
Then there is $\eps>0$ such that  $|f(\xb{})| \geq \eps$ for each $\xb{}\in K$. Suppose not, and choosing $\eps = 1/n$, construct a sequence $\xb{n}$ such that $f(\xb{n})\leq 1/n$ and so $f(\xb{n})\rightarrow 0$. 
By compactness of $K$, extract a convergent subsequence $\xb{n_k} \rightarrow \yb{}\in K$. Then $0 = \lim_k f(\xb{n_k}) = f(\yb{})$, a contradiction.}
Let now $\delta_i = \min_{j\neq i}\rho_{i,j}/2$, and notice that, for each $\xb{}\in \h{i,\delta}\cap K$ and $\yb{}\in B_{\delta_i}(\xb{})$, we have $d(\yb{},\h{j})> \delta_i > 0$. Let $A_{i,\delta} = \left[\bigcup_{\xb{}\in \h{i,\delta}\cap K} B_{\delta_i}(\xb{})\right]\cap D_\delta^c$, then $d(\yb{},\h{j})\geq \delta_i$ for each $\yb{}\in \overline{A_{i,\delta}}$. It follows that $h_j(\yb{}) > 0$ in $\overline{A_{i,\delta}}$, which is compact, so that we find a $\rho_i > 0$ such that $\|h_j(\yb{})\|\geq \rho_i$ for $\yb{}\in A_{i\,\delta}$ and each $j\neq i$. 

By possibly invoking uniform convergence of $h_j\N$ to $h_j$, for $j\in I_{dep}$, the property of $A_{i,\delta}$ allows us to conclude that, for $N$ large enough, $h_j\N(\yb{})\neq 0$ in $A_{i,\delta}$. 
Furthermore, by the robust activation of $h_i$, $h_i\N$ ultimately changes  sign within $A_{i,\delta}$.
 In particular, combining this with the fact that $\Yb{}\N$ is supported in $\h{}\N$ and $\Yb{}$ is supported in $\h{}$, we get that in $A_{i,\delta}$, $R\N$ coincides with $R\N_i$ and $R$ with $R_i$, hence they satisfy the continuity property $R\N(\xb{}\N)\rightarrow R(\xb{})$ as $\xb{}\N\rightarrow \xb{}$.  

We can deal similarly with size-independent activation functions $h_j$, $j\in I_{ind}$. In this case, however, we may have more than one guard robustly active in $\h{}$, so we really need to consider each possible activation profile. Let $\alpha$ be a boolean vector, $\alpha\in\{0,1\}^m$, such that $\alpha_i = 0$ for $i\in I_{dep}$. Call $J_{ind}$ the set of such vectors. Then we can define $\h{\alpha} = \h{} \cap \bigcap_{j:\alpha_j = 1} \h{j}$. In $int_{\h{}}(\h{\alpha})$, $h_i(\xb{}) \neq 0$ if and only if $\alpha_i = 0$, hence we can reason as for the size-dependent case to construct an open neighbourhood $A_{\alpha,\delta}$ of $\h{\alpha}\cap D_{\delta}^c$ in which $h_i\N(\xb{})\neq 0$ for $N$ large enough and all $\xb{}\in A_{\alpha,\delta}$. Since $h_j\N = h_j$ for $\alpha_j = 1$, and since $P\N$ and $P$ are supported in $\h{\alpha}$, when restricted to $A_{\alpha,\delta}$, we can conclude that $R\N(\xb{}\N)\rightarrow R(\xb{})$ as $\xb{}\N\rightarrow \xb{}$ in $A_{\alpha,\delta}$.

Recall the definition of $J_{ind}$, and let $J_{dep}$ be the set of boolean vectors $\alpha\in\{0,1\}^m$ equal to one only for a single $i\in I_{dep}$, and zero elsewhere, and $J = J_{ind}\cup J_{dep}$. For each compact $K$ and $\delta$, we have constructed an open set $A_{\delta} = \bigcup_{\alpha\in J} A_{\alpha,\delta}$ such that $R\N$ and $R$ behave nicely in it. 

Now, fix $\eps>0$, and, invoking the uniform tightness of $P\N$ and $P$, choose $K_\eps$ compact such that $P\N(K_\eps)\geq 1 - \eps/4 \|g\|_{\infty}$. Furthermore, pick $\delta>0$ such that $P(\overline{D_\delta})\leq \eps/4 \|g\|_{\infty}$. Then we have
\begin{eqnarray*}
|\bbE[g(R\N(\Yb{}\N))]  & - & \bbE[g(R(\Yb{}))]|   \leq 
\underbrace{\left|\integral{A_{\delta}}{}{R\N g(\yb{})P\N(\yb{})}{\yb{}} -  
\integral{A_{\delta}}{}{R g(\yb{})P(\yb{})}{\yb{}}\right|}_{(a)}\\
& + &  \underbrace{\left|\integral{K_\eps\setminus (A_{\delta}\cup D_\delta)}{}{R\N g(\yb{})P\N(\yb{})}{\yb{}} -  
\integral{K_\eps\setminus (A_{\delta}\cup D_\delta)}{}{R g(\yb{})P(\yb{})}{\yb{}}\right|}_{(b)}\\
& + & \underbrace{\left|\integral{D_{\delta}}{}{R\N g(\yb{})P\N(\yb{})}{\yb{}} -  
\integral{D_{\delta}}{}{R g(\yb{})P(\yb{})}{\yb{}}\right|}_{(c)}\\
& + & \underbrace{\left|\integral{K_\eps^c}{}{R\N g(\yb{})P\N(\yb{})}{\yb{}} -  
\integral{K_\eps^c}{}{R g(\yb{})P(\yb{})}{\yb{}}\right|}_{(d)}
\end{eqnarray*}
Now, term (a) goes to zero  invoking Lemma \ref{lemma:convergenceAfterReset}, given the continuity of resets in $A_\delta$. To deal with term (b), notice that $K_\eps\setminus (A_{\delta}\cup D_\delta)$ is closed and $P(K_\eps\setminus (A_{\delta}\cup D_\delta)) = 0$, so that $\limsup_{N} P\N(K_\eps\setminus (A_{\delta}\cup D_\delta)) = 0$. Term (c) is dealt with by observing that $\limsup_N P\N(\overline{D_{\delta}}) \leq P(\overline{D_{\delta}}) \leq  \eps/4 \|g\|_{\infty}$, and so $\limsup_N P\N(D_{\delta})\leq  \eps/4 \|g\|_{\infty}$. Therefore (c) is less than $\eps/2$. Finally, (d) is less than $\eps/2$ by the choice of $K_\eps$. It follows that 
\[\limsup_{N\rightarrow \infty}|\bbE[g(R\N(\Yb{}\N))] - \bbE[g(R(\Yb{}))]| \leq \eps,\]
which implies $|\bbE[g(R\N(\Yb{}\N))] - \bbE[g(R(\Yb{}))]| \rightarrow 0$. Proof of the second statement of the theorem can be copied verbatim from Lemma \ref{lemma:convergenceAfterReset}. \qed


\bigskip


We now give the proof of Theorem \ref{th:hybridInstantaneous}.

\begin{apptheorem}[\ref{th:hybridInstantaneous}]
Let $(\calA,\size{N})$ be a sequence of population-\sCCP\ models for increasing system size $\size{N}\rightarrow\infty$, as $N\rightarrow\infty$, with variables partitioned into $\X{} = (\X{d},\X{c},\X{e})$, with discrete stochastic actions satisfying scaling \ref{scaling:discreteStochastic}, instantaneous actions satisfying scaling \ref{scaling:discreteInstantaneous}, and continuous actions satisfying scaling \ref{scaling:continuousHybrid}.
Let $\Xb{}\N(t)$ be the associated sequence of normalized CTMC and $\xb{}(t)$ be the limit PDMP associated with the  normalized limit TDSHA $\norm{\calT}(\calA)$.

If $\xb{0}\N \Rightarrow \xb{0}$ (weakly) and the PDMP is \emph{non-Zeno}, \emph{robustly transversal}, has the \emph{robust activation property} and it is \emph{size-compatible}, then $\Xb{}\N(t)$ converges weakly to $\xb{}(t)$,  $\Xb{}\N \Rightarrow \xb{}$, as random elements in the space of cadlag function with the Skorohod metric.
\end{apptheorem}

\proof The argument closely follows the proof of Theorem \ref{th:hybridBasic}. The only difference is the definition of jump times $T_{i}\N$ and $T_{i}$. In this case, in fact, these are defined as the minimum of stochastic jump times $\T{i}\N$ and $\T{i}$ (conditional on having observed $i-1$ jumps) and instantaneous jump times $\zeta_i\N$ and $\zeta_i$ (conditional on having observed $i-1$ jumps). Now, as $\T{i}\N\Rightarrow \T{i}$ by Lemma \ref{lemma:convergenceJumpTimes} and $\zeta_i\N\Rightarrow\zeta_i$ by Lemma \ref{lemma:convergenceExitTimes}, by the continuous mapping theorem it follows that $T_i\N=\min\{\T{i}\N,\zeta_i\N\}\Rightarrow\min\{\T{i},\zeta_i\}=T_i$.

This allows us to extend  Corollary \ref{cor:weakKurtz}, by replacing $\T{}\N\Rightarrow\T{}$ with $T\N\Rightarrow T$ in point 2, and then showing $\Xb{}\N(T\N) \Rightarrow \xb{}(T)$ (use Proposition \ref{prop:app:convergenceValuesAtJumpTimes} conditional on $T<\infty$). 
As for convergence of the state after the reset, 
notice that $\Yb{}\N=\Xb{}\N(T\N)$ and $\yb{}=\xb{}(T)$ satisfy the conditions of Lemma \ref{lemma:convergenceAfterResetInstantaneousTransitions}. 
Moreover, we have
$R\N(\Yb{}\N) = R_s\N(\Yb{}\N)\ind{\T{i}\N < \zeta_i\N} + R_i\N(\Yb{}\N)\ind{\T{i}\N > \zeta_i\N}$, and $R(\yb{}) = R_s(\yb{})\ind{\T{i} < \zeta_i} + R_i(\yb{})\ind{\T{i} < \zeta_i}$, where
$R\N_s(\Yb{}\N)$ and $R_s(\yb{})$ are the resets kernels for  stochastic jumps (constructed from instantaneous transitions) and $R\N_i(\Yb{}\N)$ and $R_i(\yb{})$  are the resets kernels for the instantaneous jumps. 
\\
Both satisfy $R\N_s(\Yb{}\N)\Rightarrow R_s(\yb{})$ and $R\N_i(\Yb{}\N)\Rightarrow R_i(\yb{})$, as $\Yb{}\N\rightarrow\yb{}$. 
Now, as $\bbP\{\zeta_i = \T{i}\}=0$, we can apply the continuous mapping theorem first to the indicator functions $\ind{\T{}<\zeta}$ and $\ind{\T{}>\zeta}$, to show that $\ind{\T{i}\N<\zeta_i\N}\Rightarrow \ind{\T{i}<\zeta_i}$ and $\ind{\T{i}\N>\zeta_i\N}\Rightarrow \ind{\T{i}>\zeta_i}$, and then to the definition of $R\N$ and $R$, to show that $R\N(\Yb{}\N)\Rightarrow R(\yb{})$. 
\\
Reasoning similarly to Lemma \ref{lemma:convergenceAfterReset}, we have then proved the equivalent of point 4 and 5 of  Corollary \ref{cor:weakKurtz}.  Then, the proof of the theorem works as in Theorem \ref{th:hybridBasic}. \qed


\subsection{Guards in discrete stochastic transitions}
\label{app:proof:discreteStocGuards}

We turn now to prove convergence in the presence of guarded discrete stochastic transitions. As discussed in the paper, there are two main issues to deal in this case, caused by the introduction of discontinuities in the rate functions. The first is the convergence of jump times, the second is the convergence of states after the resets. Both points require an additional regularity property of the PDMP, namely the \emph{robust compatibility} with respect to guards of discrete stochastic transitions.

We start by showing convergence of exit times.
Recall that we have $m$, say, discrete stochastic transitions, with rate functions $\srateb{i}\N$, $\srateb{i}$, and activation functions $h_i\N$ and $h_i$ associated with guards, such that $\srateb{i}$ and $h_i$ are continuous, and $\srateb{i}\N$, $h_i\N$ converge uniformly on compact sets to $\srateb{i}$ and $h_i$, respectively. Let $\srateb{}\N(\xb{}) = \sum_{i=1}^m \ind{h_i\N(\xb{}) \geq 0}\srateb{i}\N(\xb{})$ and 
$\srateb{}(\xb{}) = \sum_{i=1}^m \ind{h_i(\xb{}) \geq 0}\srateb{i}(\xb{})$. 
Furthermore, we consider the following discontinuity surfaces: $\h{i}\N = \{h_i\N(\xb{}) = 0\}$ and $\h{i} = \{h_i(\xb{}) = 0\}$, $\h{}\N = \bigcup_{i=1}^m \h{i}\N$, and $\h{} = \bigcup_{i=1}^m \h{i}$.

We then can prove the following 
\begin{lemma}
\label{lemma:app:discontiuousJumpTimes}
Let $\Xb{}\N\Rightarrow \Xb{}$, as random variables in the space of cadlag functions, with $\Xb{}$ a.s. continuous and such that, with probability 1, $\{t\in \bbR^+~|~\Xb{}(t)\in\h{}\}$ has Lebesgue measure 0. Let $\tau\N$, $\tau$ be the jump times associated with rates $\srateb{}\N$ and $\srateb{}$ defined above. Then $\tau\N \Rightarrow \tau$, as $N\rightarrow\infty$.
\end{lemma}
\proof We first prove that, if $\xb{}\N\rightarrow\xb{}$ as elements in the space of cadlag functions, $\xb{}$ is continuous, and $\srateb{}(\xb{}(t))$  is almost everywhere continuous, then $\Lambda\N(T) = \integral{0}{T}{\srateb{}\N(\xb{}\N(t))}{t} \rightarrow \integral{0}{T}{\srateb{}(\xb{}(t))}{t} = \Lambda(T)$ for every $T>0$. In fact, for each continuity point $\yb{}$ of $\srateb{}$ we have that $\srateb{}\N(\yb{}\N)\rightarrow \srateb{}(\yb{})$ as $\yb{}\N\rightarrow\yb{}$. It follows that, for each $t>0$ such that $\srateb{}(\xb{}(t))$ is continuous, then 
$\srateb{}\N(\xb{}\N(t))\rightarrow \srateb{}(\xb{}(t))$, as $\xb{}\N(t)\rightarrow \xb{}(t)$ by continuity of $\xb{}$. Therefore, $\srateb{}\N(\xb{}\N)\rightarrow \srateb{}(\xb{})$ pointwise almost everywhere in $[0,T]$. 
Furthermore, by continuity of $\srateb{i}$, we have $\srateb{i}\N(\xb{}\N)\rightarrow\srateb{i}(\xb{})$, hence $\{\srateb{i}\N(\xb{}\N),\srateb{i}(\xb{})\}$ is relatively compact in the space of cadlag functions, and so bounded uniformly. It means that there is $M_i>0$ such that $\|\srateb{i}\N(\xb{}\N(t))\| \leq M_i$ and $\|\srateb{i}(\xb{}(t))\| \leq M_i$. 
But as $\srateb{}\N(\xb{}\N(t))\leq \sum_i \srateb{i}\N(\xb{}\N(t))$ and $\srateb{}(\xb{}(t))\leq \sum_i \srateb{i}(\xb{}(t))$, it follows that  $\srateb{}\N(\xb{}\N),\srateb{}(\xb{})$ are bounded by $\sum_i M_i$. Hence, by the bounded convergence theorem, $\integral{0}{T}{\srateb{}\N(\xb{}\N(t))}{t} \rightarrow \integral{0}{T}{\srateb{}(\xb{}(t))}{t}$ for every $T>0$.

Now, the statement follows by applying the Skorohod representation theorem, as in Lemma \ref{lemma:convergenceJumpTimes}. Let $\asrep{\X{}}\N$, $\asrep{\X{}}$ be representations of $\Xb{}\N$, $\Xb{}$ on a probability space $(\Omega,\calA,\bbP)$ such that $\asrep{\X{}}\N\rightarrow\asrep{\X{}}$ almost surely. 
Hence, due to the hypothesis, for $\omega$ in a subset of probability 1 of $\Omega$, we have that  $\asrep{\X{}}\N(\omega)\rightarrow\asrep{\X{}}(\omega)$,   $\asrep{\X{}}(\omega)$ is continuous, and $\srateb{}(\asrep{\X{}}(\omega))$ is almost everywhere continuous. Then we can apply the previous argument to $\asrep{\X{}}\N(\omega)$, $\asrep{\X{}}(\omega)$, and conclude that $\Lambda\N(\omega,T)$ converges pointwise to $\Lambda(\omega,T)$ for each $T$, from which we get a.s. convergence of the  representation of jump times $\asrep{\T{}\N}$ and $\asrep{\T{}}$. Hence $\T{}\N\Rightarrow\T{}$, as desired. \qed

\bigskip

We turn now our attention to reset kernels. We will extend Lemma \ref{lemma:convergenceAfterReset} to deal with the discontinuities in the limit kernel using the hypothesis that there is zero probability of being in a discontinuous state when we jump. 
Recall the definition of $\srateb{}\N$, $\srateb{}$, $h_i\N$ and $h_i$, and further let $R_i\N$, $R_i$ be the reset kernels, satisfying  $R_i\N(\xb{}\N)\Rightarrow R_i(\xb{})$, for each $\xb{}\N\rightarrow\xb{}$. 
Then the full reset kernels are
$R\N(\xb{},\cdot) = \sum_{i=1}^m \ind{h_i\N(\xb{})\geq 0}(\srateb{i}\N(\xb{})/\srateb{}\N(\xb{}))R_i\N(\xb{},\cdot)$, and 
$R(\xb{},\cdot) = \sum_{i=1}^m \ind{h_i(\xb{})\geq 0}(\srateb{i}(\xb{})/\srateb{}(\xb{}))R_i(\xb{},\cdot)$.

Equipped with these definitions, we can prove the following lemma.
\begin{lemma}
\label{lemma:convergenceAfterDiscontinuousReset}
Let $R\N(\yb{})$ and $R(\yb{})$ defined as before and 
let $\Yb{}\N \Rightarrow \Yb{}$, where $\Yb{}\N$, $\Yb{}$ are random elements in $\sspace{}$ such that $\bbP\{\Yb{}\in \calH\} = 0$. Then $R\N(\Yb{}\N)\Rightarrow R(\Yb{})$ and $(\Yb{}\N,R\N(\Yb{}\N))\Rightarrow(\Yb{},R(\Yb{}))$.
\end{lemma}

\proof The proof is based Lemma \ref{lemma:convergenceAfterReset}, with additional arguments taking care of the discontinuities in $R\N$ and $R$. Fix $\eps >0$ and a bounded and uniformly continuous function $g:\sspace{}\rightarrow\bbR$.  By the same argument of Lemma \ref{lemma:convergenceAfterReset}, $\{\Yb{}\N,\Yb{}\}$ is uniformly tight, and so there is a compact set $K_\eps$ such that $P\N(K_\eps)\geq 1- \eps/4\|g\|_{\infty}$ for each $N$, and $P(K_\eps)\geq 1- \eps/4\|g\|_{\infty}$. 
Furthermore, for $\delta\geq 0$, let $\h{i,\delta}$ be the closed $\delta$-neighbourhood of $\h{i}$, defined by $\h{i,\delta} = \overline{\bigcup_{\xb{}\in\h{i}}B_{\delta}(\xb{})}$, where $B_{\delta}(\xb{})$ is the ball of radius $\delta$ centred in $\xb{}$. Let also $\h{\delta}=\bigcup_i\h{i,\delta}$. Clearly $\h{\delta}\downarrow\h{}$ for  $\delta\downarrow 0$, and so $P(\h{\delta})\downarrow 0$. Choose $\delta$ such that $P(\h{\delta})<\eps/4\|g\|_{\infty}$.
We have that
\begin{eqnarray*}
|\bbE[g(R\N(\Yb{}\N))] - \bbE[g(R(\Yb{}))]|  & \leq &
\underbrace{\left|\integral{K_\eps\cap\h{\delta}^c}{}{R\N g(\yb{})P\N(\yb{})}{\yb{}} -  
\integral{K_\eps\cap\h{\delta}^c}{}{R g(\yb{})P(\yb{})}{\yb{}}\right|}_{(a)}\\
& + & \underbrace{\left|\integral{\h{\delta}}{}{R\N g(\yb{})P\N(\yb{})}{\yb{}} -  
\integral{\h{\delta}}{}{R g(\yb{})P(\yb{})}{\yb{}}\right|}_{(b)}\\
& + & \underbrace{\left|\integral{K_\eps^c}{}{R\N g(\yb{})P\N(\yb{})}{\yb{}} -  
\integral{K_\eps^c}{}{R g(\yb{})P(\yb{})}{\yb{}}\right|}_{(c)}
\end{eqnarray*}

Now,the reset kernels $R\N$ and $R$ in $K_\eps\cap\h{\delta}^c$ satisfy the continuity property $R\N(\xb{}\N)\Rightarrow R(\xb{})$ as $\xb{}\N\rightarrow\xb{}$, for $\xb{}\N,\xb{}\in K_\eps\cap\h{\delta}^c$. 
This follows because, by uniform convergence of $h_i\N$ to $h_i$ on the compact set $K_\eps$
and the fact that $K_\eps\cap\h{\delta}^c$ does not contain any discontinuity surface, if $\xb{}\N\rightarrow \xb{}$, then $\xb{}\N$ will ultimately satisfy the same  guards as $\xb{}$, i.e. $\ind{h_{i}\N(\xb{}\N)\geq 0} \rightarrow \ind{h_{i}(\xb{})\geq 0}$. Then convergence of the reset kernels follows as in the unguarded case. This means that we can apply Lemma \ref{lemma:convergenceAfterReset} and conclude that term (a) goes to zero.

As for term (b), notice that $\h{\delta}$ is closed, hence $\limsup_{N\rightarrow\infty} P\N(\h{\delta})\leq P(\h{\delta})\leq \eps/4\|g\|_{\infty}$ by the Portmanteau theorem. Finally, term (c) is less than $\eps/2$ by the choice of $K_\eps$ and the fact that $Rg$ and $R\N g$ are both bounded by $\|g\|_{\infty}$. Hence we have that $\limsup_{N\rightarrow\infty} |\bbE[g(R\N(\Yb{}\N))] - \bbE[g(R(\Yb{}))]| \leq \eps $, 
which implies convergence to zero by the arbitrariness of $\eps$. This proves $R\N(\Yb{}\N)\Rightarrow R(\Yb{})$. The second part of the statement, instead, follows as in Lemma \ref{lemma:convergenceAfterReset}. \qed

\bigskip

We are finally ready to prove proposition \ref{prop:hybridDiscontinuousJumps}.

\begin{appproposition} [\ref{prop:hybridDiscontinuousJumps}]
Let $(\calA,\size{N})$ be a sequence of population-\sCCP\ models for increasing systems size $\size{N}\rightarrow\infty$, as $N\rightarrow\infty$, with variables partitioned into $\X{} = (\X{d},\X{c},\X{e})$, with discrete stochastic actions satisfying either scaling \ref{scaling:discreteStochastic} or scaling \ref{scaling:DiscreteStochasticGuarded}, no instantaneous actions, and continuous actions satisfying scaling \ref{scaling:continuousHybrid}.
Let $\Xb{}\N(t)$ be the associated sequence of normalized CTMC and $\xb{}(t)$ be the limit PDMP associated with the  normalized limit TDSHA $\norm{\calT}(\calA)$.

If $\xb{0}\N \Rightarrow \xb{0}$ (weakly) and the PDMP is \emph{non-Zeno} and \emph{robustly compatible}, then $\Xb{}\N(t)$ converges weakly to $\xb{}(t)$,  $\Xb{}\N \Rightarrow \xb{}$, as random elements in the space of cadlag function with the Skorohod metric. 
\end{appproposition} 

\proof The proof proceeds essentially as that of Theorem \ref{th:hybridBasic}. The only difference is that we have to replace Lemma \ref{lemma:convergenceJumpTimes} with Lemma \ref{lemma:app:discontiuousJumpTimes} and Lemma \ref{lemma:convergenceAfterReset} with Lemma \ref{lemma:convergenceAfterDiscontinuousReset} in corollary \ref{cor:weakKurtz} and in the proof of Theorem \ref{th:hybridBasic}. To do this, we just need to show that the robust compatibility of the PDMP guarantees the satisfaction of the hypothesis of the two lemmas. This is trivial for Lemma \ref{lemma:app:discontiuousJumpTimes},  as robust compatibility is an explicit hypothesis. The condition of Lemma \ref{lemma:convergenceAfterDiscontinuousReset}, instead, holds because robust compatibility of the PDMP $\xb{}$ and the absolute continuity of the exponential distribution with respect to the Lebesgue measure imply that  the event $\{\xb{}(\T{})\in\h{}\}$ has probability zero. \qed

\end{document}